% ...
\documentclass[amsmath,preprintnumbers,showpacs,aps,nofootinbib,tightenlines,floatfix,superscriptaddress]{revtex4-1}
\usepackage{amsfonts}
\usepackage{cancel}
\usepackage{graphicx}
\usepackage{slashed}
\usepackage{xcolor}
\usepackage{mathtools}
\usepackage[hypertexnames=false]{hyperref}
\hypersetup{
    colorlinks=true,       % false: boxed links; true: colored links
    linkcolor=blue,          % color of internal links
    citecolor=blue,        % color of links to bibliography
    filecolor=blue,      % color of file links
    urlcolor=blue           % color of external links
}

\newcommand{\nnbar}{\ensuremath{n\text{-}\overline{n}}}

\newcommand{\ket}[1]{\left| #1 \right>} % for Dirac bras
\newcommand{\bra}[1]{\left< #1 \right|} % for Dirac kets
\newcommand{\mbraket}[3]{\left< #1 \vphantom{#2#3} \right|
 #2 \left| #3 \vphantom{#1#2} \right>} % for Dirac matrix elements
\newcommand{\avg}[1]{\left< #1 \right>}

\newcommand{\lbsm}{\ensuremath{\Lambda_\text{BSM}}}
\newcommand{\lqcd}{\ensuremath{\Lambda_\text{QCD}}}
\newcommand{\MSbar}{\ensuremath{\overline{\text{MS}}}}
\newcommand{\Tr}{\mathrm{Tr}}
\newcommand{\EucConv}{{\mathcal Euc}}

\newcommand{\MinkConv}{{{\mathcal M}}}

\long\def\/*#1*/{}
\definecolor{red}{rgb}{1.0, 0, 0}

\newcommand{\mcC}{{\mathcal C}}
\newcommand{\mcP}{{\mathcal P}}
\newcommand{\mcT}{{\mathcal T}}

\newcommand{\mcO}{{\mathcal O}}
\newcommand{\mcM}{{\mathcal M}}

\newcommand{\mcL}{{\mathcal L}}
\newcommand{\neut}[2]{{n^{(#1)}_{#2 1/2}}}
\newcommand{\nbar}[2]{{\bar{n}^{(#1)}_{#2 1/2}}}
\newcommand{\tsep}{t_\text{sep}}
\newcommand{\np}{\phantom{+}}

\newcommand{\wtC}{\widetilde{C}}

\begin{document}
\title{Lattice QCD determination of neutron-antineutron matrix elements with physical quark masses} %$m_\pi=135$ MeV.}
%\title{Neutron-antineutron matrix elements from lattice QCD with physical quark masses} %$m_\pi=135$ MeV.}
%\title{$\nnbar$ Long Paper}

\author{Enrico Rinaldi}
\email{erinaldi@bnl.gov}
\affiliation{RIKEN BNL Research Center, Brookhaven National Laboratory, Upton, NY 11973, USA}
\affiliation{Nuclear Science Division, Lawrence Berkeley National Laboratory, Berkeley, CA 94720, USA}
\author{Sergey Syritsyn}
\email{sergey.syritsyn@stonybrook.edu}
\affiliation{RIKEN BNL Research Center, Brookhaven National Laboratory, Upton, NY 11973, USA}
\affiliation{Department of Physics and Astronomy, Stony Brook University, Stony Brook, NY 11794, USA}
\author{Michael L. Wagman}
\email{mlwagman@mit.edu}
\affiliation{Center for Theoretical Physics, Massachusetts Institute of Technology, Cambridge, MA 02139, USA}
\author{Michael I. Buchoff}
\affiliation{Lawrence Livermore National Laboratory, Livermore, California 94550, USA}
\author{Chris Schroeder}
\affiliation{Lawrence Livermore National Laboratory, Livermore, California 94550, USA}
\author{Joseph Wasem}
\affiliation{Lawrence Livermore National Laboratory, Livermore, California 94550, USA}

%%%%%%%%%%%%%%%%%%%%%%%%%%%%%%%%%%%%%%%%%%%%%%%%%%%%%%%%%%%%%%%%%%%%%%%%%%%%%%
\begin{abstract}
Matrix elements of six-quark operators are needed to extract new physics constraints from
experimental searches for neutron-antineutron oscillations.
This work presents in detail the first lattice quantum chromodynamics calculations of the
necessary neutron-antineutron transition matrix elements including calculation methods and
discussions of systematic uncertainties.
Implications of isospin and chiral symmetry on the matrix elements, power counting in the
isospin limit, and renormalization of a chiral basis of six-quark operators are discussed.
Calculations are performed with a chiral-symmetric discretization of the quark action and
physical light quark masses in order to avoid the need for chiral extrapolation.
Non-perturbative renormalization is performed, including a study of lattice cutoff effects.
Excited-state effects are studied using two nucleon operators and multiple values of
source-sink separation.
Results for the dominant matrix elements are found to be significantly larger compared to
previous results from the MIT bag model.
Future calculations are needed to fully account for systematic uncertainties associated with
discretization and finite-volume effects but are not expected to significantly affect this
conclusion.

\end{abstract}

\preprint{MIT-CTP/5099, LLNL-JRNL-766337, RBRC-1305}

\maketitle

%%%%%%%%%%%%%%%%%%%%%%%%%%%%%%%%%%%%%%%%%%%%%%%%%%%%%%%%%%%%%%%%%%%%%%%%%%%%%%
%%%%%%%%%%%%%%%%%%%%%%%%%%%%%%%%%%%%%%%%%%%%%%%%%%%%%%%%%%%%%%%%%%%%%%%%%%%%%%
\section{Introduction}
\label{sec:intro}

In the contemporary theory of particles and fields, there is no fundamental reason for baryon 
number $B$ to be conserved.
Quantum effects in the Standard Model (SM) can lead to $B$ violation, and at temperatures above
the electroweak phase transition sphaleron processes can efficiently convert baryons into
antileptons while preserving $(B-L)$, where $L$ is lepton number.
Low-temperature $B$-violating effects have not been observed experimentally, and their existence
would have significant implications for the stability of nuclear matter.
However, the observed baryon-antibaryon asymmetry of the universe cannot be explained within the
SM, which fulfills Sakharov's conditions for baryogenesis~\cite{Sakharov:1967dj} but does not
contain enough baryon number and CP violation to reproduce the observed baryon asymmetry of the
universe~\cite{Cohen:1993nk,Rubakov:1996vz,Dine:2003ax,Canetti:2012zc}.
Moreover, while $(B-L)$ symmetry is preserved in the SM, it likely has to be violated in its
extensions (BSM theories) aimed at explaining baryogenesis, since electroweak sphaleron transitions
would otherwise ``wash out'' any net baryon number generated by $(B-L)$-conserving interactions
in the early universe.

Baryon number violation might be experimentally observed in proton decays~\cite{Miura:2016krn} 
or neutron-antineutron
oscillations~\cite{BaldoCeolin:1994jz,Chung:2002fx,Abe:2011ky,Bergevin:2011phd}.
The implications of these two hypothetical processes are fundamentally different: proton
decay changes baryon number by $|\Delta B|=1$ unit and involves (anti)leptons, while
neutron-antineutron oscillations change baryon number by $|\Delta B|=2$ units and do not involve leptons.
Proton decay, even if observed, does not necessarily violate $(B-L)$ and may be insufficient to
explain baryogenesis.

Despite decades of searches, neither process has been observed, constraining the strength of
$B$-violating interactions.
In particular models of baryogenesis, this may require higher level of $CP$ violation, which is
in turn constrained by searches for the electric dipole moments of neutrons, nuclei, and atoms.
However, excluding theories of baryogenesis using results from these experiments requires knowledge of nucleon matrix
elements of $B$- and $CP$-violating effective interactions expressed in terms of fundamental
fields, quarks and gluons.
For neutron-antineutron transitions, these calculations have previously been performed using
nucleon models~\cite{Rao:1982gt}.
Modern lattice QCD methods permit model-independent calculation of these matrix elements.
This paper reports the first completely nonperturbative calculation of the neutron-antineutron
transition matrix elements computed in lattice QCD with physical quark masses and chiral
symmetry.
In particular, we find that lattice QCD calculations result in substantially larger $\nnbar$ 
matrix elements compared to nucleon model calculations.
Our findings imply that $\nnbar$ oscillation experiments should observe 1-2 orders of magnitude
more oscillation events than was previously expected for the same BSM physics parameters.

This paper describes in detail our methodology for computing neutron-antineutron matrix elements
of operators changing baryon number by $|\Delta B|=2$ units, which have already been reported in a 
short publication~\cite{Rinaldi:2018osy}. 
In particular, the operator definitions, symmetry properties of their matrix elements, and their
impact on phenomenology within $SU(2)_L\times U(1)$-symmetric extensions are discussed in
Sec.~\ref{sec:op}.
The setup for our calculation of these matrix elements on a lattice is described in
Sec.~\ref{sec:lattice_setup}.
Extraction of ground-state matrix elements from lattice correlation functions and analysis of
potential excited state contaminations are performed in Sec.~\ref{sec:analysis}.
Nonperturbative renormalization and matching to the $\MSbar$ scheme are described in
Sec.~\ref{sec:npr}.
The final results for $\nnbar$ matrix elements and their uncertainties are provided in
Sec.~\ref{sec:results}.
In Section~\ref{sec:conclusion}, we discuss briefly the impact of our results in light of other 
potential sources of systematic uncertainties that are not controlled in our present
calculation.

%%%%%%%%%%%%%%%%%%%%%%%%%%%%%%%%%%%%%%%%%%%%%%%%%%%%%%%%%%%%%%%%%%%%%%%%%%%%%%
%%%%%%%%%%%%%%%%%%%%%%%%%%%%%%%%%%%%%%%%%%%%%%%%%%%%%%%%%%%%%%%%%%%%%%%%%%%%%%
\section{Effective $\nnbar$ interactions}
\label{sec:op}
% vim: tw=100 sw=2 sts=2 et

%%%%%%%%%%%%%%%%%%%%%%%%%%%%%%%%%%%%%%%%%%%%%%%%%%%%%%%%%%%%%%%%%%%%%%%%%%%%%%
\subsection{Chiral basis of $\nnbar$ operators}
\label{sec:ops}

%\textbf{need to relate $Q_1\ldots Q_7$ operators to the previously used defs by  Rao\&Shrock and others}

A complete basis of color-singlet, electrically-neutral six-quark operators 
with $uudddd$ flavor structure can be constructed from operators of the
form~\cite{Chang:1980ey,Kuo:1980ew,Rao:1982gt,Rao:1983sd,Caswell:1982qs}
\begin{equation}
\label{eqn:op_nnbar_orig}
  \begin{split}
  \mcO^1_{\chi_1 \chi_2 \chi_3} 
    &= (u_i^T CP_{\chi_1} u_j) (d_k^T CP_{\chi_2} d_l) (d_m^T CP_{\chi_3} d_n)
        T_{\{ij\}\{kl\}\{mn\}}^\text{(symm)}\,, \\
  \mcO^2_{\chi_1 \chi_2 \chi_3} 
    &= (u_i^T CP_{\chi_1} d_j) (u_k^T CP_{\chi_2} d_l) (d_m^T CP_{\chi_3} d_n)
        T_{\{ij\}\{kl\}\{mn\}}^\text{(symm)}\,, \\
  \mcO^3_{\chi_1 \chi_2 \chi_3} 
    &= (u_i^T CP_{\chi_1} d_j) (u_k^T CP_{\chi_2} d_l) (d_m^T CP_{\chi_3} d_n)
        T_{[ij][kl]\{mn\}}^\text{(asym)}\,
  \end{split}
\end{equation}
where quark spinor indices are implicitly contracted in the parentheses,
the $P_{L,R} = \frac{1}{2}(1 \mp \gamma_5)$ are chiral projectors,
and the quark color tensors $T$ are 
\begin{align}
\label{eqn:coltensor_symm}
T^\text{(symm)}_{\{ij\}\{kl\}\{mn\}} 
  &= \varepsilon_{ikm}\varepsilon_{jln} + \varepsilon_{jkm}\varepsilon_{iln}
   + \varepsilon_{ilm}\varepsilon_{jkn} + \varepsilon_{jlm}\varepsilon_{ikn}
  = T^{S_1 S_2 S_3} \,,
\\
\label{eqn:coltensor_asym}
T^\text{(asym)}_{[ij][kl]\{mn\}} 
  &= \varepsilon_{ijm}\varepsilon_{kln} + \varepsilon_{ijn}\varepsilon_{klm}
  = T^{A_1 A_2 S_3}\,,
\end{align}
with $S_i$, $A_i$ standing for the symmetrized and antisymmetrized pairs of color indices, 
respectively.
These operators are identical in Euclidean and Minkowski spaces with the charge-conjugation
spin matrix $C$,\footnote{
  To avoid confusion, throughout the paper we use Euclidean $\gamma$-matrices 
  $(\vec\gamma, \gamma_4)_\EucConv=
  (\vec\gamma, \gamma_4)_\EucConv^\dag = (-i\vec\gamma,\gamma_0)_\MinkConv$
  satisfying $\gamma_\mu^\dag = \gamma_\mu$.
}
\begin{equation}
\label{eqn:conjmatr}
C=\gamma_2\gamma_4 = C^* = -C^T = -C^\dag\,,
\end{equation} 
that satisfies the usual condition $C\gamma_\mu C^\dag = - \gamma_\mu^T $.
Operators involving vector diquarks $(q^T C P_\chi \gamma_\mu q)$ or tensor diquarks $(q C P_\chi
\sigma_{\mu\nu} q)$ are redundant and can be related to linear combinations of the operators in
Eq.~(\ref{eqn:op_nnbar_orig}) by spin Fierz relations.
The two choices of chirality for each $(q C P_\chi q)$ diquark above in $\mcO^{1,2,3}$
provide an overcomplete basis of 18 operators.
Fierz relations $\mcO^2_{\chi\chi\chi^\prime} - \mcO^1_{\chi\chi\chi^\prime} =
3\mcO^3_{\chi\chi\chi^\prime}$ reduce the number of independent operators to 14.\footnote{
  These Fierz relations are valid in four spacetime dimensions but are violated in dimensional
  regularization at two-loop order~\cite{Buchoff:2015qwa}.
  The $\MSbar$ scheme defined in Ref.~\cite{Buchoff:2015qwa} includes evanescent operator counterterms
  that ensure that renormalized matrix elements obey these Fierz relations.
  Provided that matching between BSM theory and SM effective operators is consistently performed in
  this $\MSbar$ scheme or is performed at a high enough scale that one-loop QCD corrections are
  negligible, these Fierz relations can be assumed for $\MSbar$ renormalized matrix elements.
}

All 14 independent effective six-quarks operators are electrically neutral and change the baryon
number by $\Delta B = -2$ units.
However, they are not independent under isospin symmetry transformations.
The electroweak (EW) symmetry $SU(2)_L \times U(1)_Y$ requires that all interactions are
$SU(2)_L$-singlet, which may be achieved with with additional factors of the Higgs field 
(see Sec.\ref{sec:eft}).
Furthermore, since the chiral symmetry $SU(2)_L\otimes SU(2)_R$ is preserved exactly in the 
massless perturbation theory, and preserved with good precision on a lattice with chiral fermions,
it is more convenient to use a basis made of operators having definite values of chiral $L,R$-isospin.

The operators in Eq.~(\ref{eqn:op_nnbar_orig}) are built from color-symmetric ($\mathbf{6}_c$) 
and antisymmetric ($\mathbf{\bar{3}}_c$) chiral diquarks, which can be denoted as
\begin{equation}
\label{eqn:diquarks_chiral}
(q_1^T C q_2)^{S,A}_\chi = (q_1^T C P_\chi q_2)^{S,A} = \pm (q_2^T C P_\chi q_1)^{S,A}
\end{equation}
where $q_{1,2}=u$ or $d$ and the relative signs upon quark permutation come from their
anticommutation, $C^T=-C$, and color (anti)symmetry.
Using the isospin doublet $\psi=(u,d)$ and its conjugate $\tilde\psi=(\psi^T C\,i\tau^2)$, 
the chiral isoscalar and isovector diquarks can be written as
\begin{equation}
(\tilde\psi \psi)^A_\chi\,,
\quad
(\tilde\psi \tau^a \psi)^S_\chi\,,
\end{equation}
where $\tau^a$ are the Pauli matrices, $[\tau^a,\tau^b]=2i\epsilon^{abc}\tau^c$.
The details of isospin classification were given in Ref.~\cite{Buchoff:2015qwa}, and here we list
only the chiral-basis operators and their relation to the conventional
basis~(\ref{eqn:op_nnbar_orig}).
All the $SU(2)_L$-singlet operators can be constructed from some $R$-diquarks and $L$-isoscalar
diquarks, resulting in three operators belonging to the $(\mathbf{1}_L,\mathbf{3}_R)$ 
irreducible representation of the chiral isospin,%\footnote{
%  Although operators $Q_{1,2,3}$ have the same chiral isospin, they do not mix in a 
%  chirally-symmetric theory because they have different number of $L,R$-quarks.
%  \textbf{what about the $U(1)_A$ anomaly?}
%},
\begin{equation}
\label{eq:Q1-3def}
\begin{split}
Q_1 &=(\tilde\psi \psi)^{A_1}_R \, (\tilde\psi \psi)^{A_2}_R \, 
      (\tilde\psi \tau^+\psi)^{S_3}_R \, T^{A_1 A_2 S_3}
    = -4\mcO^3_{RRR} \,, \\
Q_2 &= (\tilde\psi \psi)^{A_1}_L \, (\tilde\psi \psi)^{A_2}_R \, 
      (\tilde\psi \tau^+\psi)^{S_3}_R \, T^{A_1 A_2 S_3} 
    = -4\mcO^3_{LRR} \,, \\
Q_3 &= (\tilde\psi \psi)^{A_1}_L \, (\tilde\psi \psi)^{A_2}_L \, 
      (\tilde\psi \tau^+\psi)^{S_3}_R \, T^{A_1 A_2 S_3} 
    = -4\mcO^3_{LLR} \,, \\
\end{split}
\end{equation}
and one $(\mathbf{1}_L, \mathbf{7}_R)$ operator
\begin{equation}
\label{eq:Q4def}
\begin{split}
Q_4 &= \Big[(\tilde\psi \tau^3\psi)^{S_1}_R \, (\tilde\psi \tau^3\psi)^{S_2}_R 
      - \frac{1}{5}(\tilde\psi \tau^a \psi)^{S_1}_R \, (\tilde\psi \tau^a \psi)^{S_2}_R
        \Big] \, (\tilde\psi \tau^+\psi)^{S_3}_R \, T^{S_1 S_2 S_3}
    = -\frac{4}{5}\mcO^1_{RRR} - \frac{16}{5}\mcO^2_{RRR}\,,
\end{split}
\end{equation}
where $\tau^{\pm} = \frac12(\tau^1\pm i\tau^2)$.
The remaining 10 independent $\nnbar$ transition operators are not $SU(2)_L$ singlets.
Of these additional operators, three belong to the $(\mathbf{5}_L,\mathbf{3}_R)$ 
irreducible representation,
\begin{equation}
\label{eq:Q5-7def}
\begin{split}
Q_5 &= (\tilde\psi \tau^- \psi)^{S_1}_R \, (\tilde\psi \tau^+ \psi)^{S_2}_L \,
      (\tilde\psi \tau^+\psi)^{S_3}_L \, T^{S_1 S_2 S_3} = \mcO^1_{RLL}\,, \\
Q_6 &= (\tilde\psi \tau^3 \psi)^{S_1}_R \, (\tilde\psi \tau^3 \psi)^{S_2}_L \,
      (\tilde\psi \tau^+\psi)^{S_3}_L \, T^{S_1 S_2 S_3} = -4 \mcO^2_{RLL}\,, \\
Q_7 &= \Big[ (\tilde\psi \tau^3 \psi)^{S_1}_L \, (\tilde\psi \tau^3 \psi)^{S_2}_L 
        - \frac{1}{3}(\tilde\psi \tau^a \psi)^{S_1}_L \, (\tilde\psi \tau^a \psi)^{S_2}_L 
       \Big] \,(\tilde\psi \tau^+ \psi)^{S_3}_R \, T^{S_1 S_2 S_3} 
    = -\frac{4}{3}\mcO^1_{LLR} - \frac{8}{3}\mcO^2_{LLR} \,.
\end{split}
\end{equation}
The remaining seven independent operators $Q_1^\mcP,\cdots,Q_7^\mcP$ are obtained from  
$Q_1,\cdots,Q_7$ by parity transformation discussed below~(\ref{eqn:op_Ptransf})
and belong to the $(\mathbf{3}_L,\mathbf{1}_R)$, $(\mathbf{7}_L,\mathbf{1}_R)$, and
$(\mathbf{3}_L,\mathbf{5}_R)$ irreducible representations.
The operators $Q_1,\cdots,Q_7,Q_1^P,\cdots,Q_7^P$ form a complete basis of 14 linearly
independent $SU(3)_C\times U(1)_{EM}$-invariant dimension-9 operators with baryon number 
$\Delta B = -2$ and isospin $\Delta I_3 = -1$\footnote{
  The isospin of operators $\Delta I_Q$ is defined here as $[Q, \vec I] = \Delta \vec I_Q Q$,
  leading to the selection rule $I_i - I_f = \Delta I_Q$ for the isospins of initial and final
  states.
}.

\begin{table}
\centering
\caption{Summary of operator properties and relations to notations used in other papers.
  The last column shows 1-loop QCD anomalous dimensions of the operators (see
  Sec.~\ref{sec:npr}).
  \label{tab:nnbar_op_def}}
\begin{tabular}{r|l|l|r|c|r}
\hline\hline
$Q_I$ & 
  Ref.~\cite{Syritsyn:2016ijx} & 
  Ref.~\cite{Rao:1982gt} & 
  Ref.~\cite{Grojean:2018fus} &
  $(I,I_3)_R\otimes(I,I_3)_L$ & $\gamma_\mcO^{(0)}$ \\
\hline
$-\frac34 Q_1$  &        
  $\left[(RRR)_{\mathbf1}\right]$ &
  $3\mcO^3_{\{RR\}R} = \mcO^2_{\{RR\}R} - \mcO^1_{\{RR\}R}$ & 
  $12\mcO_1$ &
  $(1,-1)_R \otimes(0,0)_L$ &
  $4$ \\
$-\frac34 Q_2$ &        
  $\left[(RR)_{\mathbf1} L_{\mathbf0}\right]$ & 
  $3\mcO^3_{\{LR\}R}= \mcO^2_{\{LR\}R} - \mcO^1_{\{LR\}R}$ & 
  $6\mcO_2$ &
  $(1,-1)_R \otimes (0,0)_L$ &
  $-4$ \\
$-\frac34 Q_3$ &
  $\left[R_{\mathbf1} (LL)_{\mathbf0}\right]$ & 
  $3\mcO^3_{\{LL\}R}= \mcO^2_{\{LL\}R} - \mcO^1_{\{LL\}R}$ & 
  $12\mcO_3$ &
  $(1,-1)_R \otimes (0,0)_L$ &
  $0$ \\
\hline
$-\frac54 Q_4$ & 
  $[(RRR)_{\mathbf3}]$ & 
  $\mcO^1_{R\{RR\}} +4\mcO^2_{\{RR\}R}$ & 
  --- &
  $(3,-1)_R \otimes (0,0)_L$ &
  $+24$ \\
\hline
$-Q_5^\mcP$ &
  $\left[(RR)_{\mathbf2} L_\mathbf1\right]_{(1)}$ & 
  $\mcO^1_{L\{RR\}}$ &
  $-4\mcO_4^{\mcP}$ &
  $(2,-2)_R \otimes (1,1)_L$ &
  $+12$ \\
  $\frac14 Q_6^\mcP$ &
  $\left[(RR)_{\mathbf2} L_{\mathbf1}\right]_{(2)}$ & 
  $\mcO^2_{\{LR\}R}$ &
  $-2\mcO_5^{\mcP}$ &
  $(2,-1)_R \otimes (1,0)_L$ &
  $+12$ \\
  $\frac34 Q_7^\mcP$ &
  $\left[(RR)_{\mathbf2} L_{\mathbf1}\right]_{(3)}$ & 
  $\mcO^1_{R\{RL\}} +2\mcO^2_{\{RR\}L}$ &
  $-4\mcO_6^{\mcP}$ &
  $(2,0)_R \otimes (1,-1)_L$ &
  $+12$ \\
\hline\hline
\end{tabular}
\end{table}

Isospin properties of the $\nnbar$ operators are summarized in Tab.~\ref{tab:nnbar_op_def},
together with relations to notations used in other papers.
In the following sections, we will discuss nucleon matrix elements only of the operators
$Q_{1,2,3,5}$, and the other matrix elements can be easily obtained using symmetries discussed
below.

%%%%%%%%%%%%%%%%%%%%%%%%%%%%%%%%%%%%%%%%%%%%%%%%%%%%%%%%%%%%%%%%%%%%%%%%%%%%%%
\subsection{Operator mixing}

In this work, we study lattice regularized operators that have to be nonperturbatively renormalized
and then perturbatively converted to the $\MSbar$ scheme using the one-loop matching results of
Ref.~\cite{Buchoff:2015qwa}, as described in Sec.~\ref{sec:npr},
%The $Q_I$ and $Q_I^P$ above provide a complete basis of 14 dimension-9 operators with 
%$\Delta B =-2$, $\Delta I_z = 2$ 
%Any renormalized operator $Q_I^R(\mu)$ can be expressed as a linear combination of the ``bare''
%lattice regularized operators $Q_I^{\text{lat}}$,
\begin{equation}
   \begin{split}
      Q_I^{R}(\mu) = Z_{IJ}^{R}(\mu) Q_I^{\text{lat}}.
   \end{split}\label{eq:Qbaredef}
\end{equation}
The renormalization matrix $Z_{IJ}^R$ takes especially simple form in the ``chiral basis''
consisting of elements $Q_{I=1\ldots7}^{(\mcP)}$, because they belong to different
chiral multiplets and cannot mix with each other due to chiral $SU(2)_L\times SU(2)_R$ symmetry 
of massless QCD.
Although some chiral representations appear in Tab.~\ref{tab:nnbar_op_def} more than once, they are
actually also prevented from mixing.
Specifically, operators $Q^{(\mcP)}_{5,6,7}$ consist of different components (``rows'') of chiral
$\mathbf{3}$- and $\mathbf{5}$-multiplets, and transform differently under $SU(2)_L\times SU(2)_R$.
Operators $Q^{(\mcP)}_{1,2,3}$ cannot mix with each other for a more subtle reason.
Even though they belong to the same chiral representation, they contain different numbers of left-
and right-handed diquarks.
While the $U(1)_A$ symmetry is violated in QCD by the ABJ anomaly, operators $Q_{1,2,3}$ do not mix
in perturbative QCD because perturbative gluon exchanges preserve the $U(1)_A$ transformation
properties of external quark fields in their respective Green's functions.
At the diagram level, there are only quark (and no antiquark) external fields, which cannot
be contracted into closed loops, and thus penguin-like diagrams do not appear.
This point is discussed and illustrated by an explicit two-loop perturbative calculation in
Ref.~\cite{Buchoff:2015qwa}.

In order to avoid mixing of renormalized operators, one has to define renormalization matrix
$Z_{IJ}^R$ in a scheme respecting chiral symmetry, such as $\MSbar$, and perform perturbative 
matching calculations in massless QCD.
Likewise, to avoid mixing of bare lattice operators $Q_I^{\text{lat}}$, chiral symmetry must be 
preserved in lattice QCD regularization, which requires [M\"obius] domain wall ([M]DWF) or overlap
fermion discretization.
The MDWF action that we use in this work has been shown to have good chiral
properties~\cite{Blum:2014tka} (see Sec.~\ref{sec:lattice_setup}), and our lattice results may be
safely matched to perturbative QCD in the UV regime.
Finally, nonperturbative effects such as spontaneous chiral symmetry breaking and
$U(1)_A$-violating topological fluctuations (instantons) in the QCD vacuum could lead to operator
mixing in nonperturbative renormalization (NPR).
Mixing can be also induced by the light quark masses and residual chiral symmetry violation.
However, as we study NPR numerically in Sec.~\ref{sec:npr}, we find that this mixing is
negligible ($\approx O(10^{-3})$) and can be safely neglected at our level of precision.

%%%%%%%%%%%%%%%%%%%%%%%%%%%%%%%%%%%%%%%%%%%%%%%%%%%%%%%%%%%%%%%%%%%%%%%%%%%%%%
\subsection{Isospin relations between matrix elements}\label{sec:isospin}

Since the chiral symmetry of QCD is spontaneously broken $SU(2)_L\otimes SU(2)_R\to SU(2)_{L+R}$, 
the isospin selection rules for $\nnbar$ matrix elements constrain only the total isospin $I_{L+R}$ 
of the effective operators $Q^{(\mcP)}_{1,\dotsm,7}$.
The $\nnbar$ transition changes the isospin by $\Delta I_3=-1$, therefore the $L,R$ isospins must add
as 
\begin{equation}
\label{eqn:nnbar_vec_isospin}
(I,I_3)_L\otimes (I,I_3)_R\to (I,I_3)_{L+R}=(1,-1)\,.
\end{equation}
The operator in the $(\mathbf{1}_L, \mathbf{7}_R)$ representation~(\ref{eq:Q4def}) with
the total isospin $I_{L+R}=3$ cannot couple a neutron to an antineutron ($I_{L+R}=\pm\frac12$) in
our calculation that is performed with $SU(2)_f$-symmetric QCD with $m_u=m_d$, therefore
\begin{equation}
\langle\bar{n}|Q_4|n\rangle|_{m_u=m_d}=0\,.
\end{equation}
Even if the isospin-breaking effects $\sim(m_u -  m_d)\ne0$ are included, such $SU(2)_f$-violating
matrix elements will be suppressed with powers of $(m_u-m_d)/\Lambda_\text{QCD}$ relative to those of 
other operators.

Similarly, while the $Q_{5,6,7}$ operators introduced in Eqs.~(\ref{eq:Q5-7def}) are linearly 
independent, isospin symmetry leads to additional relations between their $\nnbar$ matrix
elements that make two of them redundant.
The relations between them are determined by the $(I,I_3)_{L+R}=(1,-1)$ component
in the product of their chiral factors,
\begin{equation}
\begin{aligned}
\label{eqn:Q567irrep}
Q_5 \sim  (2,-2)_L\otimes(1,1)_R\,,
\quad
Q_6 \sim  (2,-1)_L\otimes(1,0)_R\,,
\quad
Q_7 \sim  (2,0)_L\otimes(1,-1)_R\,,
\end{aligned}
\end{equation}
as well as their normalization.
To find the latter, one can use $SU(2)$ ladder operators
\begin{equation}
\langle I,I_3|\hat I_+|I,I_3-1\rangle 
  = \sqrt{(I+I_3)(I-I_3+1)}
  = \langle I,I_3-1|\hat I_-|I,I_3\rangle
\end{equation}
to construct the full $\mathbf{3}_R$ and $\mathbf{5}_L$ isospin multiplets starting from
$(u^T C u)^S_R\sim(1,+1)_R$ and $(u^TCu)^{\{S_1}_L (u^TCu)^{S_2\}}_L\sim(2,+2)_L$,
respectively:
\begin{gather}
\label{eqn:chi3pletR}
\mathbf{3}_R: \; \left(\begin{array}{r} 
  (u^T C u)^S_R \\ \sqrt{2}(u^T C d)^S_R \\ (d^T C d)^S_R
\end{array}\right) 
\sim \left(\begin{array}{rr} 1,&+1 \\ 1,&0 \\ 1,&-1 \end{array} \right)_R\,,
\\
\label{eqn:chi5pletL}
\mathbf{5}_L: \; \left(\begin{array}{c}
(u^T C u)^S_L (u^T C u)^S_L \\
2(u^T C u)^S_L (u^T C d)^S_L \\
\sqrt{\frac23}\big[ (u^T C u)^S_L (d^T C d)^S_L 
  + 2 (u^T C d)^S_L (u^T C d)^S_L \big] \\
2(u^T C d)^S_L (d^T C d)^S_L \\
(d^T C d)^S_L (d^T C d)^S_L 
\end{array}\right)  
\sim \left(\begin{array}{rr} 2,&+2 \\ 2,&+1 \\ 2,&0 \\ 2,&-1 \\ 2,&-2 \end{array} \right)_L\,.
\end{gather}
Combining these components to construct $Q_{5,6,7}$ according to Eq.~(\ref{eqn:Q567irrep}) yields
their relative normalizations.
Taking into account the Clebsch-Gordan coefficients for the 
projection~(\ref{eqn:nnbar_vec_isospin}), one obtains the relations between matrix elements
\begin{equation}
\langle\bar{n}|Q_5|n\rangle = \langle\bar{n}|Q_6|n\rangle 
  = -\frac32\langle\bar{n}|Q_7|n\rangle\,,
\end{equation}
which are also fulfilled in lattice contractions up to the machine precision.
Additionally, one can check that these relations hold, e.g., for the results of the Bag-model
calculation~\cite{Rao:1982gt} in the form
\begin{equation}
\langle\bar{n}|\mcO^1_{RLL}|n\rangle = (-4) \langle\bar{n}|\mcO^2_{RLL}|n\rangle
  = (+2)\langle\bar{n}|\big(\mcO^1_{LLR}+2\mcO^2_{LLR}\big)|n\rangle\,.
\end{equation}

%%%%%%%%%%%%%%%%%%%%%%%%%%%%%%%%%%%%%%%%%%%%%%%%%%%%%%%%%%%%%%%%%%%%%%%%%%%%%%
\subsection{$\mcC$, $\mcP$, and $\mcT$ relations}\label{sec:CPT}

The discrete symmetries $\mcC$, $\mcP$, and $\mcT$, which are conserved in QCD, imply further
relations for $\nnbar$ transition matrix elements.
Since the form of $\nnbar$ operators is identical in Minkowski and Euclidean space, we study the
relations between their matrix elements in Minkowski space but using Euclidean $\gamma$-matrix
conventions.
From the usual transformations for the fermion fields, we obtain $\mcC$,$\mcP$,$\mcT$-transformation 
properties for quark bilinears and the 6-quark operators, which are summarized in
Appendix~\ref{app:CPT},
\begin{align}
\label{eqn:op_Ptransf}
Q_I^\mcP &= \mcP Q_I \mcP^{-1} = -\eta_P^6 \big[Q_I\big]_{L\leftrightarrow R}\,,\\
\label{eqn:op_Ctransf}
Q_I^\mcC &= \eta_C^6 \overline{Q}_I = \eta_C^6 \big[Q_I\big]_{\psi\to\bar\psi}
  = -\eta_C^6\eta_P^6 Q_I^{\mcP\dag} \,,\\
\label{eqn:op_Ttransf}
Q_I^\mcT &= \eta_T^6 Q_I \,,
\end{align}
where $\eta_{\mcC,\mcP,\mcT}$ are arbitrary complex phases accompanying the $\mcC,\mcP,\mcT$
transformations of fermion fields.
These factors and the relevance of Eq.~(\ref{eqn:op_Ttransf}) for $CP$-violating processes are
discussed further in Refs.~\cite{McKeen:2015cuz,Berezhiani:2015uya,Gardner:2016wov}.
The conjugated operators $Q_I^\dag$ are related to $Q_I$ by the $\mcC\mcP$ transformation,
\begin{equation}
\label{eqn:op_dag_CP}
Q_I^\dag = [Q_I]_{\psi\leftrightarrow\bar{\psi},\, L\leftrightarrow R} 
  = -\eta_C^{*6}\eta_P^{*6} \big(\mcC\mcP\big)\,Q_I\,\big(\mcC\mcP)^{-1}\,.
\end{equation}
The $\mcC\mcP$ transformation also relates the transition matrix elements 
$\bar{n}\to n$ and $n\to\bar{n}$ , which can be shown to be real.
For that, one has to use the transformation properties of the neutron and antineutron states (see
Appendix~\ref{app:CPT} for the details):
\begin{equation}
\begin{aligned}
&\left(\bra{\nbar-+} Q_I \ket{\neut++}\right)^* 
  = \bra{\neut++} Q_I^\dag \ket{\nbar-+}
\\&\quad
  = -\eta_C^{*6}\eta_P^{*6} \bra{\neut++}\big(\mcC\mcP\big)\,Q_I\,\big(\mcC\mcP)^{-1}\ket{\nbar-+}
  = \bra{\nbar-+} Q_I \ket{\neut++}\,.
\end{aligned}
\end{equation}

Parity relates $n\bar{n}$ transition matrix elements of $Q_I^P$ and $Q_I$,
\begin{equation}
\label{eq:QPME}
\begin{split}
\mbraket{\nbar-+}{Q_I}{\neut++} 
  &= \mbraket{\nbar-+}{\mcP^{-1} Q_I^\mcP \mcP}{\neut++} \\
  &= -\eta^{*6} \mbraket{\nbar-+}{Q_I^\mcP}{\neut++},
\end{split}
\end{equation}
where the phase factor is complementary to that in Eq.~(\ref{eqn:op_Ptransf}).
For the conventional choice $\eta_P=1$, it is clear that only the pseudoscalar combination 
$(Q_I-Q_I^\mcP)$ has nonzero matrix elements, since $n\to\bar{n}$ transition changes parity.
Note that in all the cases, the arbitrary phase factors $\eta_{C,P,T}$ arising from the
transformations of $Q_I$ cancel with the phase factors arising from the transformations of the
states.

Finally, with the help of the $\mcT$-reflection, one can also show that the matrix elements do not
depend on the direction of the (anti)neutron spin.
Using the transformation properties of the neutron and antineutron states,
\begin{equation}
\begin{aligned}
&\mbraket{\nbar-\pm}{Q_I}{\neut+\pm}
  = \mbraket{\nbar-\pm}{\mcT^{-1} Q_I^\mcT \mcT }{\neut+\pm}
\\&\quad
  = (\mp\eta_T^{*3})\mbraket{\nbar-\mp}{Q_I^\mcT}{\neut+\mp} (\mp\eta_T^{*3})
  = \mbraket{\nbar-\mp}{Q_I}{\neut+\mp}\,.
\end{aligned}
\end{equation}
All spin-flip matrix elements of $Q_I$ are trivially zero because $Q_I$ are (pseudo)scalars.

Denoting the ground-state $n\bar{n}$ transition matrix elements for each $Q_I$ by
\begin{equation}
\label{eq:MEbasisdef}
\mcM_I = \mbraket{\nbar-+}{Q_I}{\neut++} = \mbraket{\nbar--}{Q_I}{\neut+-},
\end{equation}
the matrix element results derived above can be summarized as
\begin{equation}
\label{eq:MCPT}
\mcM_I^* = \mcM_I, 
\quad \mcM_I^P = -\mcM_I.
\end{equation}
In conjunction with the results from Sec.~\ref{sec:isospin},
\begin{equation}
\mcM_4 = 0, 
\quad \mcM_5 = \mcM_6 = -\frac{3}{2}\mcM_7,
\label{eq:M567}
\end{equation}
this implies that in the isospin limit where Eq.~(\ref{eq:M567}) is valid, $n\bar{n}$ transition
rates involving the 14 operators $Q_I^{(\mcP)}$ are given in terms of 4 real $n\bar{n}$ transition
matrix elements $\mcM_{1,2,3,5}$.

%%%%%%%%%%%%%%%%%%%%%%%%%%%%%%%%%%%%%%%%%%%%%%%%%%%%%%%%%%%%%%%%%%%%%%%%%%%%%%
\subsection{$n\bar{n}$ Effective Field Theory}
\label{sec:eft}

The $|\Delta B|=2$ effective interactions discussed above must be generated by some extension of the
Standard model at yet unknown scale $\lbsm$.
It is generally assumed that such extensions have higher symmetry, which is broken at scales below
$\lbsm$ to the electroweak symmetry $SU(2)_L\times U(1)_Y$, and thus the effective interactions
must be EW-symmetric.
From the discussion above it follows that only $Q_{1,2,3}$ are $SU(2)_L\times U(1)_Y$-singlets,
while $Q_4$, $Q_{5(67)}$, and all $Q_I^{\mcP}$ operators are not.
These latter operators require additional EW-charged factors to make them EW-symmetric, which affect
the power counting and result in higher suppression by the $\lbsm$ scale.

Such factors can be easily constructed from the Higgs field doublet $\phi$ and its conjugate 
$i\tau_2\phi^*$ to compensate for the $SU(2)_L$- and hypercharge of the operators $Q_I^{(\mcP)}$.
The Higgs v.e.v. $v$ in unitary gauge leads to nonzero effective $\nnbar$ interaction in the 
form
\begin{equation}
\label{eqn:Leff_general}
\mcL_{\nnbar} = \sum_{I(\mcP)} \frac{\widetilde{C}_I^{(\mcP)}(\mu)}{\lbsm^5}\,
  \left(\frac{v^2}{\lbsm^2}\right)^{I_L[Q_I^{(\mcP)}]} \, Q_I^{(\mcP)}(\mu) 
  \,+\,\text{h.c.} \,,
\end{equation}
where $\widetilde{C}_I(\mu)$ are dimensionless Wilson coefficients and $I_L[Q_I^{(\mcP)}]$ is the
left-handed isospin of the operator $Q_I^{(\mcP)}$. 
%\footnote{
%  \textbf{Need to avoid confusion between $I$-index enumerating operators and $I$-isospin}}
In addition to Eq.~(\ref{eqn:Leff_general}), the full $|\Delta B|=2$ Lagrangian must also include
combinations of electrically charged $|\Delta B|=2$ operators with oppositely charged Higgs
fields to assure the EW symmetry above the EW scale.
Such interactions can lead to $n\leftrightarrow\bar{p}$ and $p\leftrightarrow\bar{n},\bar{p}$
transitions, and the emitted charged Higgs bosons (e.g., decaying into leptons) would compensate for
the change in the electric charge.
These transitions are suppressed by at least one factor of $(v^2/\lbsm^2)$.~\footnote{
  Isospin breaking effects in QCD may result in suppression in powers of   $(m_u - m_d)/\lqcd$ 
  instead of $v/\lbsm$, which is beyond the scope of the present paper.
  Also, while additional higher-dimensional operators suppressed at the same level may be
  constructed using field derivatives, they are less relevant and not considered here.}

Using the effective Lagrangian~(\ref{eqn:Leff_general}) and the relations derived in the previous
sections, the full $\nnbar$ matrix element can be written as
\begin{equation}
\mcM_{\nnbar} = \frac1{\lbsm^5}\Big[ \sum_{I=1,2,3} \widetilde{C}_I \mcM_I 
  + \frac{v^2}{\lbsm^2}\sum_{I=1,2,3,5} \widetilde{C}_I^\mcP \mcM_I^\mcP
  + \frac{v^4}{\lbsm^4} \widetilde{C}_5 \mcM_5 \Big]\,,
  \label{eqn:Meft}
\end{equation}
where $\mcM^{(\mcP)}_I$ are the nucleon matrix elements of operators $Q^{(\mcP)}_I$.
The dimensionless low-energy constants $\widetilde{C}^{(\mcP)}_I(\mu)$ depend on the scale $\mu$ 
only logarithmically and can be computed perturbatively 
by using $C_I(\lbsm)\sim O(1)$ given by a particular BSM scenario as an initial condition for renormalization group evolution.
A nonperturbative calculation of the matrix elements $\mcM_I$ is presented in the following
sections.

%%%%%%%%%%%%%%%%%%%%%%%%%%%%%%%%%%%%%%%%%%%%%%%%%%%%%%%%%%%%%%%%%%%%%%%%%%%%%%
%%%%%%%%%%%%%%%%%%%%%%%%%%%%%%%%%%%%%%%%%%%%%%%%%%%%%%%%%%%%%%%%%%%%%%%%%%%%%%
\section{Lattice setup}
\label{sec:lattice_setup}
% lattice section

In this section, we fist recount the details of the lattice QCD gauge configurations and
propagators used in this study, and then describe the construction of (anti)neutron correlation
functions with the $\nnbar$ operators.
The QCD gauge field configurations were generated with the Iwasaki gauge action on a
$48^3\times96$ lattice and $N_f=2+1$ flavors of dynamical M\"obius Domain Wall fermions.
The fermion masses are tuned to be almost exactly at the physical point~\cite{Blum:2014tka},
such that the pion mass is approximately $m_\pi=139.2(4)\text{MeV}$ and the scale (the lattice
spacing) is $a=0.1141(3)\text{ fm}$.
The residual mass $m_{res}$, which encapsulates the residual violation of chiral symmetry, is smaller 
than 50\% of the input quark mass.
The physical lattice size $L\approx5.45\text{ fm}$ and $m_\pi L = 3.86$ should be sufficient to
suppress finite volume effects of the $\nnbar$ matrix elements to a level below our target
precision.
In particular, according to chiral perturbation theory, these finite size effects are expected
to be $\lesssim 1\%$~\cite{Bijnens:2017xrz}.

The three-point functions needed to evaluate the matrix elements of the operators $Q^{(\mcP)}_I$
require six quark propagators for the $u$ and $d$ quarks flavors; they result from Wick
contractions of the six-quark operators with the (anti)neutron interpolating fields.
There are no disconnected quark-loop diagrams because the operators $Q_I$($Q_I^\dagger$) contain
only quarks (antiquarks).
For both two- and three-point lattice correlation functions we compute propagators on
30 independent gauge field configurations separated by 40 molecular dynamics time steps.
All the quark propagators required for a single sample are computed from a point source
located at the operator insertion point, which is identified in the analysis with the origin 
$x_0 =(0,0,0,0)$ using translational invariance.
To reduce stochastic uncertainty, sampling of the neutron correlation functions is enhanced by
\emph{all-mode-averaging}~\cite{Shintani:2014vja}, in which we compute 1 exact and 81
low-precision samples evenly distributed over the 4D volume on each gauge configuration.
The low-precision quark propagators are computed with low-mode deflation and the conjugate
gradient algorithm truncated at 250 iterations.

The propagators are contracted at the sink into intermediate 
\emph{baryon blocks}~\cite{Doi:2012xd,Detmold:2012eu} with polarized nucleon and antinucleon
quantum numbers to minimize the time spent in the contraction step of the calculation.
(Anti)neutron source and sink interpolating operators are constructed with either point or
Gaussian-smeared (anti)quarks and are denoted with $n^{J=P,S}$, respectively.
Final contraction at the propagator source yields an (anti)neutron two-point correlation
function sample with a point source at $x_0$.
Thus, the polarized neutron two-point correlation function with zero spatial momentum for
positive time $t>0$ is
\begin{equation}
\label{eq:2ptspec_neut}
\begin{split}
G_{nn(\sigma)}^{JJ^\prime}(t>0) &= 
\sum_{\mathbf{x}} \bra{\text{vac}} n^{(+)J^\prime}_{\sigma}(\mathbf{x},t) \,
      {n}^{(+)J \dag}_{\sigma}(0) \ket{\text{vac}}
  =\Gamma^{\sigma(+)}_{\alpha\alpha^\prime} \sum_{\mathbf{x}} 
      \langle n^{J^\prime}_{\alpha^\prime}(\mathbf{x},t) \, \bar{n}^{J}_{\alpha}(0) \rangle \\
%    &= \sum_{\mathfrak{n}} \sqrt{ Z_\mathfrak{n}^{J^\prime} Z_\mathfrak{n}^{J} } e^{-E_\mathfrak{n} t},
  \end{split}
\,,
\end{equation}
and, similarly, for the polarized antineutron
\begin{equation}
\label{eq:2ptspec_nbar}
\begin{split}
G_{\bar{n}\bar{n}(\sigma)}^{JJ^\prime}(t>0) &= 
  \sum_{\mathbf{x}} \bra{\text{vac}} n^{(-)J^\prime\dag}_{\sigma}(\mathbf{x},t) \,
          {n}^{(-)J}_{\sigma}(0) \ket{\text{vac}}
  =\Gamma^{\sigma(-)}_{\alpha\alpha^\prime} \sum_{\mathbf{x}} 
      \langle n^{J}_{\alpha}(0) \, \bar{n}^{J^\prime}_{\alpha^\prime}(\mathbf{x},t) \rangle \\
%    &= \sum_{\mathfrak{n}} \sqrt{ Z_\mathfrak{n}^{J^\prime} Z_\mathfrak{n}^{J} } e^{-E_\mathfrak{n} t},
  \end{split}
\,,
\end{equation}
where the polarization matrix 
$\Gamma^{\sigma(\pm)}=\frac{1\pm\gamma_4}2 \frac{1+\sigma\gamma_3\gamma_5}2$
projects on the selected parity ($\pm$) and spin $\sigma=\pm\frac12$,
and the interpolating operator at the source is $J=P$ and the one at the sink can be either
$J^\prime=P$ or $J^\prime=S$.
Neutron/antineutron two-point functions have the spectral representation
\begin{equation}
  \begin{split}
     G_{nn(\sigma)}^{JJ^\prime} (t) = G_{\bar{n}\bar{n}(\sigma)}^{JJ^\prime}(-t)  
   &= \sum_{m} \sqrt{ Z_m^J Z_m^{J^\prime} } \  e^{-E_{m} t},
  \end{split}
\label{eqn:twopt_spectral}
\end{equation}
where the overlap factors $Z_m^J$ are identical for neutrons and antineutrons in either spin orientations.

\begin{figure}[ht!] %  figure placement: here, top, bottom, or page
  \centering
  \includegraphics[width=.39\textwidth]{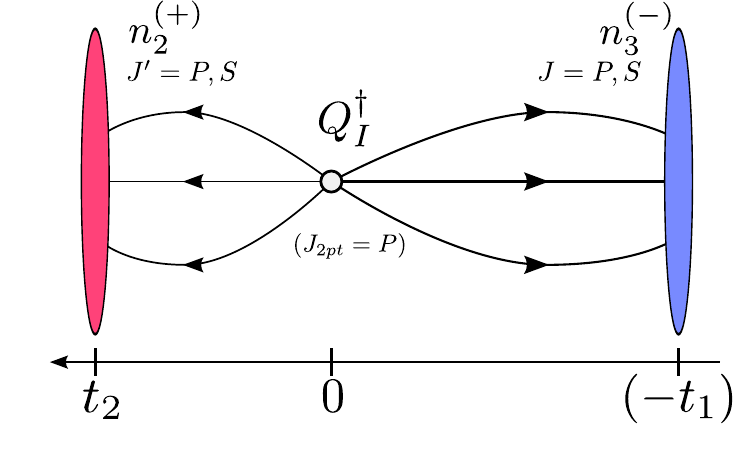}\\
  \caption{
    Contractions for the three-point correlation function of the (anti)neutrons with a
    $\bar{n}\leftarrow n$ transition operator.
    The indices of the neutron interpolating operators refer to the standard Dirac-Pauli
    representation (see Appendix~\ref{app:CPT}).
    \label{fig:nnbar_contract}}
\end{figure}
 
%Finally, pairs of baryon blocks extending forward and backward in time are contracted into neutron 
%(and the corresponding antineutron) 
%correlation functions.
The three-point functions involve two neutron or two antineutron fields to create and annihilate
states with opposite baryon numbers.
Using the (anti)neutron states defined in Eq.~(\ref{eqn:nnbar_states}), 
one can express the three-point correlation function containing, for example, 
the $n\leftarrow\bar{n}$ transition matrix element (see Fig.~\ref{fig:nnbar_contract}):
\begin{equation}
\label{eqn:threept_spectral}
\begin{split}
G_{n Q_I^\dag \bar{n}(\sigma)}^{JJ^\prime}(t_1,t_2) &= 
  \sum_{\mathbf{x},\mathbf{y}} 
    \bra{\text{vac}} n_\sigma^{(+)J^\prime}(\mathbf{x},t_2) Q_I^\dag(0)
    n_{-\sigma}^{(-)J}(\mathbf{y}, -t_1) \ket{\text{vac}} \\
 &= \big(C\Gamma^{\sigma(+)}\big)_{\alpha\alpha^\prime} \sum_{\mathbf{x},\mathbf{y}} 
    \avg{ n_{\alpha^\prime}^{J^\prime}(\mathbf{x},t_2) Q_I^\dagger(0) 
        n_{\alpha}^{J}(\mathbf{y},-t_1) }.
\end{split}
\end{equation}
where $n^{(-)}_{-\sigma}$ is nucleon interpolating field that creates an antineutron with spin
$\sigma$\footnote{
  Note that spin-flip of a spinor incorporates nontrivial signs in order to satisfy
  $n_{-(-\sigma)}=-n_\sigma$ similar to $\mcT$ transformation, which is 
  responsible for the relative signs of the neutron and antineutron states.
},
both the (anti)neutron operators are summed over the spatial coordinate to project on zero
momentum.
By calculating quark propagators with point sources located at the operator insertion point and
momentum-projected $P$ and $S$ sinks located on all time slices, the correlation functions 
$G_{n Q_I^\dag \bar{n}}^{JJ^\prime}$ can be accessed for all smearing combinations $PP$, $PS$, 
$SP$, and $SS$, any temporal separation between the source and the sink $\tsep = t_1+t_2$, and
any operator separation from the source $\tau=t_1$.
The same propagators are used to calculate $PP$ and $PS$ two-point correlation functions.
The spectral representation for Eq.~(\ref{eqn:threept_spectral}) analogous to
Eq.~(\ref{eqn:twopt_spectral}) is given by
\begin{equation}
  \begin{split}
     G_{n Q_I^\dag \bar{n}(\sigma)}^{JJ^\prime}(t_1,t_2) 
     &= \sum_{m,m^\prime}  \sqrt{ Z_{m}^J Z_{m^\prime}^{J^\prime} } 
      e^{-E_{m^\prime}t_2 -E_{m} t_1} (\mathcal{M}_I)_{mm^\prime},
  \end{split}\label{eq:G3ptPol}
\end{equation}
where $(\mathcal{M}_I)_{m^\prime m} = \mbraket{m^\prime}{Q_I}{m}$, the ground-state matrix
element of interest is $\mathcal{M}_I = (\mathcal{M}_I)_{00}$, and the overlap factors $Z_m^J$,
$Z_{m^\prime}^{J^\prime}$ are the same as in Eq.~\ref{eqn:twopt_spectral}.
We perform contractions for all combinations of point and smeared sources and sinks
in the three-point functions to enhance the analysis of the ground and
excited state matrix elements in the next section.
To reduce stochastic uncertainties, we also average lattice matrix elements over the spins
of the neutron and antineutron states.
%To reduce stochastic uncertainties, we also average lattice matrix elements over the spins
%of the neutron and antineutron states.
The specific combinations of (anti)neutron 4-spinor components in the three-point functions that
give matrix elements $\mcM_I$ are
\begin{equation}
\begin{aligned}
\mbraket{\nbar-+}{Q_I}{\neut++} 
  &\sim  - \langle n_4^\dag(t_2) Q_I(0) n_1^\dag(-t_1) \rangle \\
\mbraket{\nbar--}{Q_I}{\neut+-} 
  &\sim \np\langle n_3^\dag(t_2) Q_I(0) n_2^\dag(-t_1) \rangle \\
\mbraket{\neut++}{Q_I^\dag}{\nbar-+} 
  &\sim  - \langle n_1(t_2) Q_I^\dag(0) n_4(-t_1) \rangle \\
\mbraket{\neut+-}{Q_I^\dag}{\nbar--} 
  &\sim \np\langle n_2(t_2) Q_I^\dag(0) n_3(-t_1) \rangle \\
\end{aligned}
\end{equation}
where the signs correspond to the conventions listed in Appendix~\ref{app:CPT}.
As shown in Sec.~\ref{sec:CPT}, these matrix elements are real, and combining them with
the conjugated ones is also used to enhance statistics following Eq.~(\ref{eq:MCPT}).

%%%%%%%%%%%%%%%%%%%%%%%%%%%%%%%%%%%%%%%%%%%%%%%%%%%%%%%%%%%%%%%%%%%%%%%%%%%%%%
%%%%%%%%%%%%%%%%%%%%%%%%%%%%%%%%%%%%%%%%%%%%%%%%%%%%%%%%%%%%%%%%%%%%%%%%%%%%%%
\section{Analysis of matrix elements}
\label{sec:analysis}
% vim: sw=2 sts=2 et
%%%%%%%%%%%%%%%%%%%%%%%%%%%%%%%%%%%%%%%%%%%%%%%%%%%%%%%%%%%%%%%%%%%%%%%%%%%%%%
\begin{figure}
  \centering
  \includegraphics[width=.45\textwidth]{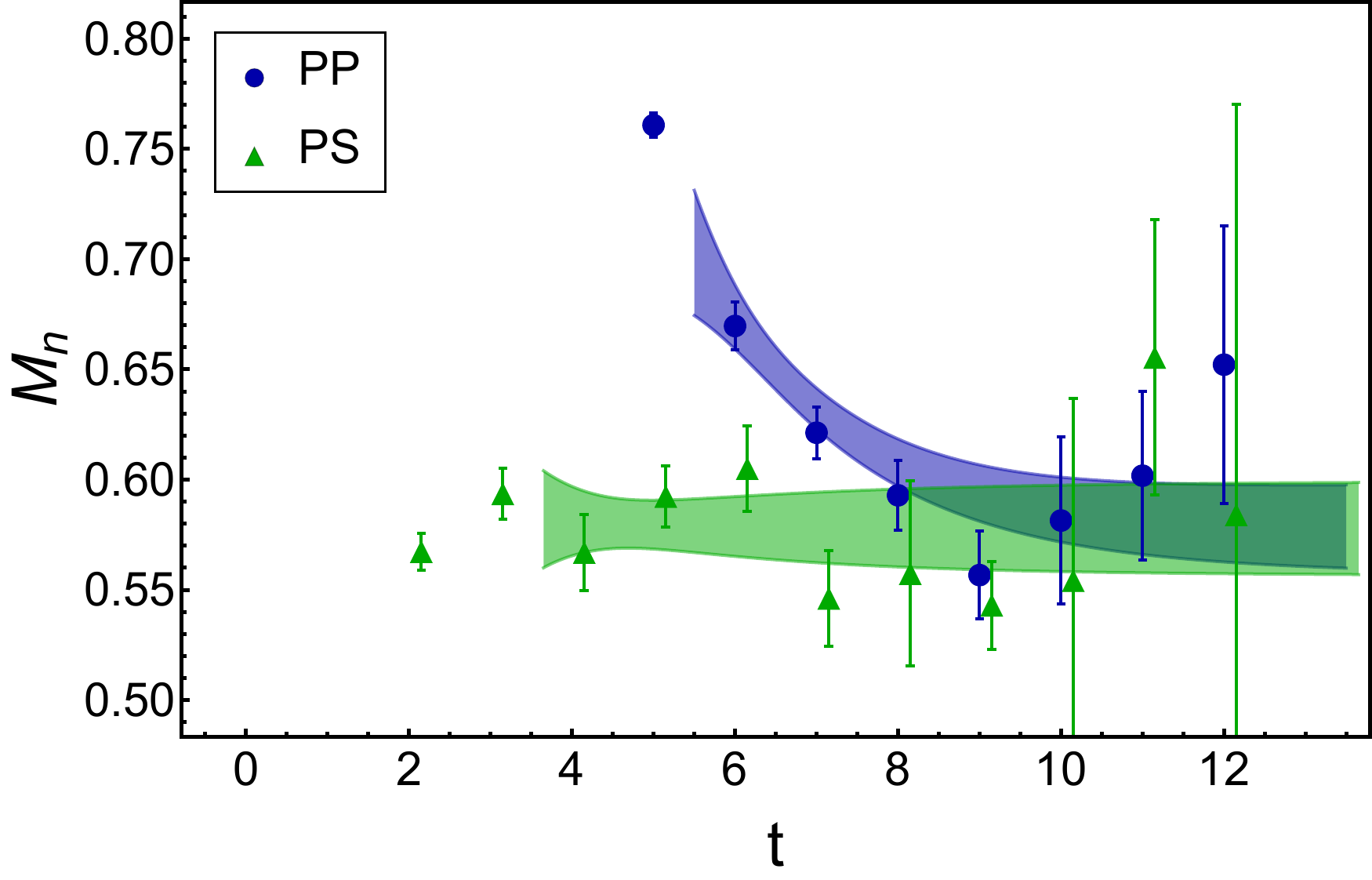}
  \caption{
    Combined correlated $\chi^2$ fits of $PP$, $PS$ two-point to Eq.~(\ref{eqn:2pt2state}) in
    the time range shown in the first row of Tab.~\ref{tab:2ptResults}.
    The covariance matrix is estimated with optimal shrinkage $\lambda^*$ as described in the main text.
    Corresponding data points show the effective masses $M_n(t) = \ln G_{nn}(t) - \ln G_{nn}(t+1)$
    with their statistical uncertainties.
    Note that  $t^{max}$ in Tab.~\ref{tab:2ptResults} indicates the largest separation for $G_{nn}$ considered
    and that the effective mass is consequently shown for $0 \leq t \leq t^{max} - 1$.
    \label{fig:2pt}}
\end{figure}

\begin{table}
   \begin{tabular}{|c|c|c|c||c|c||c|c|}
   \hline
   $t^{min}_{PP}$ & $t^{min}_{PS}$ & $t^{max}$ & $N_\text{dof}$ &
      $E_0$ & $E_1$ & $\chi^2/N_\text{dof}$ & $\lambda^*$ \\\hline
   6 &  4 &  13 &  12 &  0.578(23) &  1.23(27) &  0.50 &  0.14\\\hline
   6 &  6 &  13 &  10 &  0.556(22) &  1.11(15) &  0.42 &  0.15\\\hline
   6 &  5 &  13 &  11 &  0.560(24) &  1.13(21) &  0.40 &  0.14\\\hline
   5 &  5 &  13 &  12 &  0.566(20) &  1.26(9) &  0.40 &  0.13\\\hline
   7 &  5 &  13 &  13 &  0.554(69) &  0.98(43) &  0.42 &  0.15\\\hline\hline
    \multicolumn{4}{|l|}{Weighted Ave}  & 0.565(24)(8) & 1.21(15)(65) & \multicolumn{2}{|l|}{\ } \\\hline
  \end{tabular}
  \caption{Results of two-point function fits from different time ranges: ground- and
    excited-state energies, reduced $\chi^2/N_\text{dof}$, and optimal shrinkage parameters
    $\lambda^*$.
    The uncertainties in individual fits are statistical.
    The last line shows ``fit averages'' with statistical and systematic uncertainties computed
    as described in Appendix~\ref{app:stat_modavg}.
  \label{tab:2ptResults}}
\end{table}

To account for excited-state contributions, we perform two-state fits to a truncation of
Eq.~(\ref{eqn:threept_spectral}),
\begin{equation}
\label{eqn:3pt2state}
%\begin{split}
G_{n Q_I^\dag \bar{n}}^{JJ^\prime}(\tsep, \tau)
  = \sqrt{ Z_0^J Z_{0}^{J^\prime}  } e^{ -E_0 \tsep } \mcM_I  %\\
    + e^{-E_0 \tau -E_1(\tsep-\tau)} \mathcal{A}_I^{JJ^\prime}
    + e^{-E_1 \tau -E_0(\tsep-\tau)} \mathcal{A}_I^{J^\prime J}
    + e^{ -E_1 \tsep} \mathcal{B}_I^{JJ^\prime}\,,
%  \end{split}
\end{equation}
where $\mathcal{A}_I^{JJ^\prime}$ and $\mathcal{B}_I^{JJ^\prime}$ are products of overlap
factors and matrix elements involving only excited states, which are discarded in our
calculation.
The ground-state overlap factors $Z_0^P$ and $Z_0^S$ required to extract matrix elements of
$G_{n Q_I^\dag \bar{n}}^{JJ^\prime}$ can be obtained independently from fits of two-point
functions $G_{nn}^{PP}$ and $G_{nn}^{PS}$ to an analogous two-state model
\begin{equation}
\label{eqn:2pt2state}
\begin{split}
G_{nn(\sigma)}^{JJ^\prime}(t)
  &= \sqrt{ Z_0^J Z_{0}^{J^\prime}  } e^{ -E_0 t} + \sqrt{ Z_1^J Z_{1}^{J^\prime}  } e^{ -E_1 t},
\end{split}
\end{equation}
The energies $E_0$ and $E_1$ appear in both Eq.~(\ref{eqn:3pt2state}) and
Eq.~(\ref{eqn:2pt2state}), therefore fits of $G_{n Q_I^\dag \bar{n}}$ may be simplified by fixing
the state energies $E_0$, $E_1$ to values determined from fits of two-point functions
$G_{nn}^{JJ^\prime}$.
In principle, the overlaps with excited neutron states $Z_1^J$ are also determined from
two-point function fits, thus the number of parameters in Eq.~(\ref{eqn:3pt2state}) can be reduced
by factoring $\mathcal{A}_I^{JJ^\prime}$, $\mathcal{B}_I^{JJ^\prime}$ into excited-state matrix
elements and overlap factors $Z_{0,1}^J$, of which only the latter would depend on the neutron
interpolating operators.
It would be possible if the two- and three-point functions were saturated by contributions only
from the ground and the first excited states, or their contributions could be reliably
distinguished from higher-energy states omitted from Eqs.~(\ref{eqn:3pt2state},\ref{eqn:2pt2state}).
However, as our two-point function fits in Fig.~\ref{fig:2pt} show, there are higher excited-state
contributions to $G_{nn}$; in particular, there is large systematic uncertainty on $E_1$ in (see
Tab.~\ref{tab:2ptResults}).

These considerations lead us to adopt the following fit strategy: first, a combined fit of
$G_{nn}^{PP}$ and $G_{nn}^{PS}$ to Eq.~(\ref{eqn:2pt2state}) is used to determine the four
parameters $E_{0,1}$ and $Z_0^{P,S}$ as summarized in Fig.~\ref{fig:2pt}
and Tab.~\ref{tab:2ptResults};
then, a combined fit of $G_{n Q_I^\dag \bar{n}}^{PS}$, $G_{n Q_I^\dag \bar{n}}^{SP}$, and
$G_{n Q_I^\dag \bar{n}}^{SS}$ to Eq.~(\ref{eqn:3pt2state}) is used to determine the six parameters
$\mcM_I$, $\mathcal{A}_I^{PS}$, $\mathcal{A}_I^{SP}$, $\mathcal{A}_I^{SS}$,
$\mathcal{B}_I^{PS} = \mathcal{B}_I^{SP}$, and $\mathcal{B}_I^{SS}$.
Also, since $PP$ three-point functions would have even large excited-state contamination and $PP$
three-point/two-point ratios are not close to their plateau region for the $\tsep$ used here
(not shown), we do not include $G_{n Q_I^\dag \bar{n}}^{PP}$ in our analysis.

\begin{figure}[t]
  \centering
  \includegraphics[width=.45\textwidth]{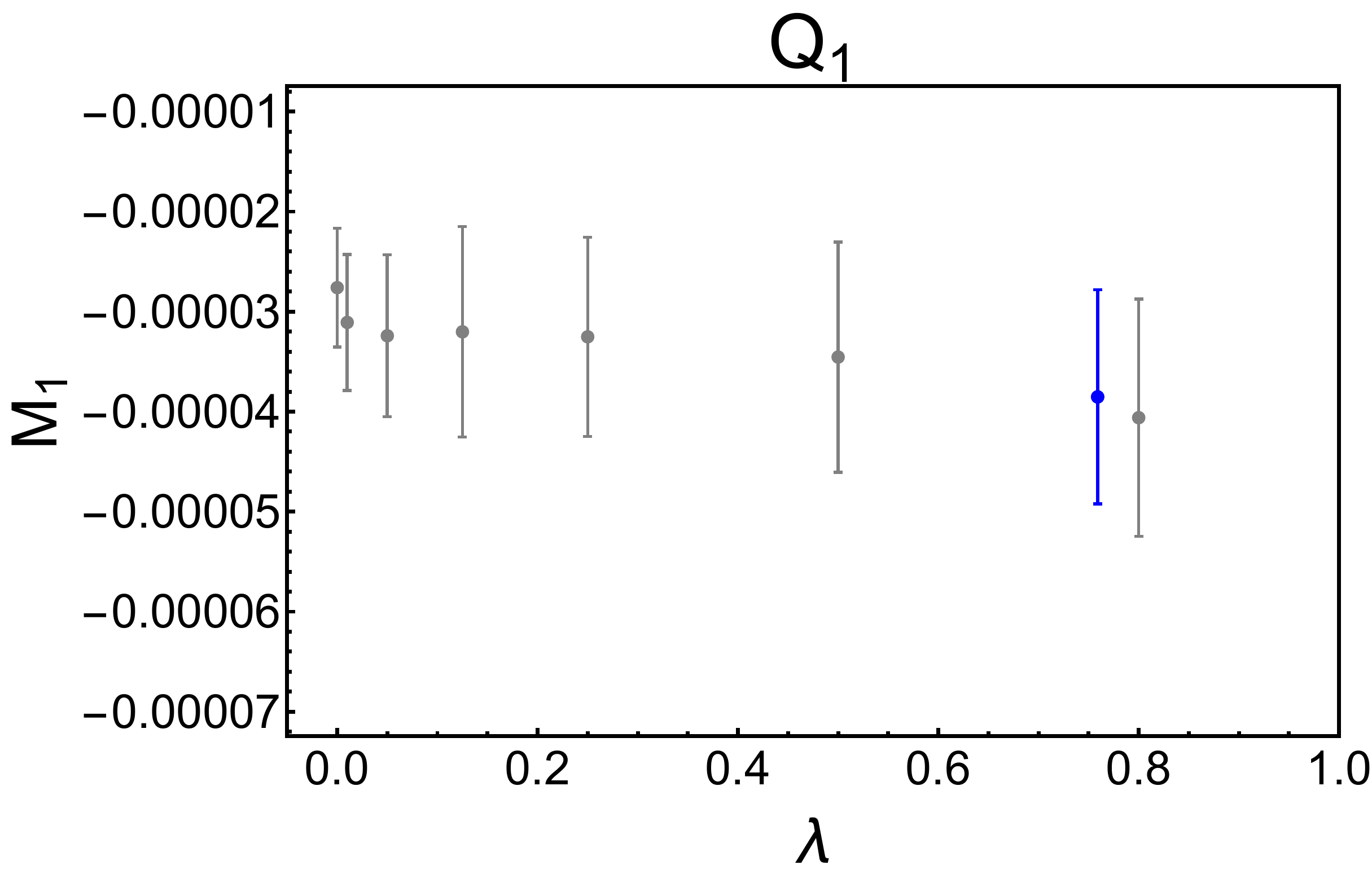}
  \includegraphics[width=.45\textwidth]{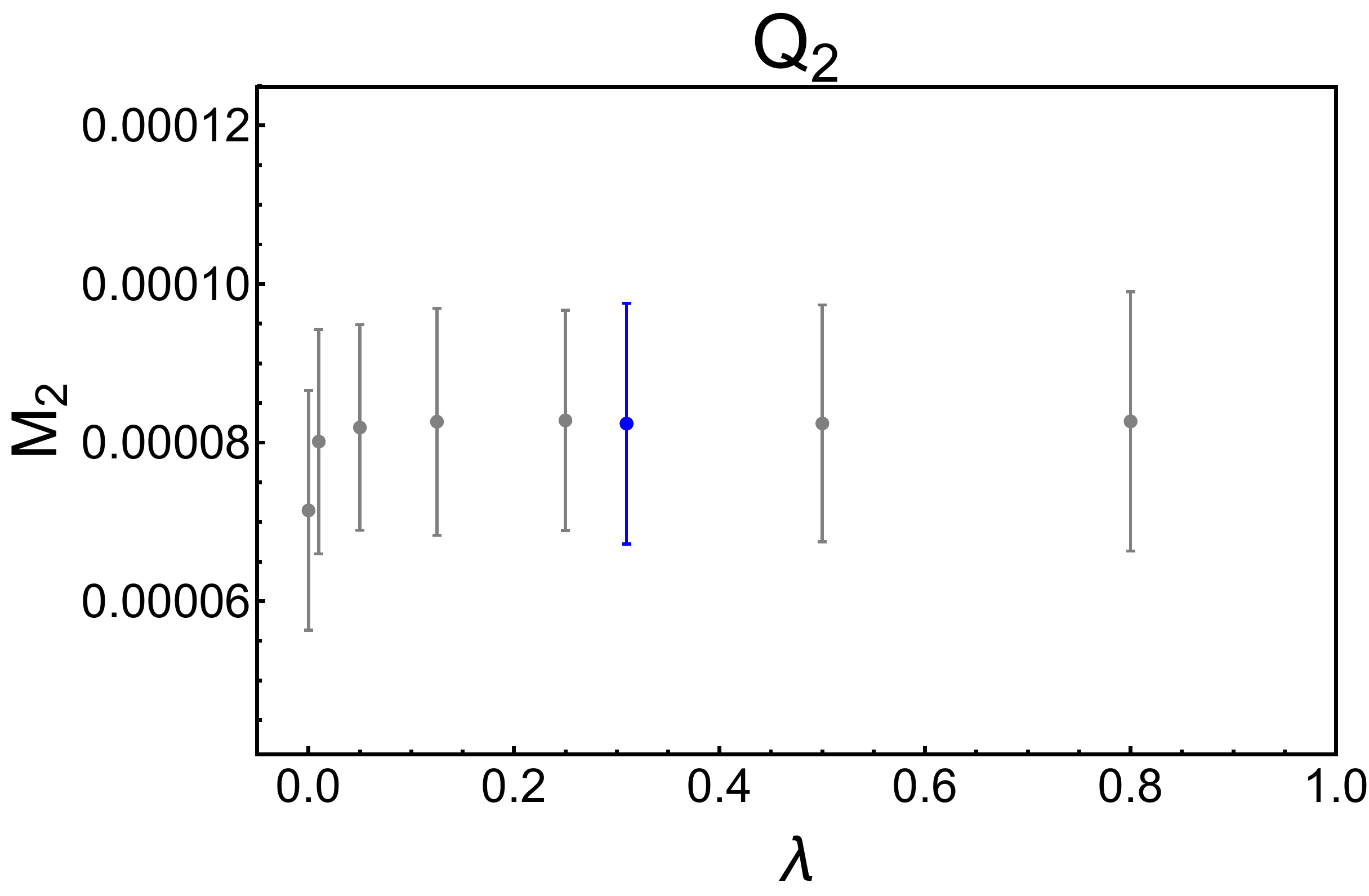}
  \includegraphics[width=.45\textwidth]{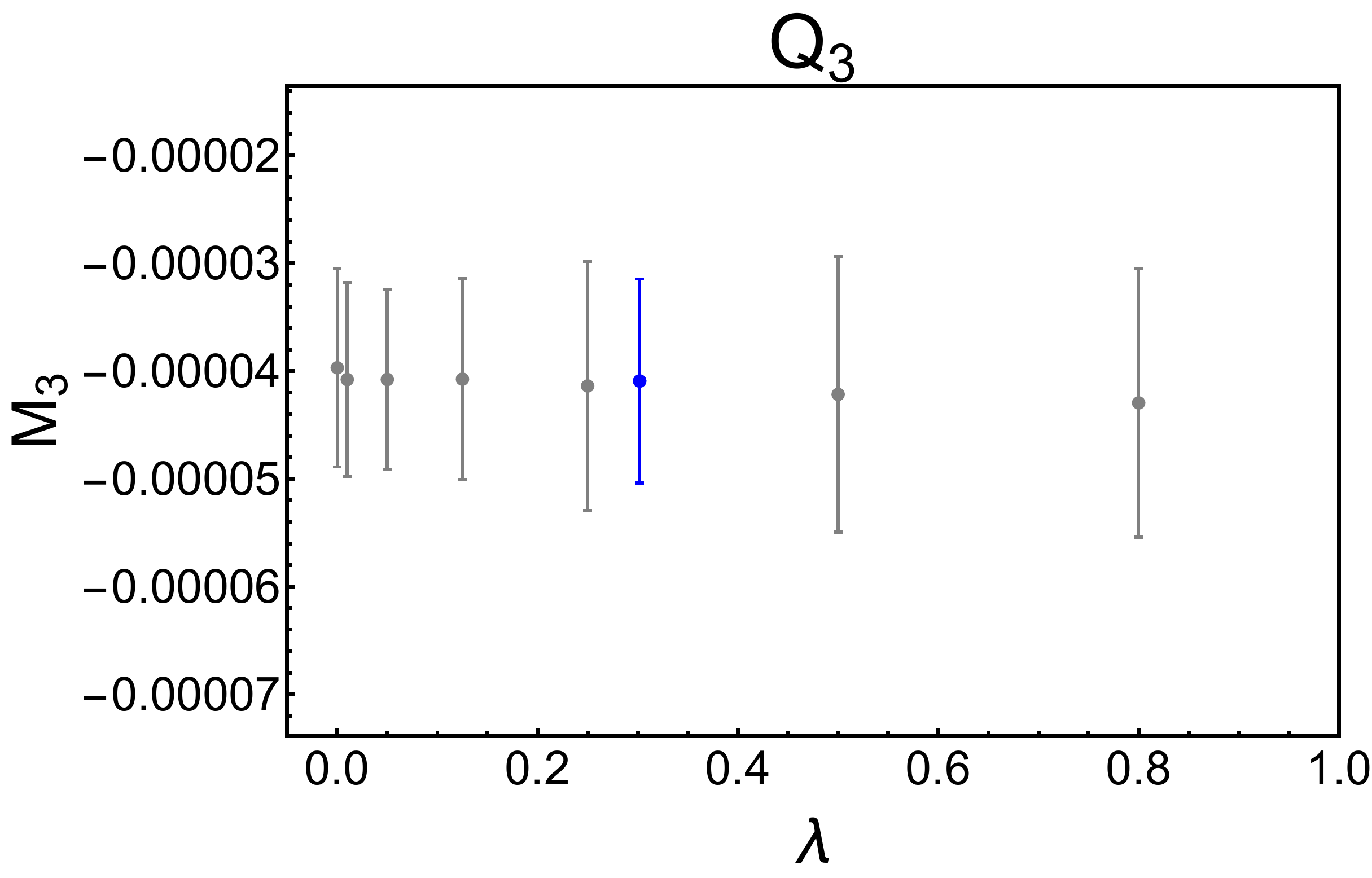}
  \includegraphics[width=.45\textwidth]{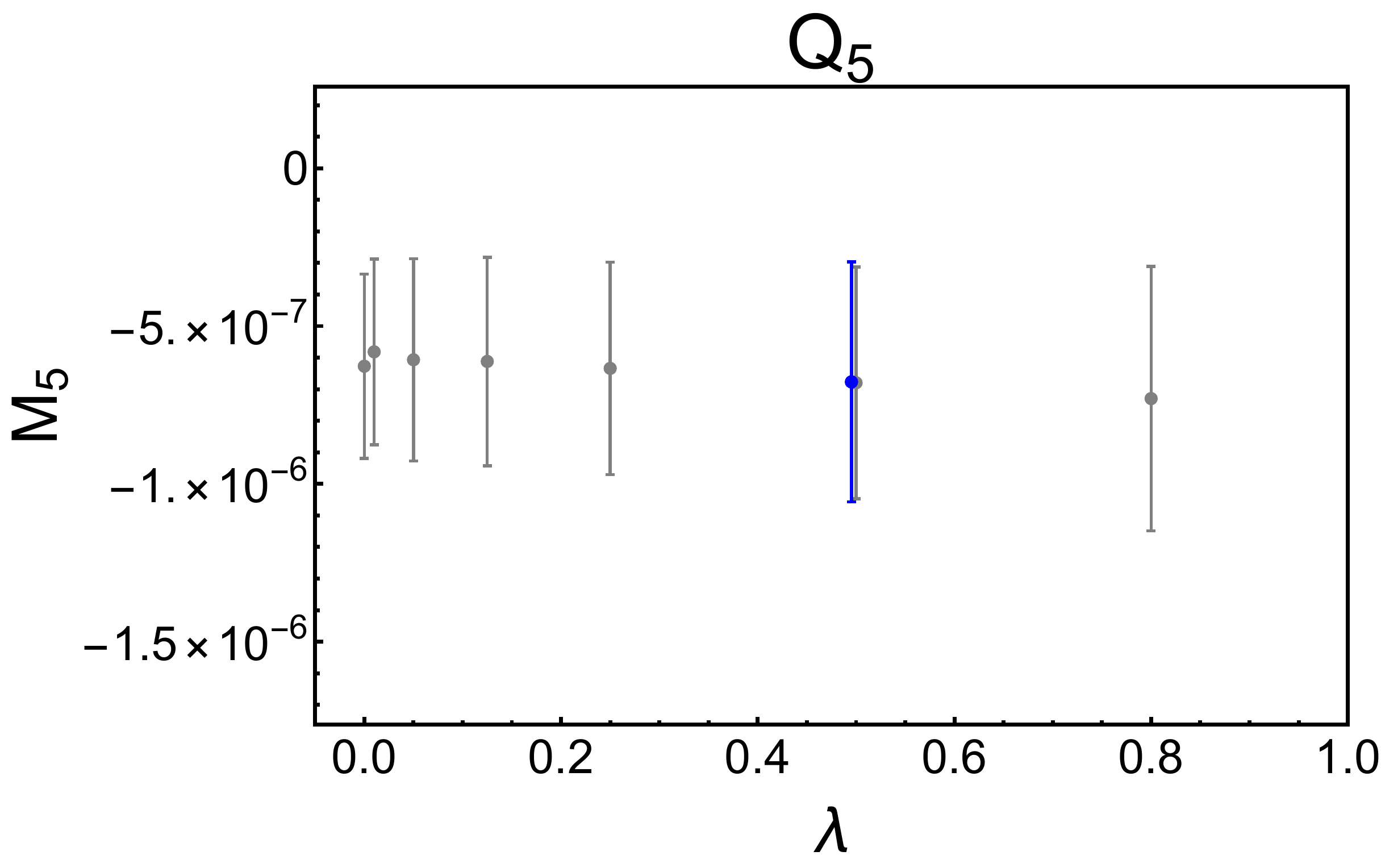}
  \caption{Sensitivity of resulting (bare) matrix elements $\mcM_I$ obtained from three-point function
    fits (Eq.~(\ref{eqn:3pt2state})) to the shrinkage parameter $\lambda$ that is used to estimate
    the covariance matrix as described in the main text (\emph{black points}).
    The \emph{dark blue points} indicate the values obtained with optimal shrinkage parameters
    $\lambda^*$ (see Eq.~(\ref{eqn:shrinkagedef}) and Appendix~\ref{app:shrinkage_cov}).
    The fit ranges are shown in the fourth row of Tab.~\ref{tab:3ptResults}.
    \label{fig:3ptShrink}}
\end{figure}

With \emph{all-mode-averaging} described in Sec.~\ref{sec:lattice_setup}, we obtain one unbiased
sample of the two- and three-point functions per gauge field configuration.
The number of gauge field configurations used in this calculation $N_\text{conf}=30$ is not
large enough to obtain nondegenerate determination of a covariance matrix for the required
number of data points $31 \leq K \leq 76$ included in the three-point correlator fits.
Therefore, spin and parity symmetries are used to increase the effective number of unbiased
samples of correlation functions.
Thus, $G_{n Q^\dagger \bar{n}}$ and $G_{n (Q_I^\mathcal{P})^\dagger \bar{n}}$ with two
polarizations are treated as four samples per gauge-field configuration, resulting in
$N=120$ samples for each data point after all-mode-averaging bias correction.
Polarized two-point functions $G_{nn(\pm1/2)}$, $G_{\bar{n}\bar{n}(\pm1/2)}$ are similarly combined
to obtain a statistical ensemble of $N=120$ two-point functions.
Although this yields an ``ensemble'' with $N > K$ samples, it is still not sufficient for
reliable determination of covariance matrix, which typically requires $N \gtrsim K^2$.

\begin{figure}
  \centering
  \includegraphics[width=.45\textwidth]{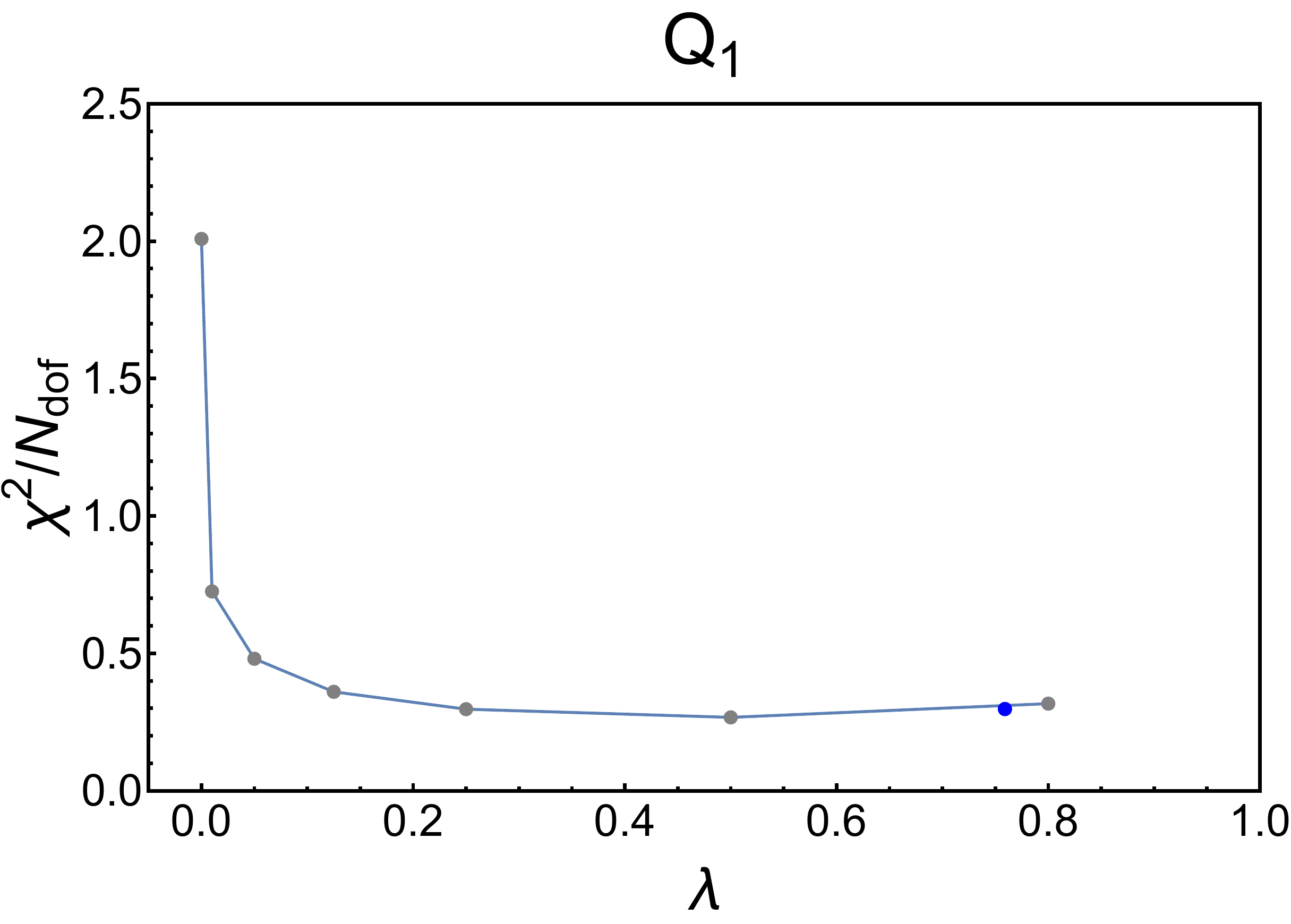}
  \includegraphics[width=.45\textwidth]{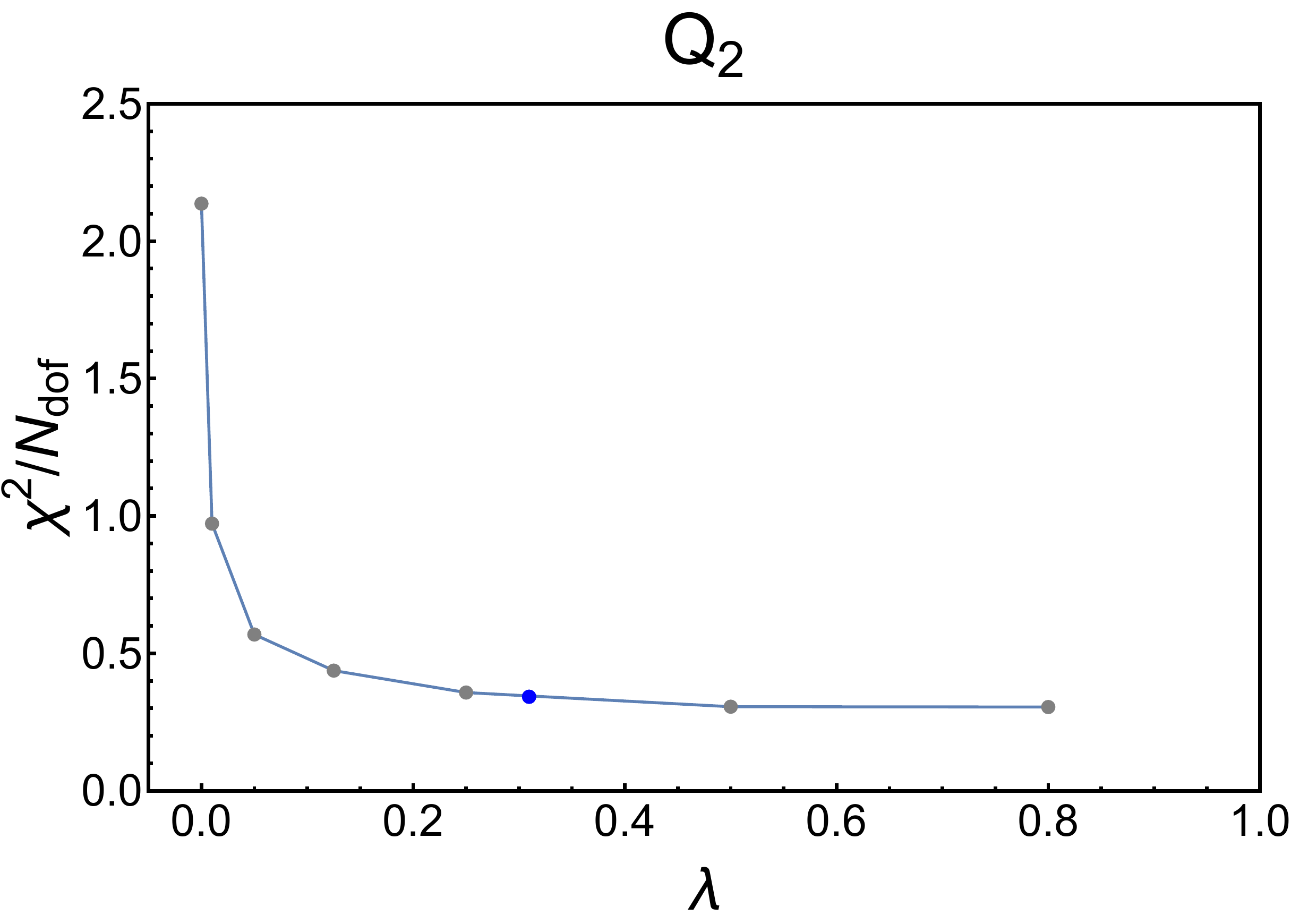}
  \includegraphics[width=.45\textwidth]{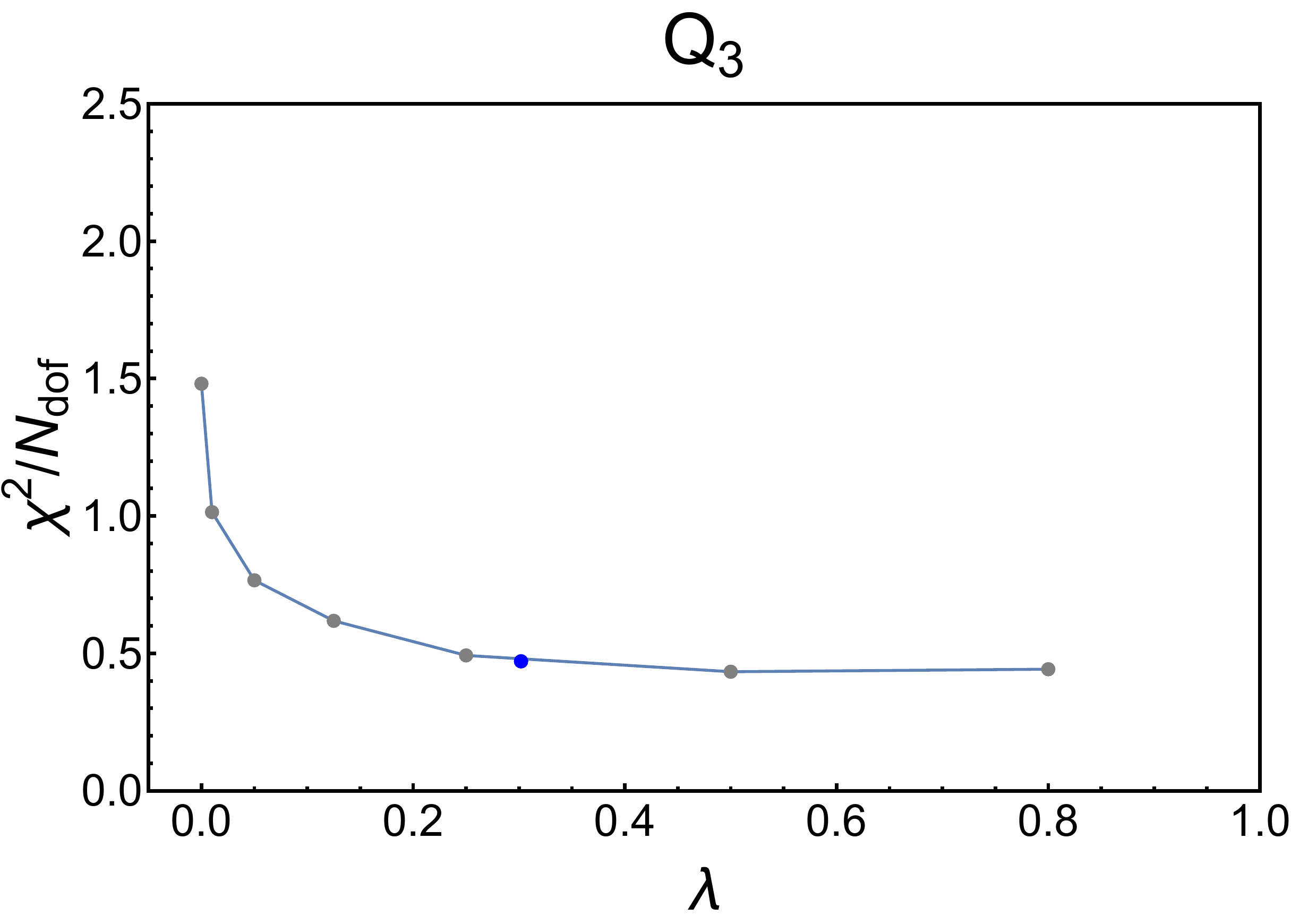}
  \includegraphics[width=.45\textwidth]{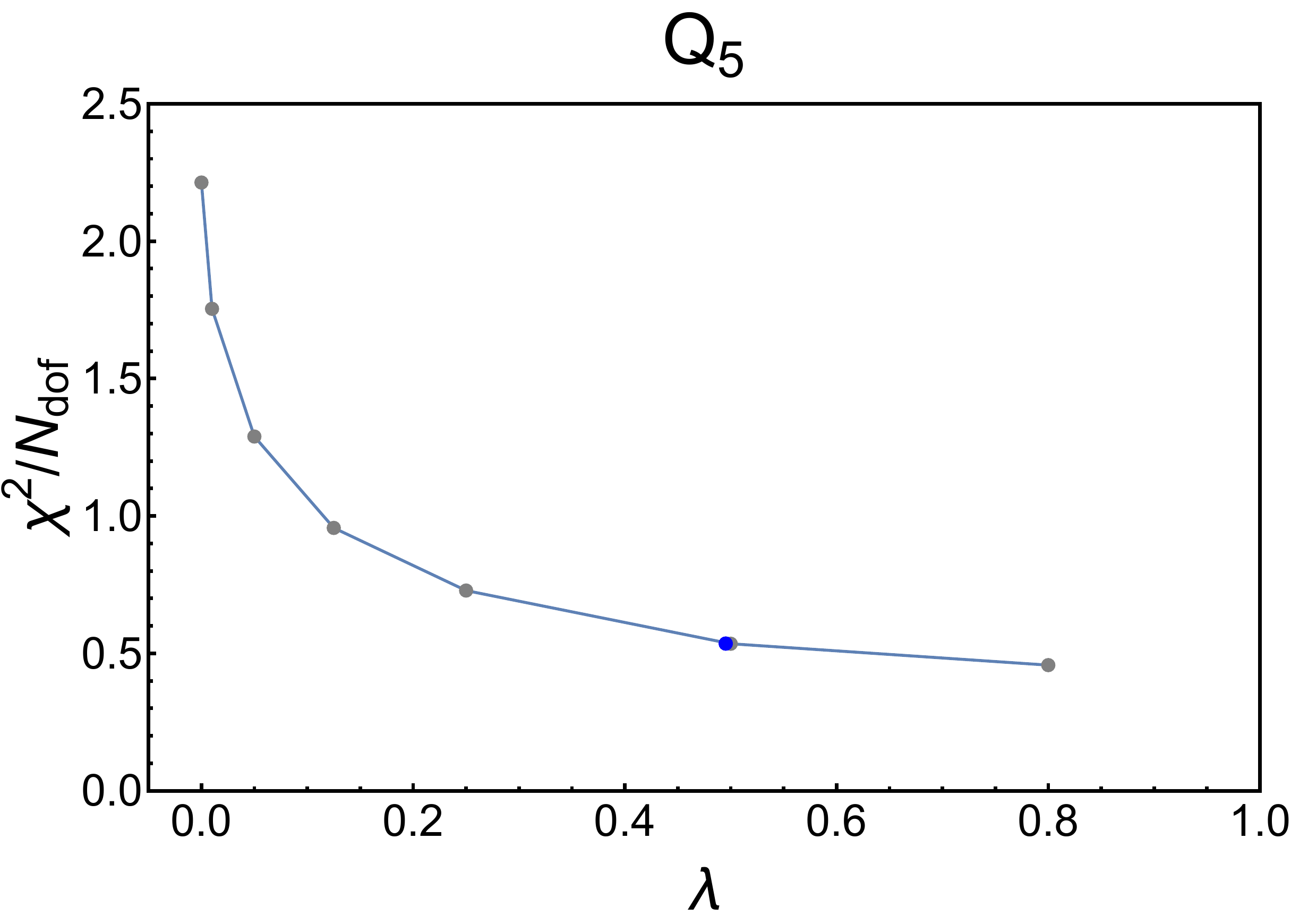}
  \caption{Dependence of correlated $\chi^2/N_{\text{dof}}$ values of three-point function fits
    (Eq.~(\ref{eqn:3pt2state})) on the shrinkage parameter $\lambda$ that is used to estimate
    the covariance matrix as described in the main text (\emph{black points}).
    The \emph{dark blue points} indicate the values obtained with optimal shrinkage parameters
    $\lambda^*$ (see Eq.~(\ref{eqn:shrinkagedef}) and Appendix~\ref{app:shrinkage_cov}).
    The fit ranges are shown in the fourth row of Tab.~\ref{tab:3ptResults}.
  \label{fig:3ptShrinkChisq}}
\end{figure}

For both two-point and three-point functions, finite sample-size fluctuations may make the
sample covariance matrix ill-determined and lead to a numerically unstable
inverse covariance matrix required for least-squares fitting.
Shrinkage~\cite{stein1956,LEDOIT2004365} has been proposed as a method of improving the
condition number of covariance matrix estimates.
Denoting the sample covariance matrix by $S$, the corresponding covariance matrix estimate with
shrinkage is given by
\begin{equation}
   \begin{split}
     \Sigma(\lambda) =  \lambda\, \text{diag}(S) + (1-\lambda)S,
   \end{split}\label{eqn:shrinkagedef}
\end{equation}
where $0 \leq \lambda \leq 1$ is the shrinkage parameter, and $\text{diag}(S)$ is a particular
``shrinkage target''.
Taking any $\lambda > 0$ ``shrinks'' the spectrum of the covariance matrix by reducing the
relative size of off-diagonal correlations compared to the diagonal covariance matrix elements.
This leads to a better-conditioned covariance matrix and a more robust estimate of the inverse
covariance matrix used for $\chi^2$-minimization.
Trivial $\lambda = 0$ corresponds to no shrinkage, while $\lambda = 1$ removes
off-diagonal correlations completely, which is equivalent to an uncorrelated fit.
Therefore, varying the parameter $0\le\lambda\le1$ interpolates continuously between correlated
(albeit with potentially poorly-determined covariance matrix) and uncorrelated fits.
A standard prescription for choosing the optimal shrinkage parameter is to minimize the
\emph{rms} difference between $\Sigma(\lambda)$ and the true covariance matrix.
A sample estimator for the optimal shrinkage parameter $\lambda^*$ is suggested in
Ref.~\cite{LEDOIT2004365} and summarized in Appendix~\ref{app:shrinkage_cov}.
Bootstrap covariance matrices with optimal shrinkage\footnote{
  $\lambda^*$ are chosen to provide optimal shrinkage for the normalized sample correlation
  matrix as described in Appendix~\ref{app:shrinkage_cov} rather than the bootstrap covariance
  matrix.
  It is possible that finite-sample-size bias will lead to differences between the optimal
  shrinkage parameters for the two matrices.
  Since $\lambda^*$ defined by this prescription vanishes in the infinite-statistics limit, the
  bootstrap covariance matrix obtained with this choice of shrinkage parameter will provide an
  unbiased (but not necessarily optimal) estimate of the true covariance matrix in the
  infinite-statistics limit.
}
$\Sigma^* = \Sigma(\lambda^*)$ are obtained by inserting $\lambda^*$ from
Eq.~(\ref{eqn:opt_lambda}) into Eq.~(\ref{eqn:shrinkagedef}) with $S$ the bootstrap covariance
matrix obtained from $N_{boot} = 10,000$ samples of two-point and three-point correlation
functions.
The effects of shrinkage on the central values, uncertainties, and goodness-of-fit of the matrix
element fits described below are explored by varying $\lambda$, and the results for one choice of
fit range are shown in Figs.~\ref{fig:3ptShrink}-\ref{fig:3ptShrinkChisq}.
For all operators, the central values and the statistical uncertainties are relatively insensitive
to the value of the shrinkage parameter once $\lambda>0$.
The $\chi^2/N_{\text{dof}}$ values decrease sharply in a small region around $\lambda=0$,
however they are much less sensitive for larger $\lambda$ values.
In all cases, the optimal values $\lambda^*$ for the shrinkage parameter are found outside of the
region of strong dependence of $\chi^2$ on $\lambda$.

\begin{figure}
  \centering
  \includegraphics[width=.45\textwidth]{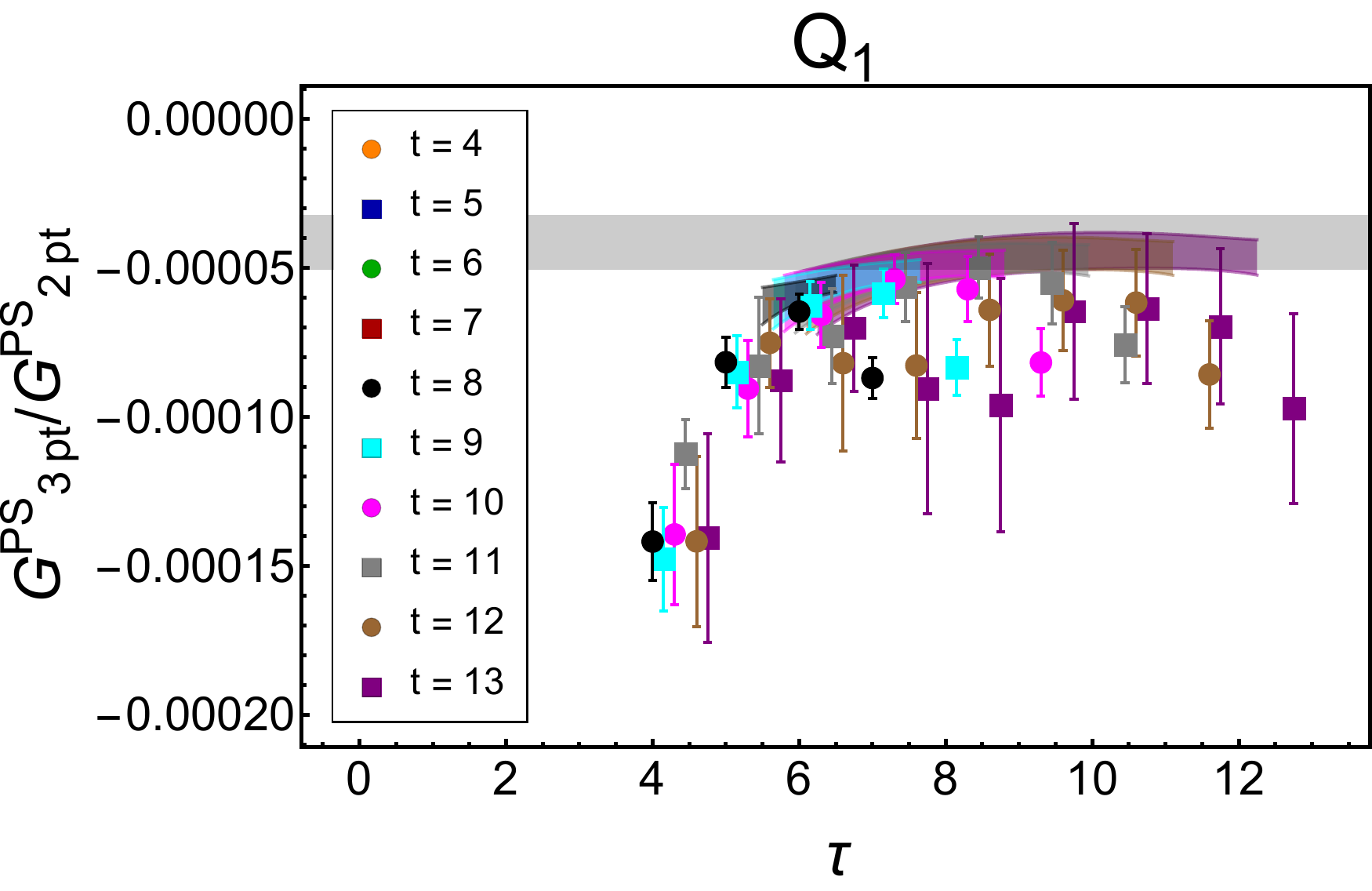}~
  \hspace{.05\textwidth}~
  \includegraphics[width=.45\textwidth]{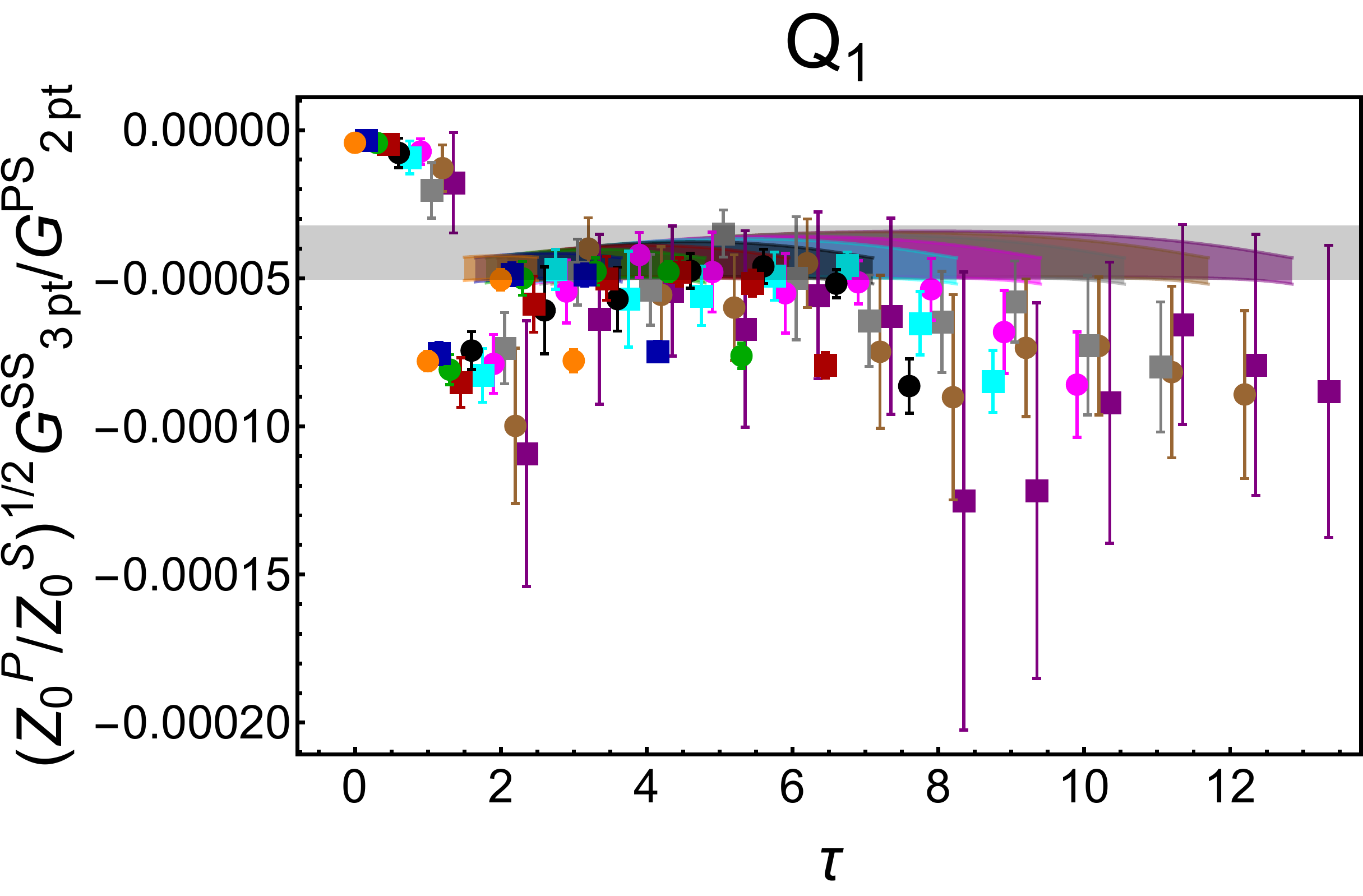}\\
  \includegraphics[width=.45\textwidth]{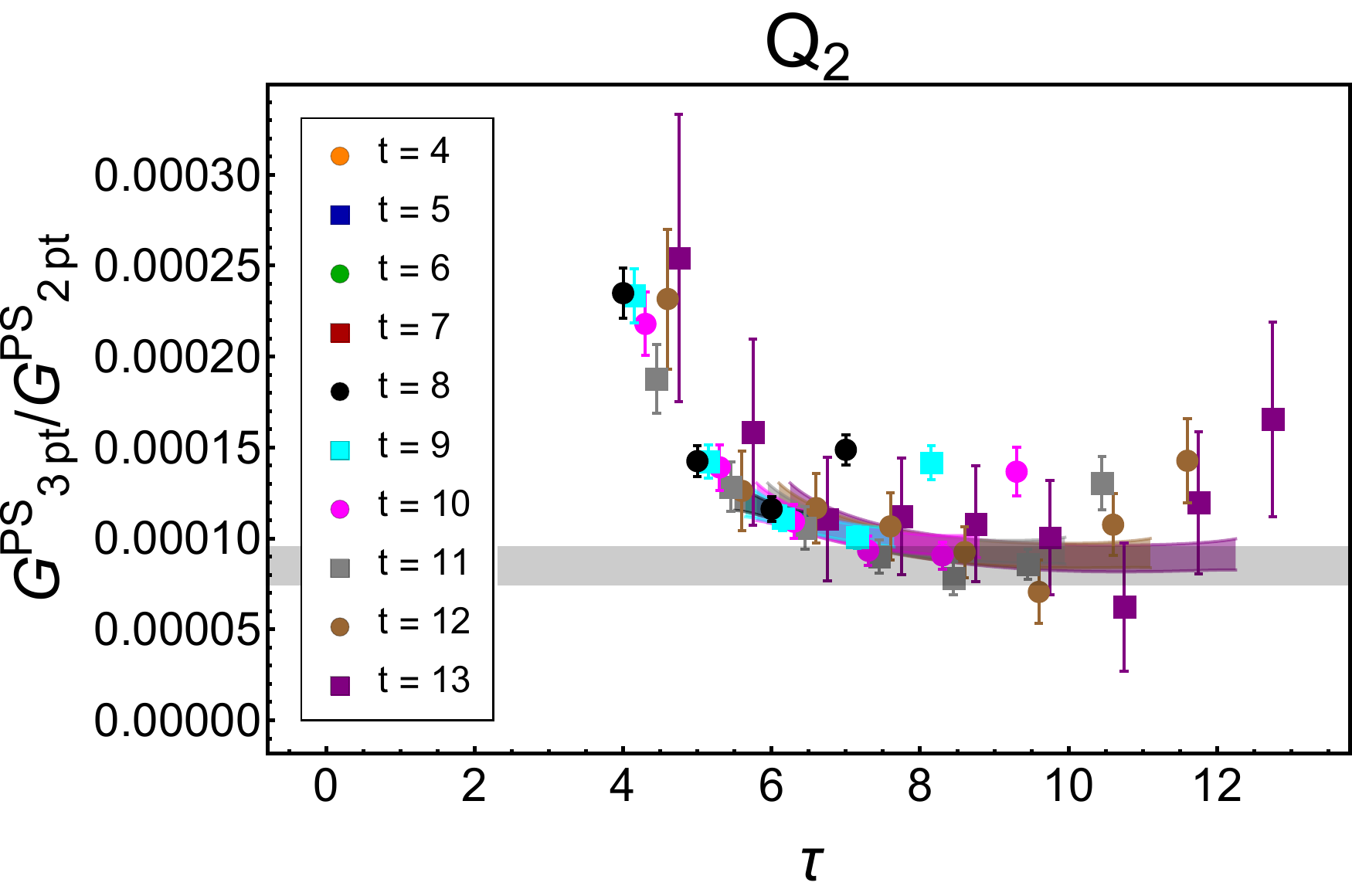}~
  \hspace{.05\textwidth}~
  \includegraphics[width=.45\textwidth]{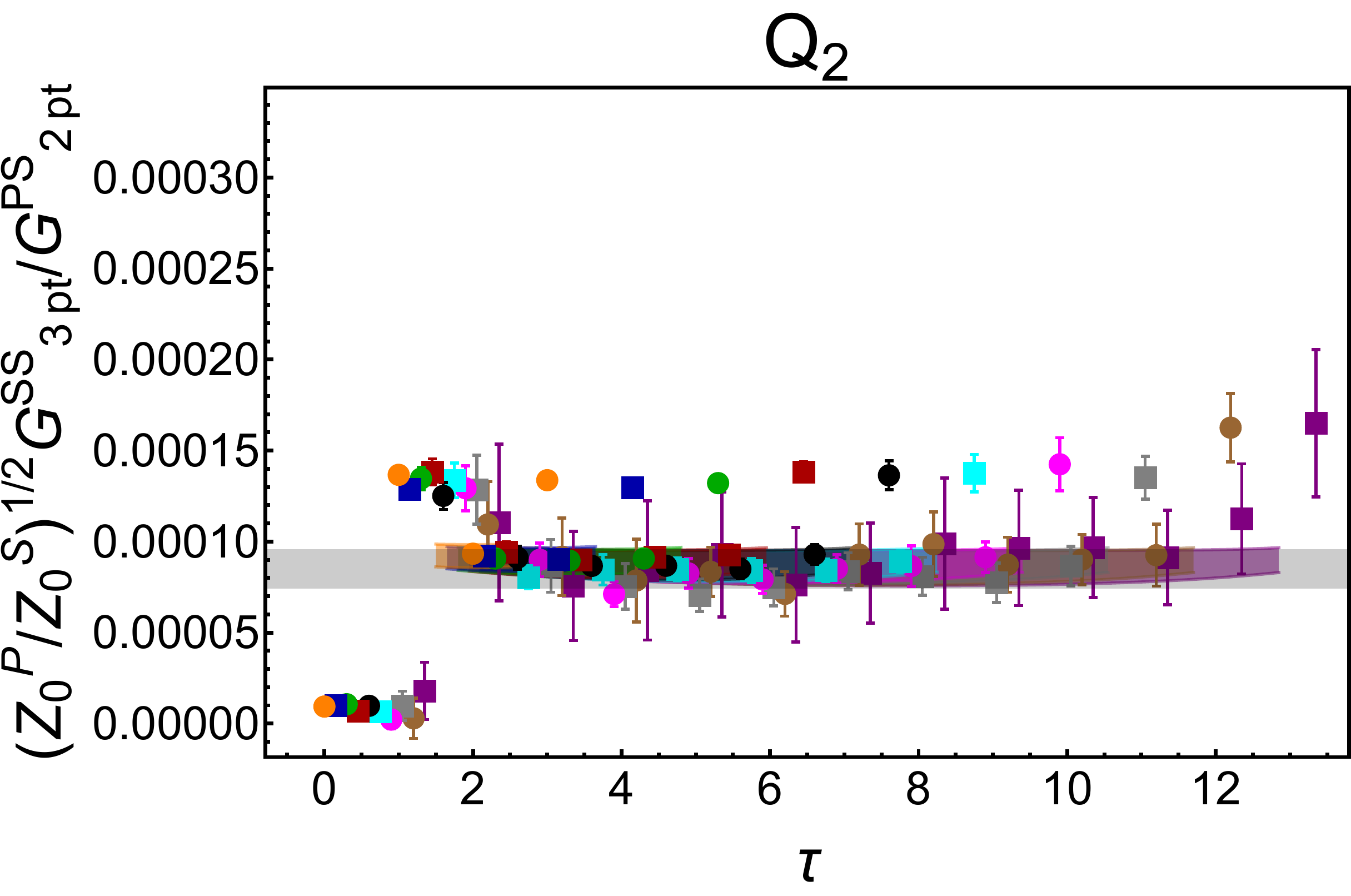}\\
  \includegraphics[width=.45\textwidth]{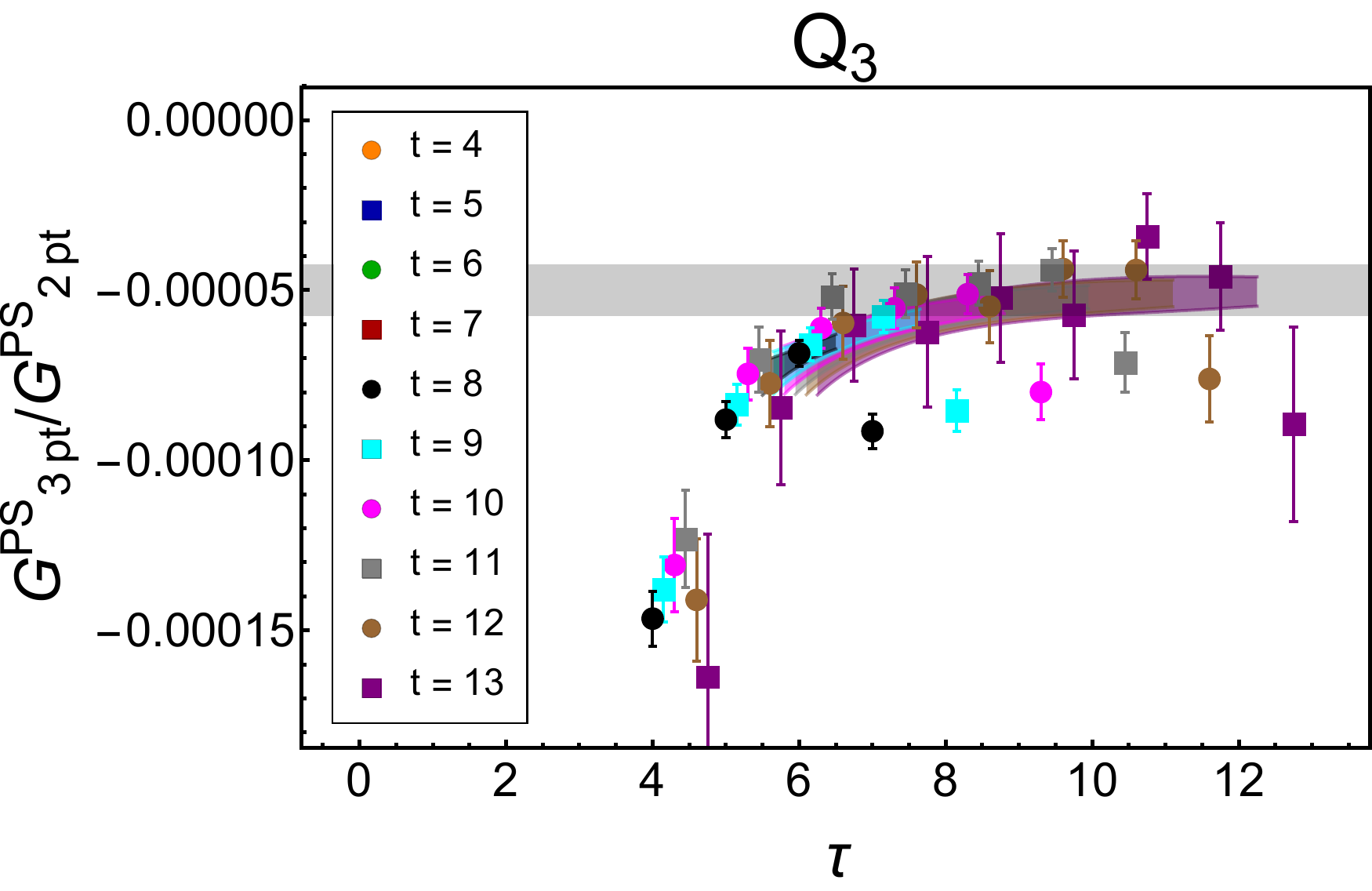}~
  \hspace{.05\textwidth}~
  \includegraphics[width=.45\textwidth]{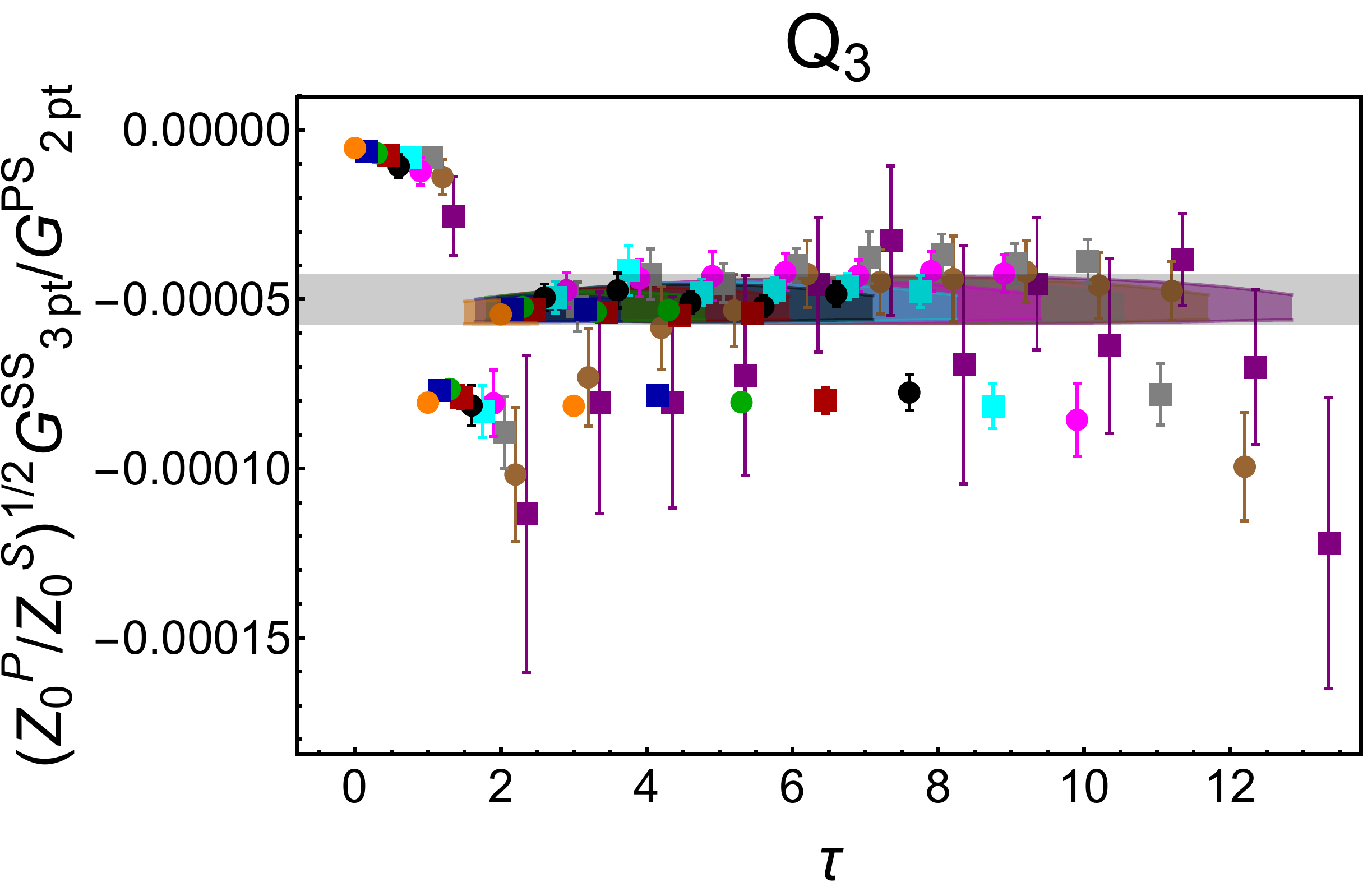}\\
  \includegraphics[width=.45\textwidth]{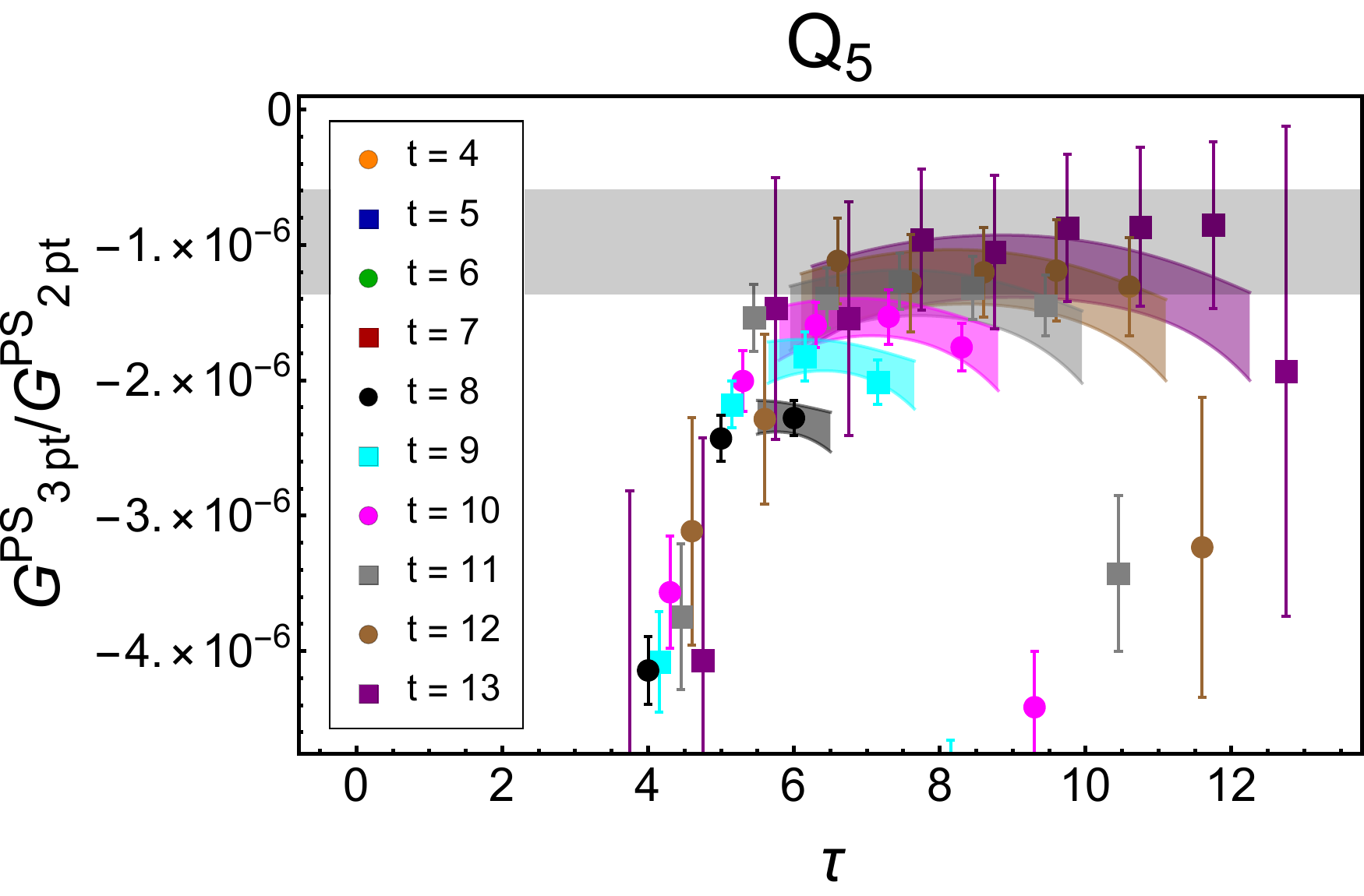}~
  \hspace{.05\textwidth}~
  \includegraphics[width=.45\textwidth]{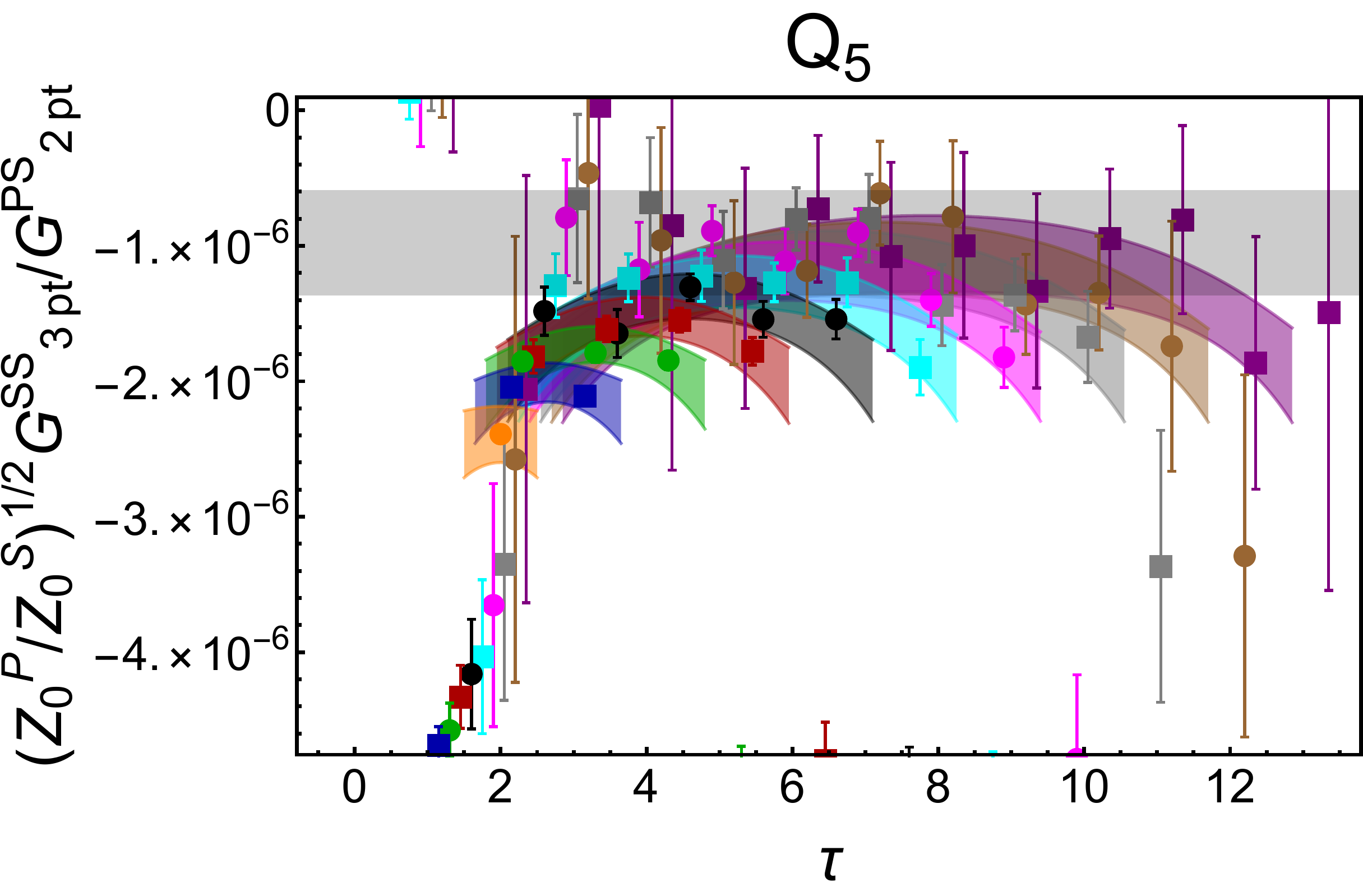}\\
  \caption{
    Combined correlated $\chi^2$ fits of $PS$, $SS$ three-point functions to
    Eq.~(\ref{eqn:3pt2state}) for operators $Q_1$, $Q_2$, $Q_3$, and $Q_5$
    for the time range shown in the first row of Tab.~\ref{tab:3ptResults} (\emph{shaded bands}).
    The state energies are determined from fits to two-point functions as shown in Fig.~\ref{fig:2pt}.
    Covariance matrix is estimated with optimal shrinkage $\lambda^*$ as described in the main text.
    Corresponding data points for ratios of three- to two-point correlation functions for all
    used source/sink separations $\tsep$ are shown with intermittent \emph{square and circle} data points.
    The central values of matrix elements $\mcM_I$ and their statistical uncertainties
    are shown with \emph{gray shaded bands}.
    All the displayed uncertainties are statistical and estimated using bootstrap.
    \label{fig:3ptQ1235}}
\end{figure}

The average two-point function and the corresponding bootstrap covariance matrix with optimal
shrinkage are used for nonlinear $\chi^2$-minimization to determine $E_0$, $E_1$,
$\sqrt{Z_0^P}$, and $\sqrt{Z_0^S}$.
$\chi^2$-minimization is reduced to a two-parameter optimization problem by \emph{variable
projection(VarPro)} technique~\cite{varpro0,varpro1} detailed in Appendix~\ref{app:var_pro}.
In VarPro, the products of overlap factors in Eq.~(\ref{eqn:2pt2state}) are found from a linear
$\chi^2$-fit for particular values of $E_{0,1}$ and the solution is substituted back into
$\chi^2$ in order to obtain a two-parameter function $\chi^2_{VP}(E_0, E_1)$, which is then
minimized using nonlinear numerical methods.
With these four parameters held fixed, the remaining six free parameters in the three-point
function fit~(\ref{eqn:3pt2state}) can also be found from a linear $\chi^2$  fit.
The parameter covariance matrix for all 10 parameters is subsequently estimated using an
additional correlated bootstrap resampling and fitting of two- and three-point function data.
The original bootstrap covariance matrices are used in all these fits in order to avoid the
possibility of ill-conditioned covariance matrices.
The bootstrap parameter covariance matrix is obtained from $N_{boot}^\prime = 200$ correlated
resampling draws.
The parameter covariance matrix is diagonalized, the eigenvalues are resampled
$N_{boot}^{\prime\prime} = 200$ times, and the resampled eigenvalues are transformed back to the
original parameter basis.
Finally, the standard deviation of the resulting resampled values of $\mcM_I$ is used to define
the marginalized uncertainty of $\mcM_I$ for each fit range shown in Tab.~\ref{tab:3ptResults}.
An analogous procedure is used to obtain the uncertainties of $E_0$ and $E_1$ shown in
Tab.~\ref{tab:2ptResults}.
Results for the matrix elements from the fits that include our smallest $\tsep$ value are compared
to the ratios of three-point to two-point functions (adjusting for proper overlap factors) in
Figs.~\ref{fig:3ptQ1235}.
In addition, the two-point function fits are compared to the corresponding effective masses in
Fig.~\ref{fig:2pt}.

\begin{figure}
  \centering
  \includegraphics[width=.45\textwidth]{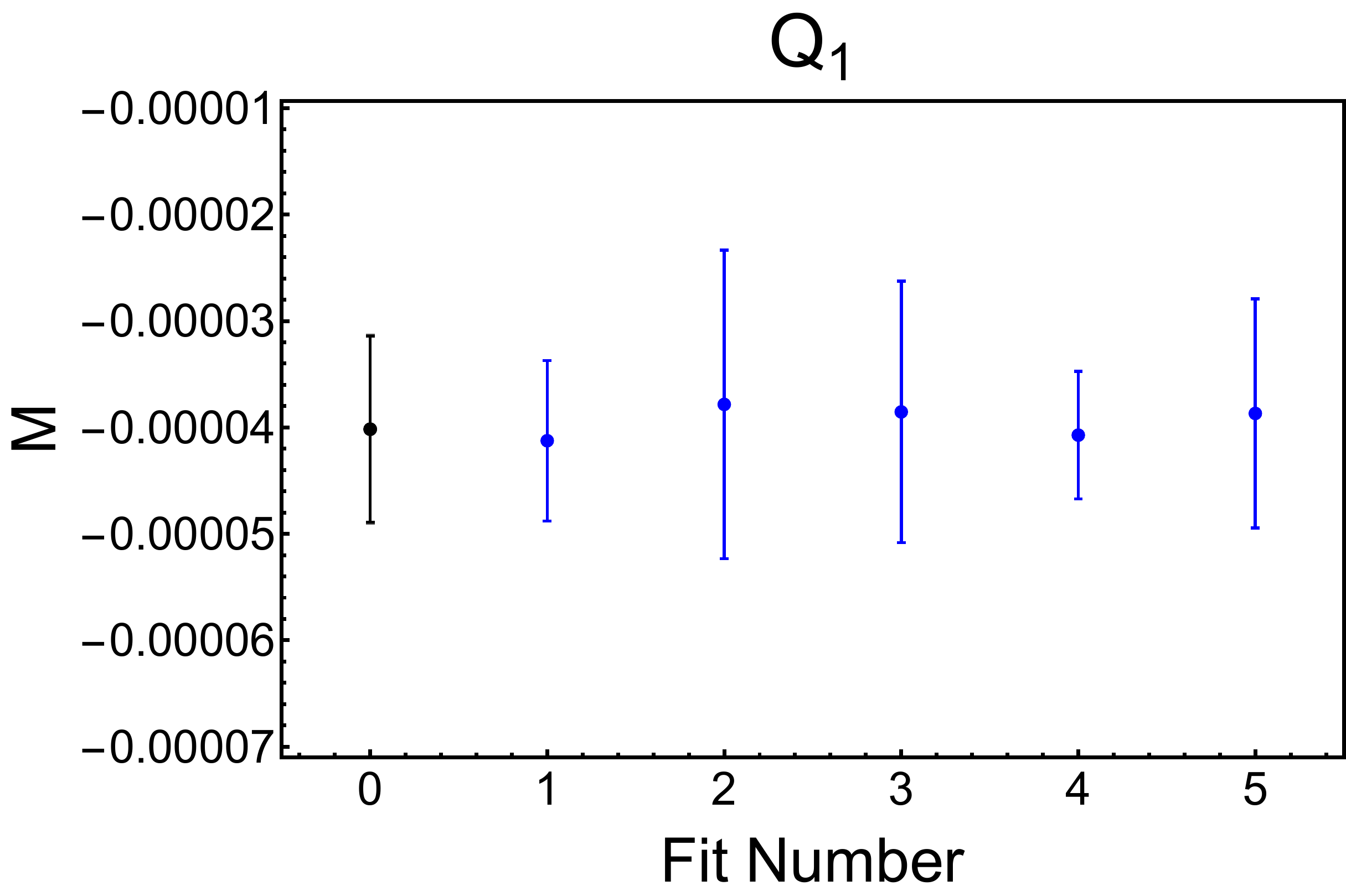}
  \includegraphics[width=.45\textwidth]{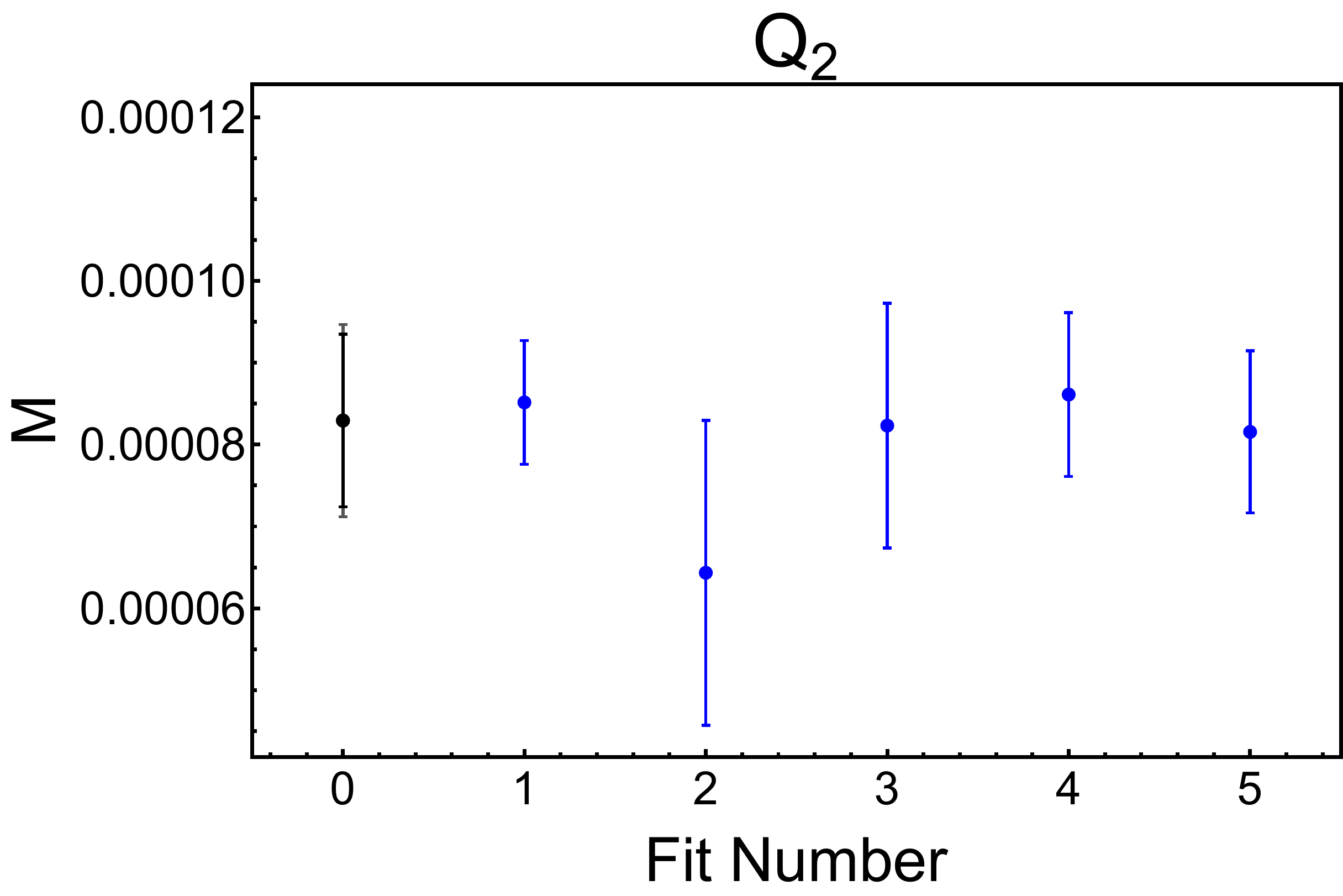}
  \includegraphics[width=.45\textwidth]{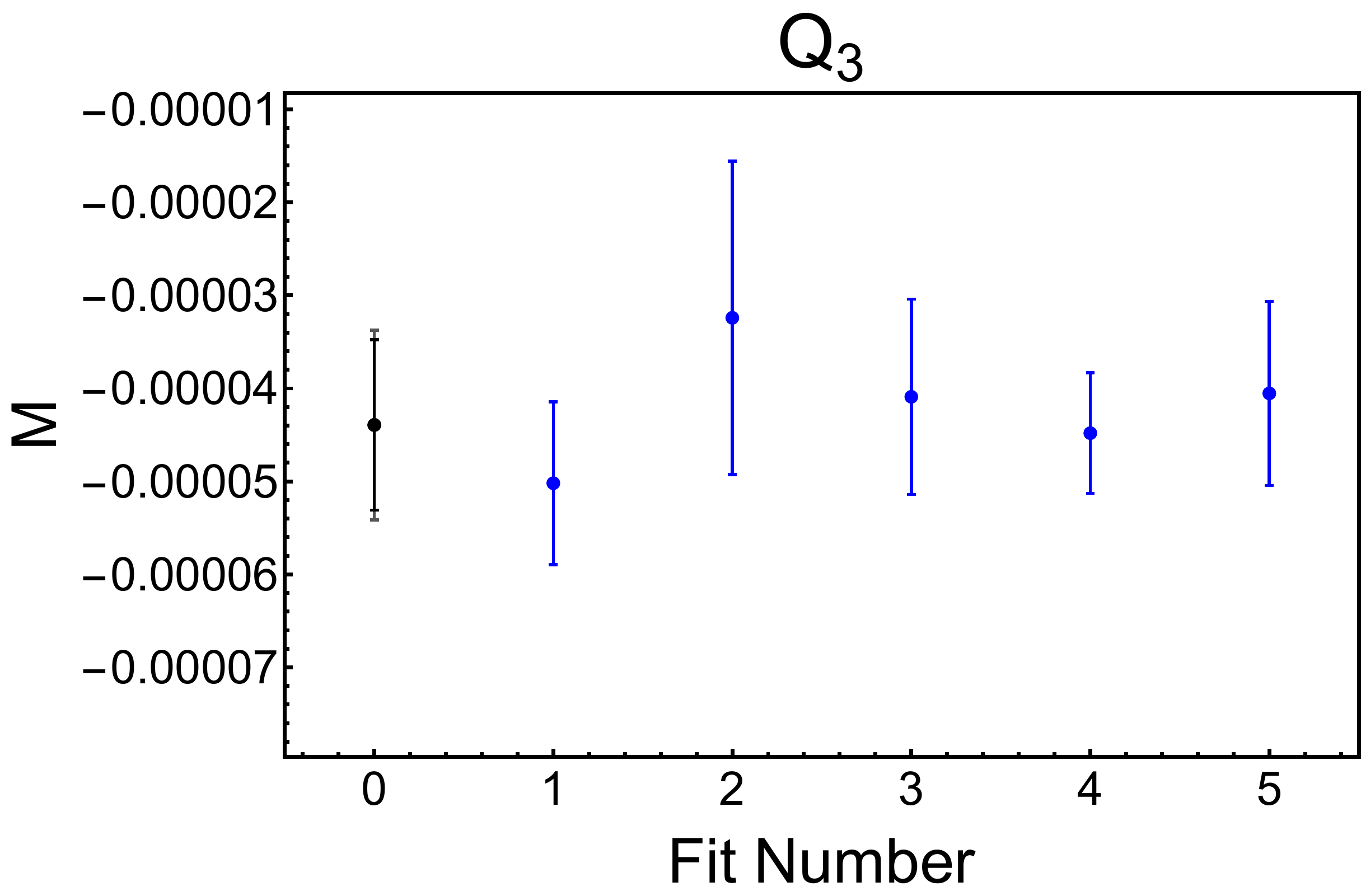}
  \includegraphics[width=.45\textwidth]{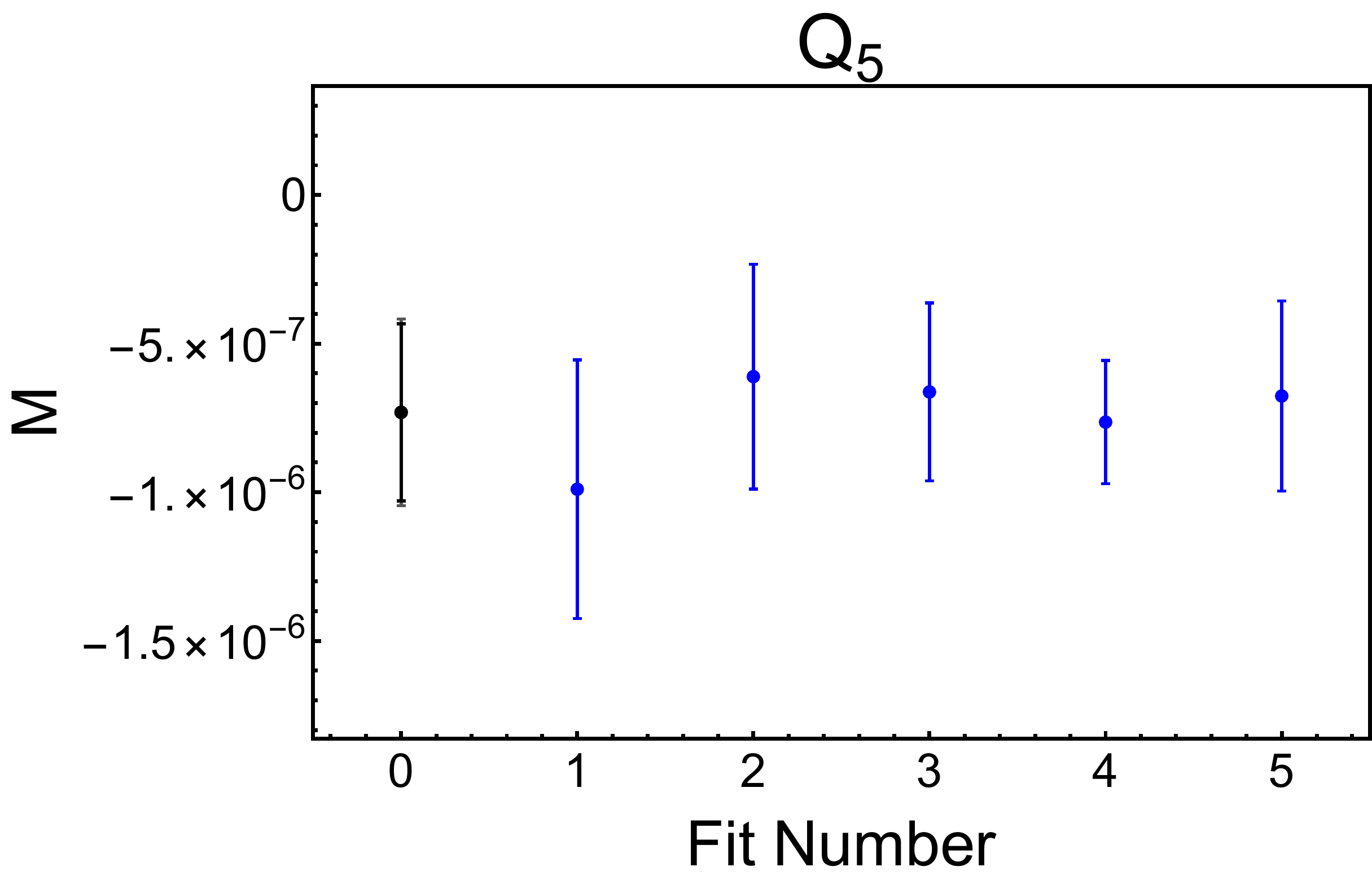}
  \caption{Comparison of ground state matrix elements (lattice units)
    extracted in fits with different fit ranges that are listed in Tab.~\ref{tab:3ptResults}.
    The black point at zero indicates the result of the weighted averaging procedure described
    in Appendix~\ref{app:stat_modavg}, and the small and large error bars indicate
    statistical and statistical-plus-systematic uncertainties, respectively.
  \label{fig:3ptFits}}
\end{figure}

Systematic uncertainties of our analysis procedure are studied by varying the time ranges of
data included in the two-state fits.
Results of fits of two-point function data
$G^{PP}_{nn(\bar{n}\bar{n})}(t^{min}_{PP} \le t \le \tsep^{max})$ and
$G^{PS}_{nn(\bar{n}\bar{n})}(t^{min}_{PS} \le t \le \tsep^{max})$
for a variety of $t^{min}_{PP}$ and $t^{min}_{PS}$ are shown in Tab.~\ref{tab:2ptResults}.
Results of corresponding fits of three-point function data
$G_{n Q_I^\dag \bar{n}}^{SS}(\tau_{min}^P + \tau_{min}^S \leq \tsep \leq \tsep^{max},
\tau_{min}^S \le \tau \le \tsep - \tau_{min}^S)$
and $G_{n Q_I^\dag \bar{n}}^{PS}(2\tau_{min}^S \leq \tsep \leq \tsep^{max},
\tau_{min}^P \le \tau \le \tsep - \tau_{min}^S)$
for a variety of $\tau_{min}^S$ and $\tau_{min}^P$ are shown in Tab.~\ref{tab:3ptResults}.
The data for $SP$ and $PS$ three-point correlation functions are averaged using relation
$G_{n Q_I^\dag \bar{n}}^{SP}(\tsep,\tau) = G_{n Q_I^\dag \bar{n}}^{PS}(\tsep,\tsep-\tau)$
to reduce the number of data points in the fits.
Results for bare ground-state matrix elements from different fits are in very good agreement
with each other, as shown in Fig.\ref{fig:3ptFits}.

The five fit range choices shown in Tabs.~\ref{tab:2ptResults}-\ref{tab:3ptResults}
result in acceptable correlated $\chi^2/N_\text{dof}$ values in fits of two- and three-point
function data.
These results are combined into final estimates of $\mcM_I$ and estimates of their
statistical and systematic uncertainties.
Since the various fits have different $N_\text{dof}$, we use a weighted-averaging procedure
defined in Appendix~\ref{app:stat_modavg}.
For a particular fit, the weight is a combination of the likelihood that the fit describes the
data (we use its $p$-value as the likelihood proxy) and its statistical precision, to penalize
both fits that fail to describe data and fits that do not constrain the relevant parameters.
The same weights are used to determine the average statistical uncertainty, which ensures that
including multiple similar fits will not lead to a spurious reduction in the final statistical
uncertainty.
The weighted mean-square difference between each fit result and the weighted average is
used to define the systematic uncertainty due to arbitrariness of choice of a fit window.
Applying this weighted averaging procedure to the ground-state energy $E_0$ of the two-point
function yields the result for the nucleon mass that agrees well with the physical value,
\begin{equation}
   \begin{split}
      E_0 = 0.565(24)(8) a^{-1} = 977(42)(13) \text{ MeV},
   \end{split}\label{eqn:E0}
\end{equation}
where we have used the scale-setting result $a = 0.1141(3)\,\text{fm}$ from
Ref.~\cite{Blum:2014tka}, which has negligible uncertainty for our purposes as it is
much smaller compared to other uncertainties in our calculation.
Applying the same procedure to the fit results in Tab.~\ref{tab:3ptResults} provides our final
estimate of the bare matrix elements including statistical and fitting systematic uncertainties,
\begin{equation}
\begin{split}
\mcM_1^{\text{lat}} &= -3.99(1.08)(0.13) \times 10^{-5} a^{-6} = -107(29)(3) \times 10^{-5} \text{ GeV}^{6}\\
\mcM_2^{\text{lat}} &= 8.28(1.29)(0.54)\times 10^{-5} a^{-6} = 221(35)(14) \times 10^{-5} \text{ GeV}^{6}\\
\mcM_3^{\text{lat}} &= -4.37(0.86)(0.52)\times 10^{-5} a^{-6} = -117(23)(14) \times 10^{-5} \text{ GeV}^{6}\\
\mcM_5^{\text{lat}} &= -0.075(32)(10)\times 10^{-5} a^{-6} = -2.01(86)(22) \times 10^{-5} \text{ GeV}^{6}.
\end{split}\label{eqn:MIbare}
\end{equation}
These lattice regularized matrix elements can be related to renormalized matrix elements through
NPR as described in the next section.

\begin{table}
   \begin{tabular}{|c|c|c|c||c|c|c||c|c|c||c|c|c||c|c|c|}
   \hline
   $\tau^{min}_{P}$ & $\tau^{min}_{S}$ & $\tsep^{max}$ & $N_\text{dof}$ &
      $\mcM_1^{\text{lat}} \times 10^5$ & $\chi^2/N_\text{dof}$ & $\lambda^*$ &
      $\mcM_2^{\text{lat}} \times 10^5$ & $\chi^2/N_\text{dof}$ & $\lambda^*$ &
      $\mcM_3^{\text{lat}} \times 10^5$ & $\chi^2/N_\text{dof}$ & $\lambda^*$ &
      $\mcM_5^{\text{lat}} \times 10^5$ & $\chi^2/N_\text{dof}$ & $\lambda^*$\\\hline
   6 & 2 & 13 & 70 & -4.13(0.92) &  0.25 & 0.77 & 8.50(1.07) & 0.40 & 0.35 &
      -5.01(0.76) & 0.44 & 0.32 &  -0.098(39) & 0.62 & 0.53 \\\hline
   6 & 4 & 13 & 25 & -3.81(1.78) &  0.44 & 0.72 & 6.46(2.15) & 0.31 & 0.31 &
      -3.21(1.25) & 0.40 & 0.29 &  -0.063(45) & 0.53 & 0.41 \\\hline
   6 & 3 & 13 & 45 & -3.85(1.07) &  0.30 & 0.76 & 8.24(1.52) & 0.34 & 0.31 &
      -4.09(0.95) & 0.47 & 0.30 &  -0.068(38) & 0.54 & 0.50 \\\hline
   5 & 3 & 13 & 51 & -4.09(0.92) &  0.28 & 0.75 & 8.61(1.06) & 0.34 & 0.29 &
      -4.50(0.67) & 0.44 & 0.29 &  -0.077(22) & 0.54 & 0.47 \\\hline
   7 & 3 & 13 & 40 & -3.87(1.13) &  0.34 & 0.76 & 8.13(1.32) & 0.37 & 0.32 &
      -4.05(1.00) & 0.50 & 0.31 &  -0.069(32) & 0.55 & 0.53 \\\hline
   \multicolumn{4}{|l|}{Weighted Ave} & \multicolumn{3}{|l|}{-3.99(1.08)(0.13)} &
      \multicolumn{3}{|l|}{8.28(1.29)(0.54)} &  \multicolumn{3}{|l|}{-4.37(0.86)(0.52)} &
      \multicolumn{3}{|l|}{-0.075(32)(10)} \\\hline
 \end{tabular}
  \caption{Fit ranges, bare matrix element results and uncertainties in lattice units, reduced
$\chi^2$ showing goodness-of-fit, and optimal shrinkage parameters used for each three-point
function fit for the electroweak-singlet operators $Q_1$, $Q_2$, $Q_3$, and $Q_5$.}
 \label{tab:3ptResults}
\end{table}

%%%%%%%%%%%%%%%%%%%%%%%%%%%%%%%%%%%%%%%%%%%%%%%%%%%%%%%%%%%%%%%%%%%%%%%%%%%%%%
%%%%%%%%%%%%%%%%%%%%%%%%%%%%%%%%%%%%%%%%%%%%%%%%%%%%%%%%%%%%%%%%%%%%%%%%%%%%%%
\section{Renormalization of lattice operators}
\label{sec:npr}
% vim: sw=2 sts=2 et
%\begin{itemize}
%\item discuss symmetry and mixing
%\item npr Green's functions: Landau gauge, momenta, amputation
%\item averaging over momenta to enforce flavor symm
%\item subtraction and definition of Z-factors
%\item off-diagonal components in chiral basis
%\end{itemize}

Since the matrix elements of the 6-quark operators are computed on a lattice, they have to be
converted to some perturbative scheme, e.g., $\MSbar$, before they can be used in BSM
phenomenology.
We calculate conversion factors between lattice-regularized operators and their perturbative
definitions nonperturbatively, by computing their Green's functions on a lattice and matching
them to perturbative calculations.
The operators $Q^{(\mcP)}_{I}$ are the lowest-dimension operators with $\Delta B=-2$, therefore
they can only either mix with each other, or get discretization corrections from
higher-dimensional operators that vanish in the continuum limit.
In the chiral basis, all the 14 operators transform differently under $U(2)_L\otimes U(2)_R$
flavor symmetry, so they can mix only due to the spontaneous chiral symmetry breaking
(S$\chi$SB) in QCD, non-perturbative $U(1)_A$ violation, or chiral symmetry violations by quark masses and discretization of
the fermion action.
Mixing due to quark masses and non-perturbative effects should be small if renormalization is carried out in the UV region $|p|\approx\mu\gg\{\Lambda_{QCD}, m_q\}$ where perturbative matching is applicable.
Furthermore, effects of the explicit chiral symmetry violation by the (M)DWF fermion action
on a lattice are suppressed as the ``residual mass'' $m_{res}\lesssim m_q$~\cite{Blum:2014tka},
and thus are also negligible.
Therefore, we don't expect that renormalization of our results will be affected by mixing
between the chiral-basis operators~\footnote{
  This holds even without taking the continuum limit, since the continuum and the chiral limits
  can be taken separately in calculations with the (M)DWF lattice fermions, see Ref.~\cite{Blum:2014tka} and references within.
}.

% renormalization section
%%%%%%%%%%%%%%%%%%%%%%%%%%%%%%%%%%%%%%%%%%%%%%%%%%%%%%%%%%%%%%%%%%%%%%%%%%%%%%
\subsection{RI-MOM amplitudes on a lattice}

The lattice renormalization constants for the 6-quark operators are defined as
\begin{equation}
\label{eqn:op6q_renorm_def}
Q_I^R(\mu) = Z_{IJ}^\text{lat}(\mu,a) Q_J^\text{lat}(a)
\end{equation}
but, as will be shown below, in the chiral-diagonal basis $|Z_{I\ne J}|\ll Z_{II}\equiv Z_I$,
so $Q_I^R(\mu) = Z_I Q_I^\text{lat}$ both on a lattice and in continuum perturbation theory.
The nonperturbative renormalization and mixing of the six-quark operators is computed using a
variant of the RI-MOM scheme~\cite{Martinelli:1994ty} with a specific choice of momenta of the
external quark states.
Since the external states are not color-singlets, the gauge is fixed to the Landau gauge using the
Fourier-accelerated conjugate gradient algorithm~\cite{Hudspith:2014oja}.
All the operators of interest with $\Delta B=2$ and $\Delta I=1$
\footnote{
  Instead of the 6-quark, we study renormalization of the 6-antiquark operators, which is equivalent
  but is more natural on a lattice since it does not require conjugating quark propagators.}
can be represented in the generic form
\begin{equation}
\label{eqn:nnbar_op_gen}
\begin{aligned}
\overline{Q}_{I}
  &= (\Gamma_I)^{a_1 a_2 a_3 a_4 a_5 a_6}_{\alpha_1\alpha_2 \alpha_3\alpha_4\alpha_5\alpha_6}
    \bar d^{a_6}_{\alpha_6} \bar d^{a_5}_{\alpha_5}
    \bar d^{a_4}_{\alpha_4} \bar d^{a_3}_{\alpha_3}
    \bar u^{a_2}_{\alpha_2} \bar u^{a_1}_{\alpha_1}
\\
  &= (\Gamma_I)^{[A_1 A_2] [A_3\ldots A_6]}
    \bar d^{[A_6} \bar d^{A_5}
    \bar d^{A_4} \bar d^{A_3]}
    \bar u^{[A_2} \bar u^{A_1]} \,,
\end{aligned}
\end{equation}
where $A_i=(\alpha_i,a_i)$ are the spin$\times$color indices.
Then their Green's functions with external plane-wave quark states,
\begin{equation}
\label{eqn:nnbar_npr_corr}
G_I^{B_1\ldots B_6} (\{p_i\})
  = \sum_{x_i}\,e^{i\sum_i p_i x_i}\,\langle \overline{Q}_I(0)
      u^{B_1}(x_1)\dotsb d^{B_6}(x_6)\rangle\,.
\end{equation}
are computed on a lattice contracting six quark propagators computed with a point source at the
operator location.
The same propagators are used as for the $\nnbar$ three-point correlators (see
Sec.~\ref{sec:lattice_setup}), with the only difference that prior to the contraction
the propagators are Fourier-transformed at the sink.
The six-quark vertex functions are obtained by ``amputating'' the Green's
functions~(\ref{eqn:nnbar_npr_corr})
\begin{equation}
\label{eqn:nnbar_npr_vertex}
\begin{aligned}
\Lambda_I^{A_1\ldots A_6} (\{p_i\})
  &= \langle \overline{Q}_I(0) u^{A_1}(p_1)\dotsb d^{A_6}(p_6)\rangle_\text{amp}\\
  &= G_I^{B_1\ldots B_6} (\{p_i\}) \cdot [S^{-1}(p_1)]^{B_1 A_1}
    \dotsm [S^{-1}(p_6)]^{B_6 A_6}\,,
\end{aligned}
\end{equation}
where contraction in $\{B_i\}$ is implied, and the momentum-projected quark propagators are
\begin{equation}
S^{AB}(p) = \sum_x \, e^{ipx} \, \langle q^A(x) \bar{q}^B(0)\rangle
\end{equation}
Note that the amputated Green's functions~(\ref{eqn:nnbar_npr_vertex}) are not symmetric
with respect to permutation of the spin$\times$color indices $A_i$, unlike the tree-level
vertex function $\Gamma_i^{[A_1 A_2] [A_3\ldots A_6]}$ in Eq.~(\ref{eqn:nnbar_op_gen}).
This is due to the fact that $G_I(\{p_i\})$ and $\Lambda_I(\{p_i\})$ depend on the non-equal
momenta $p_i$ of the external fields.
Such dependency would break the isospin symmetry and thus may mix operators from different
chiral representations.

%%%%%%%%%%%%%%%%%%%%%%%%%%%%%%%%%%%%%%%%%%%%%%%%%%%%%%%%%%%%%%%%%%%%%%%%%%%%%%
\begin{figure}
\centering
\includegraphics[width=.2\textwidth]{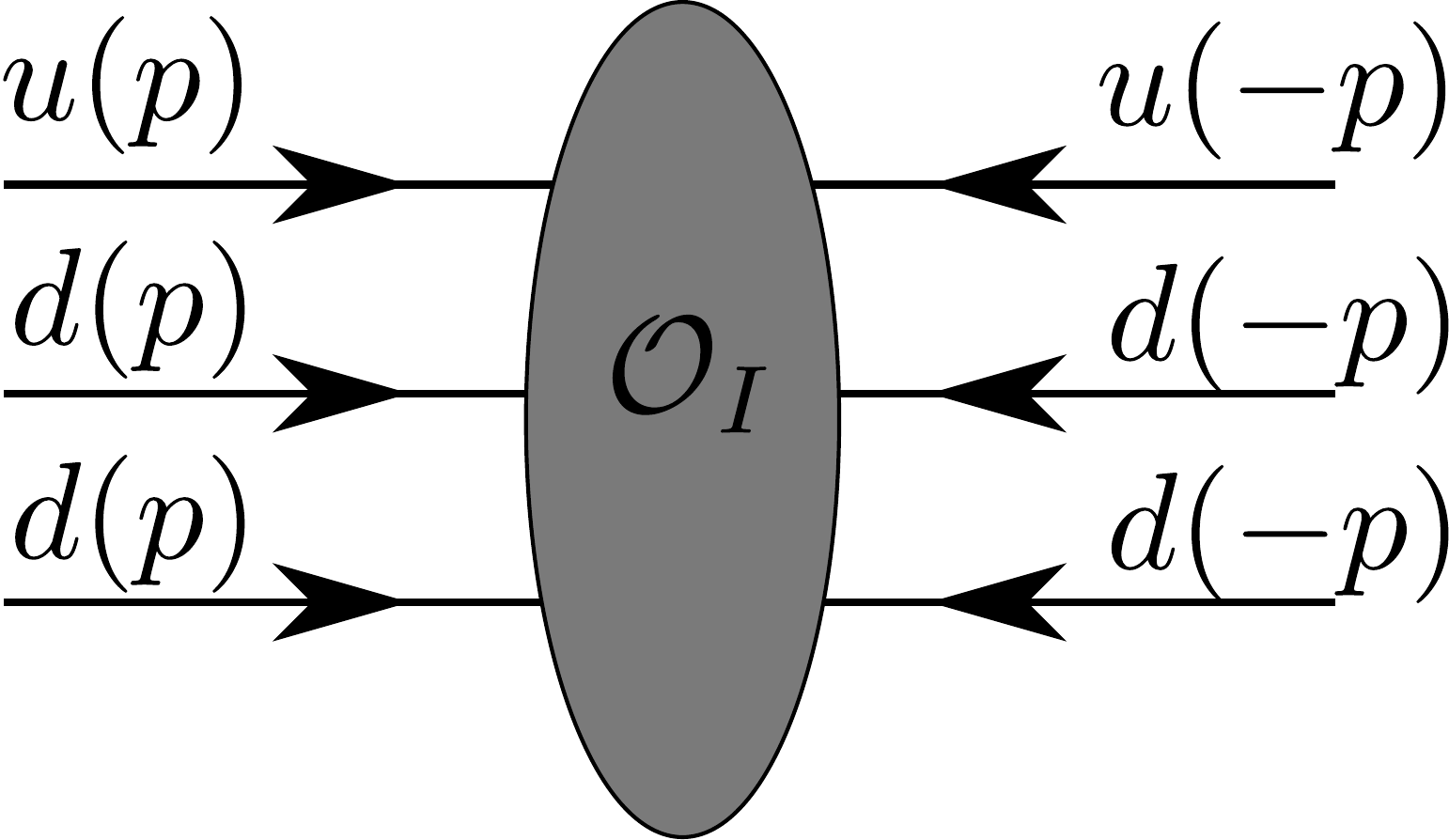}
\caption{Momentum configuration for nonperturbative renormalization of 6-quark operators using
  RI-MOM scheme (only one permutation).
  \label{fig:nnbar_npr_diag}}
\end{figure}

One must choose specific momenta for external quark fields in order to preserve the chiral
isospin symmetry.
The simplest choice $p_i=p$ would result in a large momentum $p_\mcO = 6p$ at the operator
insertion leading to large perturbative corrections in conversion to the $\MSbar$ scheme.
To avoid that, the external quark momenta are arranged so that $\sum_i p_{i}=0$ and, specifically,
$p_i=\pm p$ (see Fig.~\ref{fig:nnbar_npr_diag}), where $p^2=\mu^2$ determines the scale for
perturbative RI-MOM $\to\,\MSbar$ matching.
In addition, the amputated amplitudes~(\ref{eqn:nnbar_npr_vertex}) must be averaged over
permutations of the $\pm p$ momenta to enforce the symmetry with respect to the external quark
states~\cite{Syritsyn:2016ijx},
\begin{equation}
\label{eqn:nnbar_npr_vertex_symm}
\begin{aligned}
\Lambda_I^{[A_1 A_2] [A_3\ldots A_6]} (p)
  = \Big[
   & \frac15 \langle\overline{Q}_I\, u(+p) u(+p) d(+p) d(-p) d(-p) d(-p)\rangle_\text{amp} \\
  +& \frac35 \langle\overline{Q}_I\, u(+p) u(-p) d(+p) d(+p) d(-p) d(-p)\rangle_\text{amp} \\
  +& \frac15 \langle\overline{Q}_I\, u(-p) u(-p) d(+p) d(+p) d(+p) d(-p)\rangle_\text{amp}
  \Big]^{[A_1 A_2][A_3\ldots A_6]}\,,
\end{aligned}
\end{equation}
where the factors are determined by combinatorics.
All possible permutations of momenta are implicitly included by Wick contractions, and the
symmetries of the color$\times$spin indices are restored automatically.
Perturbative matching at the one-loop level for this particular scheme has been computed in
Ref.~\cite{Buchoff:2015qwa}.

\begin{figure}[ht!]
\centering
\includegraphics[width=.49\textwidth]{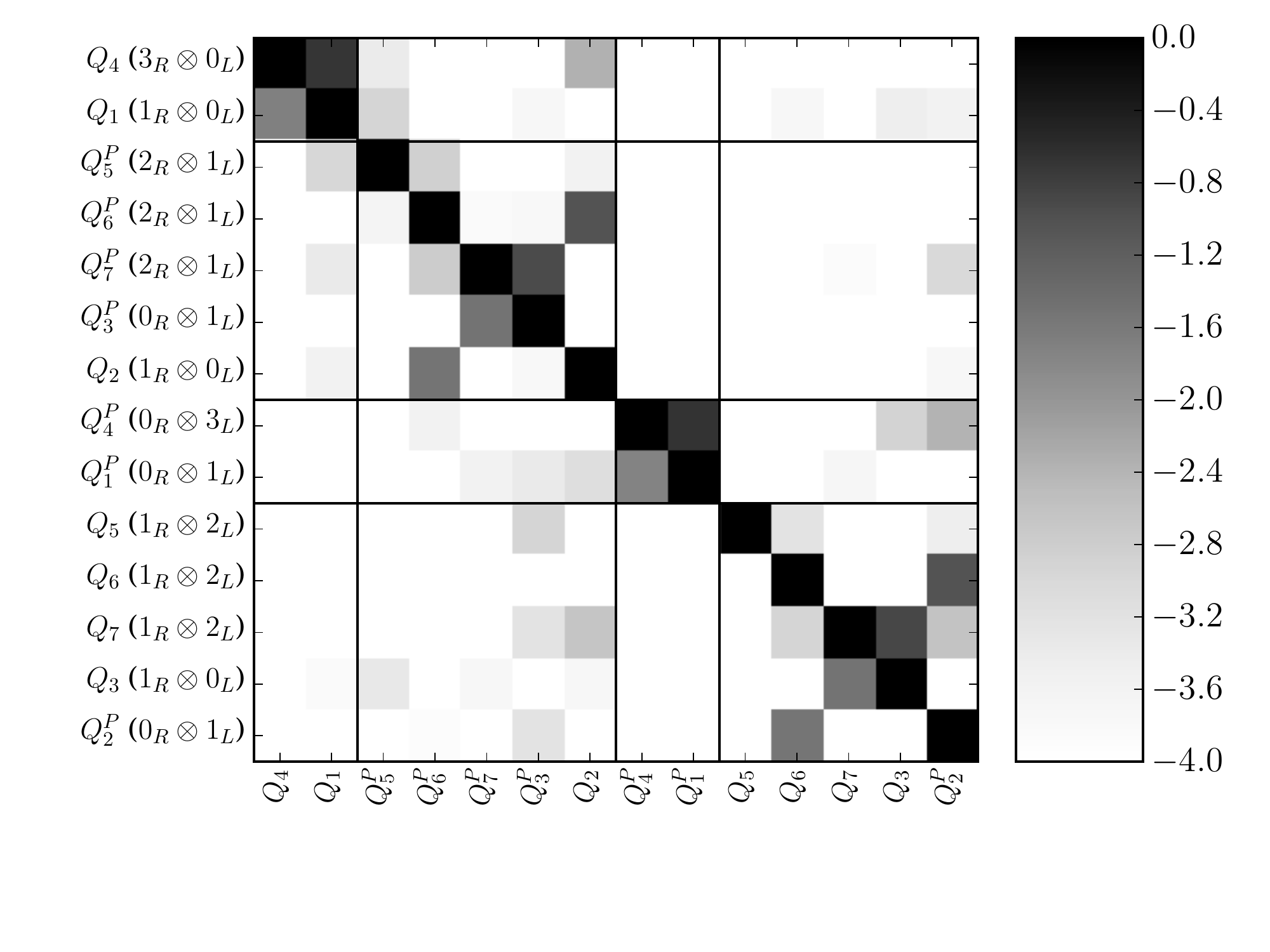}~
\hspace{.02\textwidth}~
\includegraphics[width=.49\textwidth]{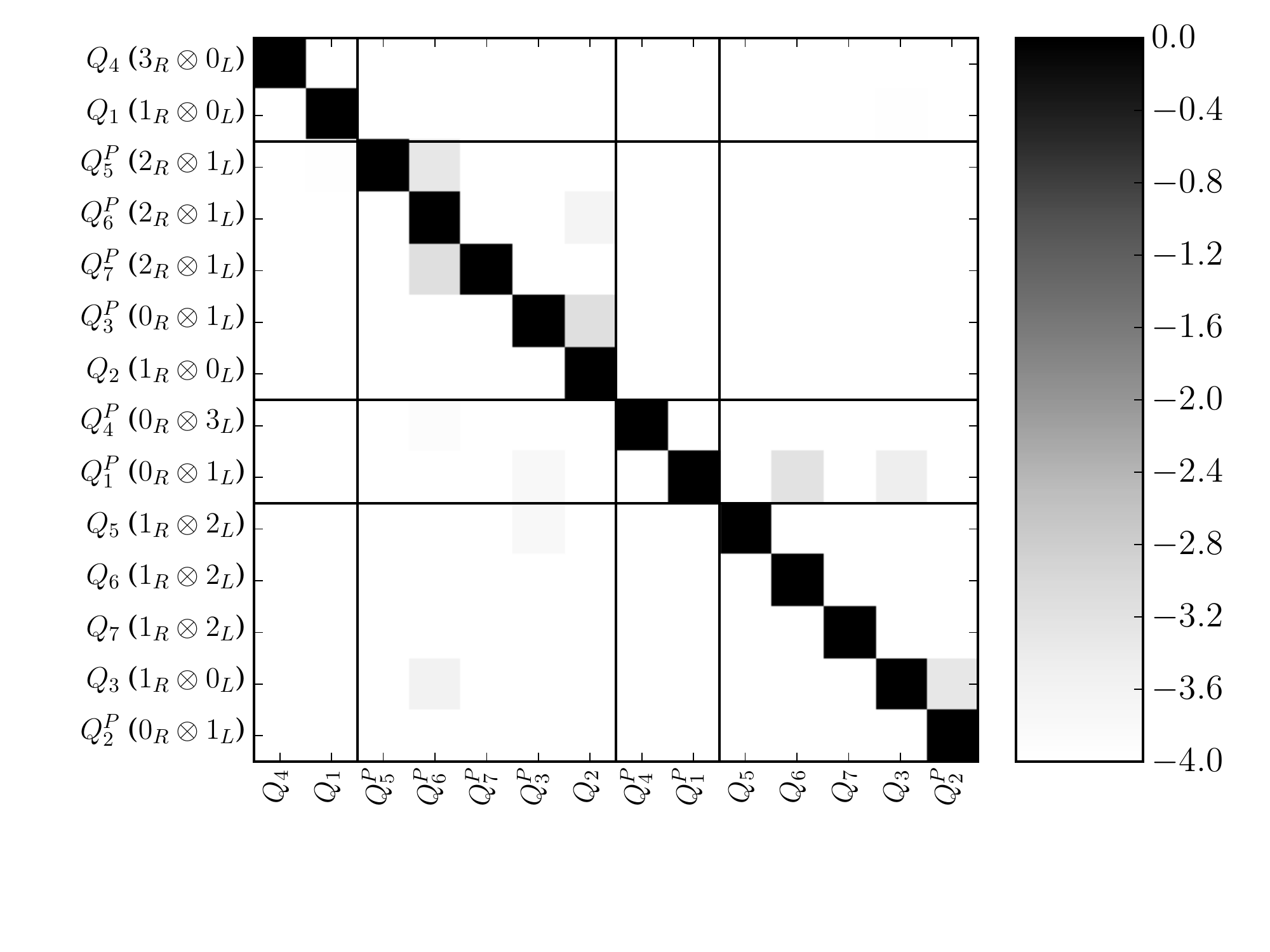}\\
\caption{Magnitude of the off-diagonal components of the lattice mixing matrix~
  (\ref{eqn:nnbar_log_mixing}) for (approximately) 4d-diagonal momentum $p^2=
  (5\,\text{GeV})^2$:
  (left) with quark external momenta shown in Fig.~\ref{fig:nnbar_npr_diag} and
  (right) averaged over their permutation~(\ref{eqn:nnbar_npr_vertex_symm}).
  Only the values $|X_{IJ}|\ge10^{-4}$ are shown.
  The operator labels show their chiral isospin structure (see Tab.~\ref{tab:nnbar_op_def}).
  The solid lines delineate operators that contain $RRR$, $RRL$, $LLL$, and $LLR$
  diquarks.
  \label{fig:nnbar_mixmap_orig_symm}
}
\end{figure}

The lattice renormalization factors $Z^\text{lat}(p^2)$~(\ref{eqn:op6q_renorm_def}) can be
computed by imposing the condition
\begin{equation}
\label{eqn:nnbar_Zop_def}
Z_q^{-3}(p) Z^\text{lat}_{IJ}(p) \, \Lambda_{J}^{\{A_i\}}(p) = \Gamma_I^{\{A_i\}}\,,
\end{equation}
where $Z_q$ is the lattice quark field renormalization
factor
\begin{equation}
\label{eqn:quark_renorm_def}
q^R(\mu) = Z_q^{1/2}(\mu,a) q^\text{lat} \,.
\end{equation}
The renormalization factors $Z_{IJ}^\text{lat}$ can be expressed in terms of the amputated and
symmetrized vertex functions $\Lambda_I^{\{A_i\}}(p)$ that are projected onto the original
tree-level structures $\Gamma_J^{\{A_i\}}$,
\begin{align}
\label{eqn:nnbar_Zop}
Z^\text{lat}_{IJ}(p)
  &= Z_q^3(p) \big[\Lambda^{-1}(p)]_{IJ}\,,\\
\label{eqn:nnbar_Zop_proj}
\Lambda_{IJ}(p)
  &= \big[\sum_{A_i} \Lambda_I^{\{A_i\}} \Gamma_K^{*\{A_i\}}\big] (g^{-1})_{KJ} \,,\\
\label{eqn:nnbar_treelev_metric}
g_{JK} &= \sum_{A_i} \Gamma_J^{\{A_i\}} \Gamma_K^{*\{A_i\}} \,,
\end{align}
where the ``metric tensor'' $g_{JK}$ is diagonal in the chiral basis $\overline{Q}_I^{(\mcP)}$.
(Approximate) chiral symmetry on a lattice is important for ensuring that $Z^{IJ}$ and
$\Lambda_{IJ}$ are also (predominantly) diagonal in this basis.
Deviations from the diagonal form are due to the nonzero quark mass and residual chiral
symmetry breaking of the DWF discretization.
The effect of symmetrization~(\ref{eqn:nnbar_npr_vertex_symm}) is evident from
the magnitude of the off-diagonal components, which is shown in the log scale as the matrix
\begin{equation}
\label{eqn:nnbar_log_mixing}
X_{IJ} = \log\Big(\frac{|\Lambda_{IJ}|}{\sqrt{\Lambda_{II} \Lambda_{JJ}}}\Big)
\end{equation}
in Fig.~\ref{fig:nnbar_mixmap_orig_symm} comparing the momentum permutation-averaged
amplitude~(\ref{eqn:nnbar_npr_vertex_symm}) to the one with a specific choice of
momentum $p_1=p_3=p_4 = -p_2=-p_4=-p_6 = p$ as in Fig.~\ref{fig:nnbar_npr_diag}.
These data are shown for the momentum
$p = \frac{2\pi}{a}\big(\frac{11}{48}, \frac{11}{48},\frac{11}{48}, \frac{22.5}{96}\big)$, which
is close to a 4d diagonal direction (up to $(\pi/L)$ along the time axis due to the
antiperiodic boundary conditions) and $p^2\approx (5\,\text{GeV})^2$.
Therefore, we conclude that in the chiral basis the renormalization matrix $Z_{IJ}$
is diagonal, $|Z_{IJ}| / \sqrt{Z_{II} Z_{IJ}} \lesssim O(10^{-3})$,  which is definitely within
our target precision, and the operators $Q^{(P)}_I$ may be renormalized multiplicatively in our
lattice calculation.
Additionally, we observe that the mixing between 6-quark operators containing different numbers of
$L,R$-diquarks is negligible, indicating that nonperturbative chirality-changing effects due to
fluctuations of topology of the QCD vacuum do not lead to mixing in excess of the $10^{-3}$ level.

\begin{figure}[ht!]
\centering
\includegraphics[width=.49\textwidth]{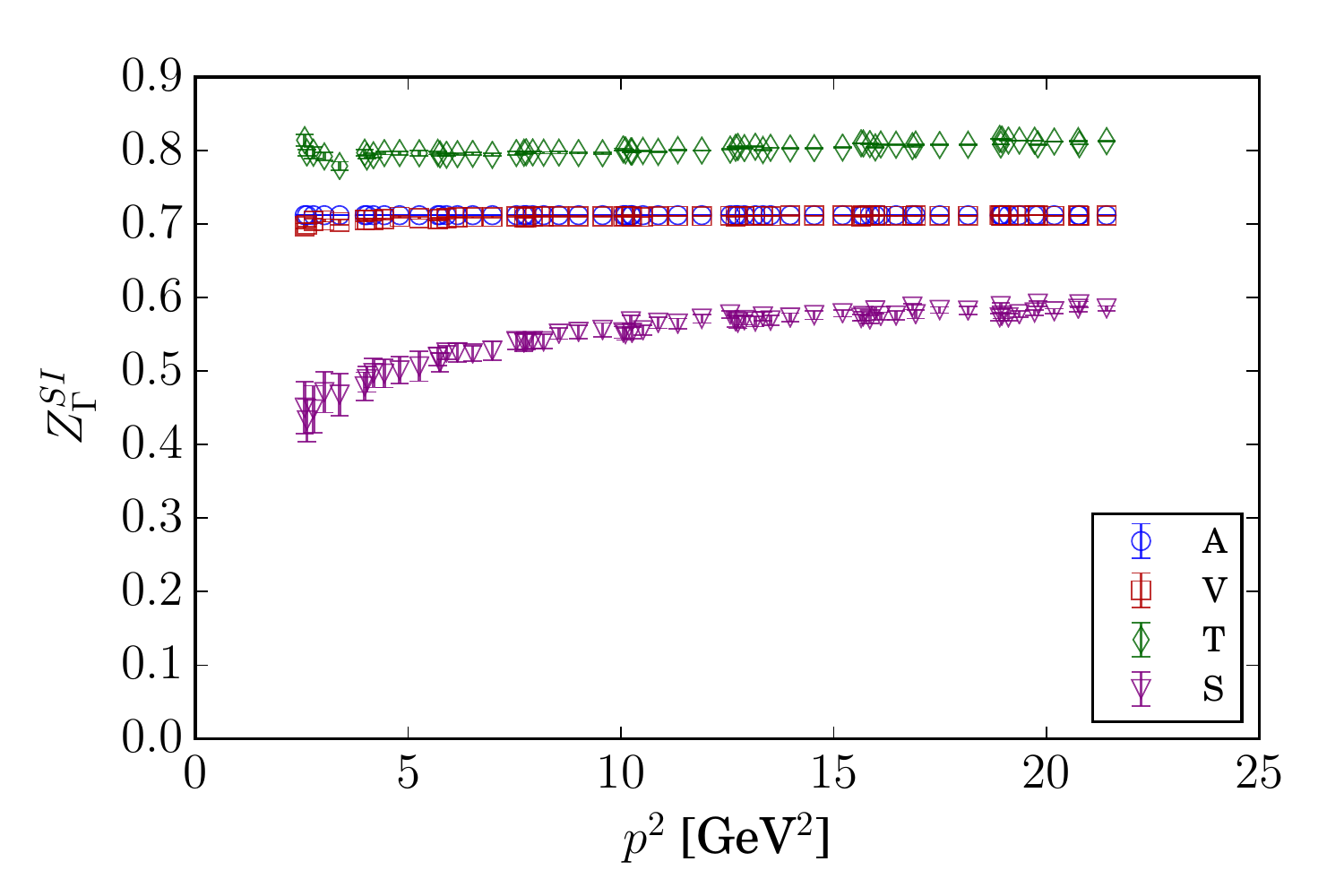}
\caption{Scale-independent renormalization factors for vector, scalar, and tensor currents.
The axial current renormalization $Z_A$ is trivially constant because its vertex is used to
eliminate the quark field renormalization $Z_q$.
The close values of the vector and axial-vector renormalization constants indicate that
chiral symmetry-breaking effects are negligible.
\label{fig:ZSI_AVTS}}
\end{figure}

We define lattice renormalization factors in the RI-MOM scheme for the $\nnbar$ operators in the
chiral basis as
\begin{equation}
\label{eqn:Zlat}
Z^{lat}_{I}(p) \doteq Z^\text{lat}_{II}(p) \approx \frac{Z_q^3(p)}{\Lambda_{II}(p)} \,.
\end{equation}
Finally, to get rid of the quark field renormalization, we use the renormalization constant
$Z_A$ for the local axial-vector current $A_\mu=\bar q\gamma_\mu\gamma_5 q$.
Using the value of $Z_A$ computed in Ref.~\cite{Blum:2014tka},
we can compute $Z_q(p)$ in the RI-MOM scheme from the condition
\begin{equation}
Z_q^{-1}(p) Z_A(p) \langle A_\mu(0) q(p) \bar q(p)\rangle_\text{amp}^\text{lat}
  = \gamma_\mu\gamma_5 \,,
\end{equation}
where $\langle A_\mu q\bar q\rangle$ is the amputated Green's function for the axial current
computed analogously to Eq.~(\ref{eqn:nnbar_npr_vertex}).
``Scale-independent'' lattice renormalization factors
$Z_\Gamma^{SI} = Z^\text{lat}_\Gamma(p) / Z^{RI,\text{pert}}(p)$ for the vector, tensor, and
scalar vertices are shown in Fig.~\ref{fig:ZSI_AVTS}.

The value of the lattice renormalization constants $Z_I(p)$ may depend on the orientation of
the momentum $p$ with respect to the lattice axes due to discretization effects.
We compute the lattice vertex functions~(\ref{eqn:nnbar_Zop_proj}) for various orientations of
lattice momenta interpolating between 3d-diagonal and 4d-diagonal orientations to study these
effects in the following sections.

%%%%%%%%%%%%%%%%%%%%%%%%%%%%%%%%%%%%%%%%%%%%%%%%%%%%%%%%%%%%%%%%%%%%%%%%%%%%%%
\subsection{Perturbative running}
\label{sec:renorm_pert}

In order to convert operator normalization from the RI-MOM scheme discussed above
to $\MSbar$, perturbative matching calculations are required.
To extract lattice renormalization factors independent from the momentum subtraction point $p$,
the lattice factors~(\ref{eqn:Zlat}) are compared to the perturbative predictions for the RI-MOM
scheme in some window $p_\text{min}\le |p|\le p_\text{max}$ where lattice artifacts are believed
to be under control.
In this section, details of relevant perturbative results are summarized.

The one-loop anomalous dimensions of the operators~(\ref{eq:Q1-3def}-\ref{eq:Q5-7def}) were
computed in Ref.~\cite{Caswell:1982qs}, and the $\MSbar$ anomalous dimensions to the
$O(\alpha_S^2)$ precision together with $O(\alpha_S)$ conversion factors were computed in
Ref.~\cite{Buchoff:2015qwa}.
In the chiral basis, the perturbative renormalization of the operators is diagonal (no mixing),
and their independent anomalous dimensions are
\begin{equation}
\label{eqn:anom_dim}
\frac1{Z_I}\,\frac{d}{d\,\ln\mu} Z_I
  = -\gamma_I(\alpha_S)
  = -\gamma_I^{(0)}\left(\frac{\alpha_S(\mu)}{4\pi}\right)
    -\gamma_I^{(1)}\left(\frac{\alpha_S(\mu)}{4\pi}\right)^2\,,
\end{equation}
with the coefficients $\gamma_I^{(0)}$ given in Tab.~\ref{tab:nnbar_op_def}.
These anomalous dimensions are substantially different, which would complicate operator
renormalization if chiral symmetry was violated by a lattice fermion action and mixing was allowed.
We integrate the equations~(\ref{eqn:anom_dim}) together with an RG equation for the coupling
constant $\alpha_S(\mu)$ using the 4-loop $\beta(\alpha_S)$-function.
Since our lattice QCD action has $N_f=2+1$ dynamical flavors, the lattice factors~(\ref{eqn:Zlat})
are matched to $Z^{RI}(\mu)$ factors computed in $N_f=3$ perturbative QCD and the coupling
constant $\alpha_S^{N_f=3}$ is matched to its physical value at $\mu\le m_c$.
The latter is obtained from a global fit~\cite{Bethke:2009jm} and matched at the $m_{b,c}$
quark mass thresholds.
For the reference point $\mu_0=2\,\text{GeV}$, its values\footnote{
  The coupling constant in the RI-MOM scheme is conveniently defined to be equal to the $\MSbar$
  coupling constant.
}
are $\alpha_S^{N_f=3}=0.2827$ and $\alpha_S^{N_f=4}=0.2948$.

\begin{figure}[ht!] %  figure placement: here, top, bottom, or page
  \centering
  \includegraphics[width=.49\textwidth]{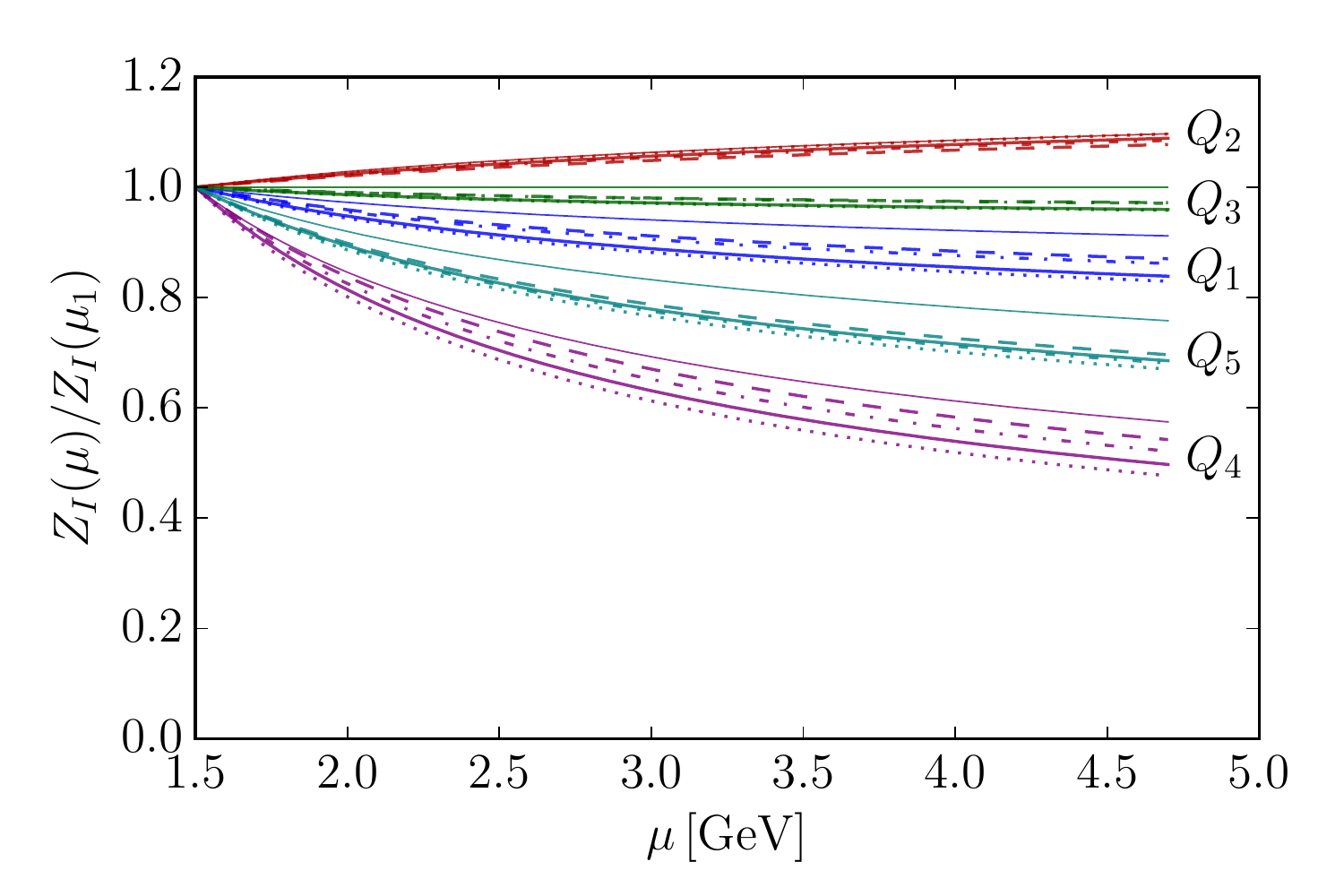}\\
  \caption{
    Perturbative running of the operators $Q_I$ at the 2-loop level~\cite{Buchoff:2015qwa} in
    the RI-MOM scheme with $N_f=3$ (solid) and $N_f=4$ (dotted) and in the $\MSbar$ scheme with
    $N_f=3$ (dashed) and $N_f=4$ (dash-dotted).
    The 1-loop results are shown with thin solid lines.
    The reference point is $\mu_1=1.5\,\text{GeV}\approx m_c$.
    \label{fig:Zpert}}
\end{figure}

The final results are converted to $N_f=4$ QCD at $\mu_0=2\,\text{GeV}$, again matching at the
$m_c$ threshold.
The final conversion factors from lattice to the $\MSbar$ scheme at scale $\mu_0$ are
\begin{equation}
\label{eqn:Zfinal}
C_I^{\MSbar(N_f=4)\leftarrow\text{lat}}(\mu_0) =
    \left[\frac{Z_I^{\MSbar(N_f=4)}(\mu_0)}{Z_I^{\MSbar(N_f=4)}(m_c)}\right]_\text{pert}
    \left[\frac{Z_I^{\MSbar(N_f=3)}(\mu_0)}{Z_I^{\MSbar(N_f=3)}(m_c)}\right]_\text{pert}^{-1}
    C_I^{\MSbar\leftarrow RI(N_f=3)}(\mu_0)
    Z^{SI}_I(\mu_0,a)\,,
\end{equation}
where $Z^{SI}_I$ is a ``scale-independent'' lattice renormalization factor with a reference
point $\mu_0$ defined in the next section.
The perturbative scale dependence in both the $\MSbar$ and RI-MOM schemes with $N_f=3$ and 4
flavors is shown in Fig.~\ref{fig:Zpert}.

%%%%%%%%%%%%%%%%%%%%%%%%%%%%%%%%%%%%%%%%%%%%%%%%%%%%%%%%%%%%%%%%%%%%%%%%%%%%%%
\subsection{Fits of nonperturbative and discretization effects}

\begin{figure}[ht!]
\centering
\includegraphics[width=.49\textwidth]{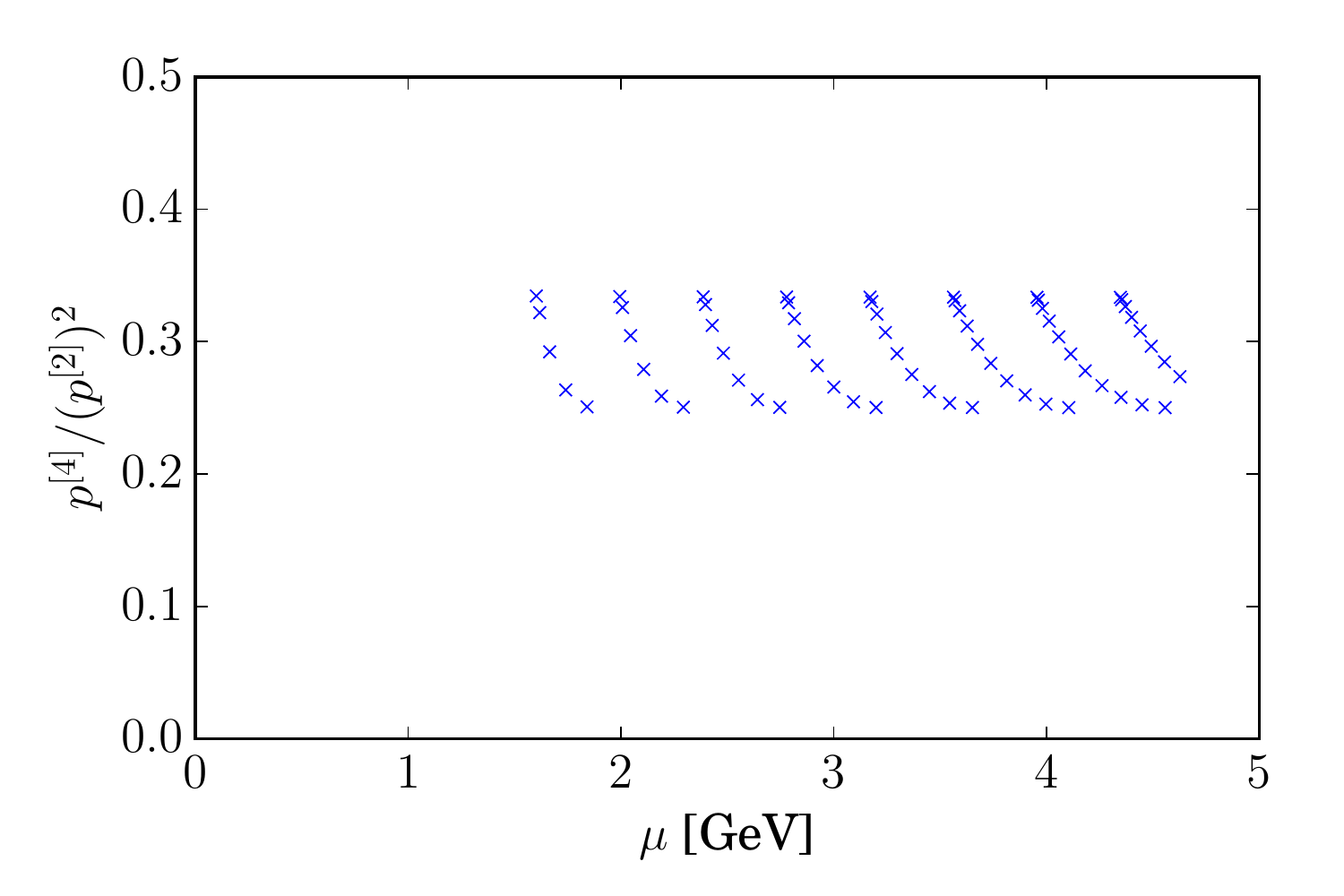}
\caption{Normalized H(4) invariant $p^{[4]}/(p^2)^2$ (see Eq.~\ref{eqn:H4inv}) for the lattice
  momenta included in the analysis.
\label{fig:H4inv}}
\end{figure}

With known perturbative running, we can separate scale-independent renormalization from lattice
artifacts and nonperturbative effects.
Correlation functions computed on a lattice are subject to discretization effects that may break
rotational symmetry at short distances, which are relevant for the large momenta used in the
nonperturbative renormalization.
In addition, they may have nonperturbative contributions that complicate matching with
perturbative calculations.
Below we follow closely the analysis performed in Ref.~\cite{Blossier:2014kta} and extract the
scale-invariant renormalization constants $Z_I^\text{SI}$ from a fit
\begin{equation}
\label{eqn:ZSIfit}
Z^\text{lat}
  = Z^\text{SI}_I\big(\mu_0,a\big) \,\left[
    \frac{Z_I^\text{RI}(|p|)}{Z_I^\text{RI}(\mu_0)}\right]^\text{pert}
    + \Delta Z_I^\text{disc}\big(a^k p^{[k]}\big) + \Delta Z_I^{NP}\big(p^2\big)\,,
\end{equation}
where $Z^{SI}(\mu_0,a)$ is the momentum-independent lattice renormalization constant,
$Z_I^{RI,pert}(\mu)$ is the perturbative running of $Q_I$ in the RI-MOM scheme
%discussed in the previous section
and $\Delta Z_I^\text{disc,NP}$ encapsulates discretization and nonperturbative corrections.
In our calculation with $O(a)$-improved action, the discretization effects must
scale as $O\big((ap)^2\big)$,
\begin{align}
\label{eqn:deltaZ_disc}
\Delta Z_I^\text{disc}(a^k p^i{[k]})
  = A (ap)^2 + \big[ B_1 (ap)^2 + B_2 (ap)^4 \big]\frac{a^4 p^{[4]}}{(ap)^4}
\end{align}
where we also include the hypercubic invariant $\propto p^{[4]}$ (see Fig.~\ref{fig:H4inv})
\begin{equation}
\label{eqn:H4inv}
p^{[2k]} = \sum_\mu p_\mu^{2k}
\end{equation}
that breaks the rotational symmetry $O(4)\to H(4)$ for $k\ne0,1$.

Although the vertex functions~(\ref{eqn:nnbar_npr_vertex}) are computed with ``exceptional''
kinematics $p_\mcO=0$ (see Fig.~\ref{fig:nnbar_npr_diag}), they do not have ``pole'' contributions
$\propto 1/p^2$ because, unlike the pseudoscalar density operator that can couple to pions,
the 6-quark operators $Q_I$ can couple only to 2-baryon ($B=2$) states with masses $M\ge2m_N$.
However, the nonperturbative contributions are added to Eq.~(\ref{eqn:ZSIfit})
\begin{equation}
\label{eqn:deltaZ_np}
\Delta Z^{NP}(p^2) = \frac{C}{p^2}\,,
\end{equation}
to account for effects of the dimension-2 gluon condensate~\cite{Boucaud:2000nd,Boucaud:2001st,
Dudal:2003vv,RuizArriola:2004en,Megias:2005ve,Dudal:2010tf,Kondo:2001nq} that may be present in
the quark propagators used to amputate the Green's functions.
Contributions of condensates to correlation functions are scale-dependent and should be
evaluated using OPE as in, e.g., Ref.~\cite{Blossier:2010vt}.
Such analysis has not been performed yet, and the correction in Eq.~(\ref{eqn:deltaZ_np})
should be regarded as a phenomenological assumption.
Another potential source of $\propto1/p^2$ effects are nonperturbative \emph{infrared}
contributions due to potential low-momentum subdiagrams, which may appear due to the same
arguments as in Ref.~\cite{Aoki:2007xm}.

\begin{figure}[ht!] %  figure placement: here, top, bottom, or page
  \centering
  \includegraphics[width=.49\textwidth]{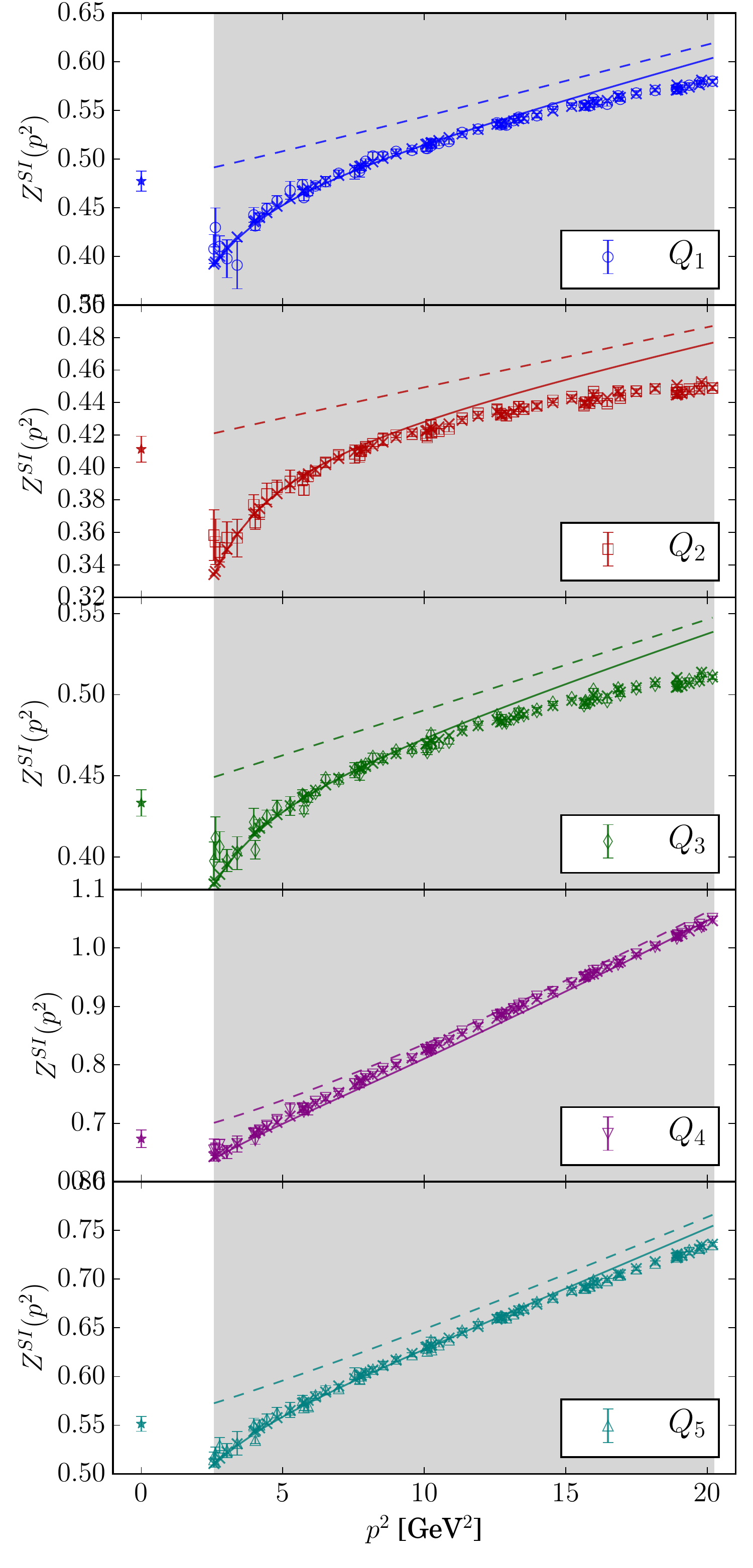}~
  \hspace{.02\textwidth}~
  \includegraphics[width=.49\textwidth]{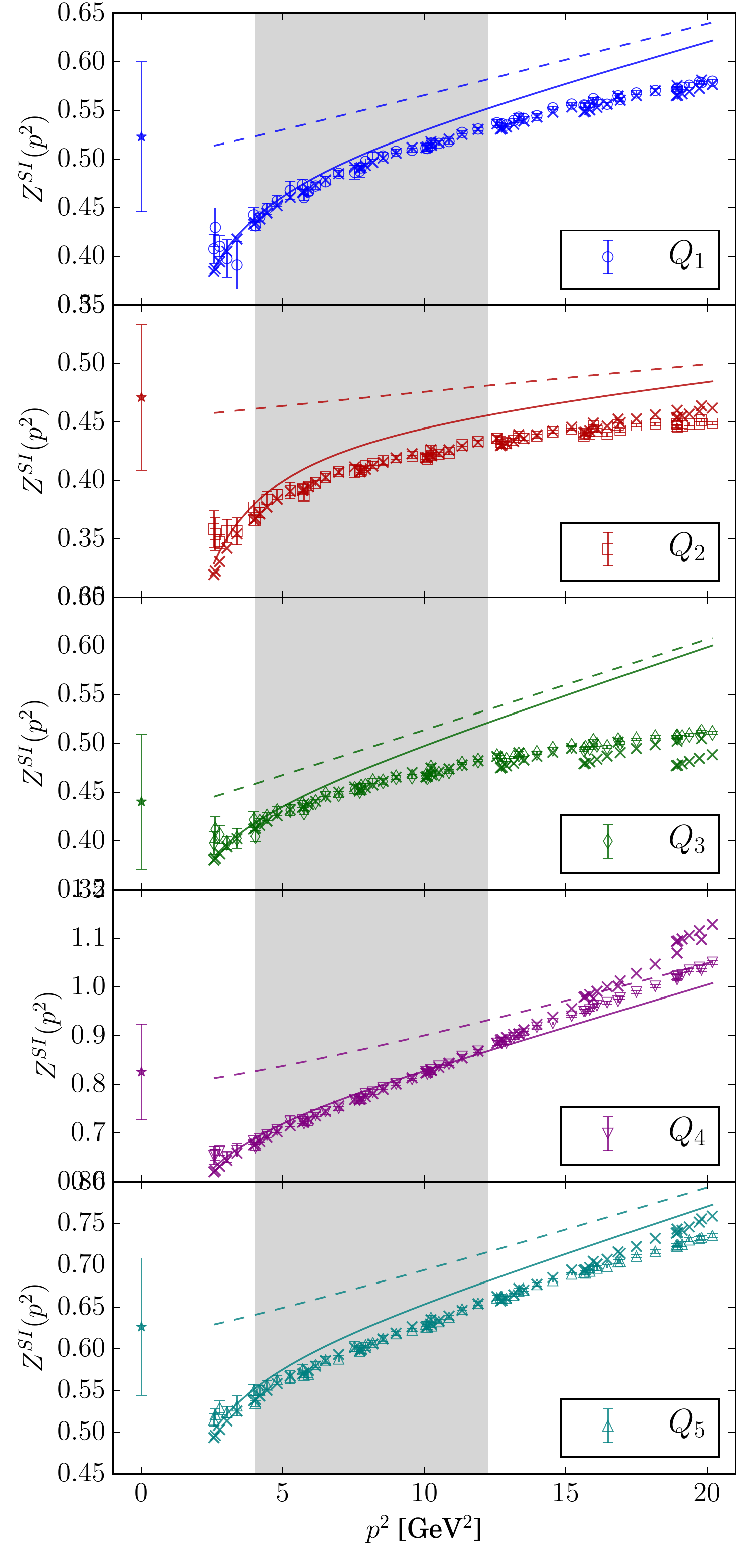}
  \caption{
    Fits of lattice renormalization constants $Z^\text{lat}(p)$ to the form~(\ref{eqn:ZSIfit}),
    for $1.6\le p \le 4.5\,\text{GeV}$ (left) and $2.0\le p \le 3.5\,\text{GeV}$ (right).
    For each operator, the figures show the $Z^{SI}$ contribution together with the
    $\propto(ap)^2$ discretization correction (dashed lines),
    plus the nonperturbative correction~(\ref{eqn:deltaZ_np}) (solid lines), plus the
    discretization corrections~(\ref{eqn:deltaZ_disc}) (crosses) vs. lattice values $Z^\text{lat}$
    (open symbols).
    The gray bands indicate the fit regions.
    The star symbols on the left of each panel show the final $Z^{SI}$ values and their
    statistical uncertainties.
    \label{fig:ZSIfits}}
\end{figure}

We perform uncorrelated fit~(\ref{eqn:ZSIfit}) with five parameters ($Z_I^\text{SI}$, and
$A,B_{1,2},C$) to the lattice data $Z_I^\text{lat}(p)$ for varying sets of momenta $p$, and two examples are shown in Fig.~\ref{fig:ZSIfits}. 
To keep discretization errors omitted from Eq.~(\ref{eqn:deltaZ_disc}) as small as possible,
we include only momenta $p$ that interpolate between the 3d- and 4d-diagonals,
\begin{equation}
\begin{aligned}
p_\mu &= \big(\pm k_1, \pm k_2, \pm k_2, k_2\big)\,,
\quad k_1 \le k_2\,, \\
k_1 &=a^{-1}\big(0\ldots \frac{\pi}2\big)=\big(0\ldots2.7\big)\text{GeV}\,, \\
k_2 &= a^{-1}\big(\frac{\pi}6\ldots\frac{\pi}2\big)=\big(0.9\ldots2.7\big)\text{GeV}\,,
\end{aligned}
\end{equation}
The lowest rotational symmetry-breaking contribution $\propto p^{[4]}/p^4$ to
Eq.~(\ref{eqn:deltaZ_disc}) is shown in Fig.~\ref{fig:H4inv}.
Values $Z_I^\text{lat}(p)$ at $H(4)$-equivalent momenta $p$ are averaged.
The fit range $p_\text{min}^2 \le p^2 \le p_\text{max}^2$ is varied with
$p_\text{min}=1.6,\,2.0\,\text{GeV}$ and $p_\text{max}=3.5,\,4.0,\,4.5\,\text{GeV}$,
resulting in $27\le n_\text{mom}\le 61$ lattice momentum data points that are distinct with
respect to $H(4)$ transformations.
We use uncorrelated $\chi^2$ values to evaluate goodness-of-fit and estimate systematic
uncertainties from variation of the results with the fit range and the order of the perturbation
theory.
Although correlated fits would be preferred, we resort to uncorrelated fits, because with a
small number of independent configurations $N_\text{cfg}=30$, it is difficult to ensure that
covariance matrices of sizes $n_\text{mom}\sim N_{\text{cfg}}$ are estimated with uniform
reliability.

The results of the fits for all fit windows using $Z^{RI,pert}$ from 1- and 2-loop perturbative
calculations $Z^{RI,pert}$ are collected in Tab.~\ref{tab:nnbar_npr_tab}, together with the
resulting \emph{uncorrelated} $\chi^2$ values.
In order to obtain the final value, we average the central values over all the fitting methods
as described in Sec.~\ref{app:stat_modavg}.
In the last row of Table~\ref{tab:nnbar_npr_tab}, we show the final conversion coefficients
between the lattice bare and $\MSbar$-renormalized operators $Q_I$ that take into account the
difference between $N_f=3$ and $N_f=4$ QCD perturbative running (see Sec.~\ref{sec:renorm_pert}
and Eq.~(\ref{eqn:Zfinal})).

\begin{table}[ht!]
  \centering
  \caption{
    Summary of renormalization constants from fits in different $q$ ranges and with
    1- and 2-loop QCD running, with statistical uncertainties and \emph{uncorrelated}
    $\chi^2$-values.
    The first two columns show the fit ranges and the perturbative orders of QCD matching.
    The last two rows show the $w_m$-weighed~(\ref{eqn:wQweight}) final values with
    statistical and systematic uncertainties for $Z^{SI}_I$~(\ref{eqn:ZSIfit}) and the
    conversion coefficients between the lattice bare values and the $\MSbar(2\,\text{GeV})$
    scheme.
    \label{tab:nnbar_npr_tab}}
  \begin{tabular}{llc|cc|cc|cc|cc|cc}
\hline\hline
$p\,[\text{GeV}$ & $Z^\text{pert}$ & ndof &
$Q_1$ & $\chi^2$ & % $(RRR)_{\mathbf1}$ & $\chi^2$ &
$Q_2$ & $\chi^2$ & % $(RR)_{\mathbf1} L_{\mathbf0}$ & $\chi^2$
$Q_3$ & $\chi^2$ & % $R_{\mathbf1} (LL)_{\mathbf0}$ & $\chi^2$ &
$Q_4$ & $\chi^2$ & % $(RRR)_{\mathbf{3}} $ \\
$Q_5$ & $\chi^2$ \\ % $(RR)_{\mathbf2} L_{\mathbf1}$ & $\chi^2$ &
\hline
1.6:3.5 & 1L & 28 & 0.425(30) & 31.7 & 0.378(36) & 25.6 & 0.369(39) & 33.8 & 0.615(55) & 22.3 & 0.509(47) & 27.4 \\
        & 2L & 28 & 0.432(31) & 32.0 & 0.380(37) & 25.5 & 0.372(39) & 33.9 & 0.646(58) & 23.0 & 0.519(48) & 27.7 \\
1.6:4.0 & 1L & 42 & 0.458(11) & 37.8 & 0.403(14) & 33.3 & 0.421(11) & 47.1 & 0.605(14) & 28.2 & 0.526(11) & 35.6 \\
        & 2L & 42 & 0.471(11) & 38.4 & 0.405(14) & 33.2 & 0.426(11) & 47.9 & 0.650(15) & 29.6 & 0.544(11) & 37.4 \\
1.6:4.5 & 1L & 56 & 0.462(10) & 69.3 & 0.409(08) & 68.4 & 0.427(08) & 83.8 & 0.622(14) & 50.2 & 0.530(07) & 62.9 \\
        & 2L & 56 & 0.477(10) & 72.1 & 0.411(08) & 68.3 & 0.433(08) & 86.1 & 0.673(15) & 55.0 & 0.551(08) & 68.5 \\
2.0:3.5 & 1L & 22 & 0.508(75) & 24.6 & 0.469(61) & 19.0 & 0.434(68) & 28.9 & 0.763(91) & 17.1 & 0.602(79) & 21.9 \\
        & 2L & 22 & 0.523(76) & 24.6 & 0.471(62) & 19.0 & 0.440(69) & 28.9 & 0.826(98) & 17.1 & 0.626(82) & 21.7 \\
2.0:4.0 & 1L & 36 & 0.476(15) & 30.7 & 0.433(17) & 25.3 & 0.452(14) & 38.0 & 0.638(32) & 24.6 & 0.559(24) & 29.2 \\
        & 2L & 36 & 0.494(16) & 30.7 & 0.435(17) & 25.3 & 0.460(14) & 38.4 & 0.698(35) & 24.7 & 0.586(25) & 29.6 \\
2.0:4.5 & 1L & 50 & 0.477(12) & 60.4 & 0.433(10) & 55.7 & 0.451(10) & 70.4 & 0.656(21) & 43.5 & 0.556(13) & 53.1 \\
        & 2L & 50 & 0.495(13) & 62.1 & 0.435(10) & 55.6 & 0.459(11) & 71.7 & 0.720(23) & 45.6 & 0.585(14) & 55.9 \\
\hline
\multicolumn{3}{l|}{$Z^{SI}_I(\mu_0)$ $(\delta^\text{stat})(\delta^\text{sys})$}
  & \multicolumn{2}{l|}{0.471(15)(15)}
  & \multicolumn{2}{l|}{0.420(17)(17)}
  & \multicolumn{2}{l|}{0.437(16)(20)}
  & \multicolumn{2}{l|}{0.644(22)(35)}
  & \multicolumn{2}{l }{0.543(16)(19)} \\
\hline
\multicolumn{3}{l|}{$C^{\MSbar\leftarrow\text{lat}}_I$ $(\delta^\text{stat})(\delta^\text{sys})$}
  & \multicolumn{2}{l|}{0.433(14)(14)}
  & \multicolumn{2}{l|}{0.429(17)(17)}
  & \multicolumn{2}{l|}{0.425(15)(20)}
  & \multicolumn{2}{l|}{0.520(18)(28)}
  & \multicolumn{2}{l }{0.527(15)(19)} \\
\hline\hline
\end{tabular}
\end{table}

Our lattice vertex functions~(\ref{eqn:nnbar_npr_vertex}) are computed with nonzero quark masses
but matched to massless perturbation theory.
Since we analyze only one ensemble, we cannot take the chiral limit $m_{u/d,s}\to0$ and our
renormalization can potentially have systematic bias due to quark mass dependence.
While the light quark masses are small and are unlikely to have significant effect, the strange
quark mass is larger and it may bias our results.
Although we cannot directly assess this quark mass dependence with data at only one combination
of quark masses, we can make a rough estimate of its magnitude from the quark mass dependence of
the axial-vector renormalization constant $Z_A$.
Since the operator renormalization constants in our analysis are multiplied by factors
$\propto Z_q^3\propto Z_A^3$, we can estimate their corresponding quark-mass correction as
\begin{equation}
\label{eqn:deltaZI_chiral}
\frac{\delta Z_I}{Z_I} = 3 \frac{\delta Z_A}{Z_A}\,,
\end{equation}
where the correction $\delta Z_A$ due to the $m_{u/d,s}\to0$ limit may be conservatively estimated
as
\begin{equation}
\delta Z_A = m_s^{phys}\cdot\text{max}\Big\{
        \Big|\frac{\partial Z_A}{\partial m_l}\Big|\,,
        \Big|\frac{\partial Z_A}{\partial m_s}\Big|
\Big\}\,.
\end{equation}
Using the data from Ref.~\cite{Aoki:2010dy} obtained with a very similar fermion action and lattice
spacing, we find from Eq.~(\ref{eqn:deltaZI_chiral}) that $\delta Z_I/Z_I\approx1.7\%$.
Considering that this (likely overestimated) correction is small compared to the uncertainties
quoted in Tab.~\ref{tab:nnbar_npr_tab}, we neglect it in the present analysis.

%%%%%%%%%%%%%%%%%%%%%%%%%%%%%%%%%%%%%%%%%%%%%%%%%%%%%%%%%%%%%%%%%%%%%%%%%%%%%%
%%%%%%%%%%%%%%%%%%%%%%%%%%%%%%%%%%%%%%%%%%%%%%%%%%%%%%%%%%%%%%%%%%%%%%%%%%%%%%
\section{Results}
\label{sec:results}
% vim: sw=2 sts=2 et

\renewcommand*{\arraystretch}{1.2}

The four indepedent non-vanishing $n\overline{n}$ matrix elements in the isospin limit are given
in terms of the above bare matrix elements and renormalization factors as
\begin{equation}
   \begin{split}
      \mcM_I^{\MSbar}(2\text{ GeV}) = C_I^{\MSbar(N_f=4)\leftarrow\text{lat}}(2\text{ GeV}) \mcM_I^{\text{lat}}
   \end{split}\label{eq:MIMSbar}
\end{equation}
Combining the uncertainties from $\mcM_I^{\text{lat}}$ and $Z_I^{SI}$ in quadrature gives the result
\begin{equation}
   \begin{split}
      \mcM_1^{\MSbar}(2\text{ GeV}) &= -46(13)(2) \times 10^{-5} \text{ GeV}^{6} \\
      \mcM_2^{\MSbar}(2\text{ GeV}) &= 95(15)(7) \times 10^{-5} \text{ GeV}^{6} \\
      \mcM_3^{\MSbar}(2\text{ GeV}) &= -50(10)(6) \times 10^{-5} \text{ GeV}^{6} \\
      \mcM_5^{\MSbar}(2\text{ GeV}) &= -1.06(45)(15) \times 10^{-5} \text{ GeV}^{6},
   \end{split}\label{eq:M2GeV}
\end{equation}
where the first uncertainty is the combined statistical uncertainty in
$\mcM_I^{\text{lat}}$ and $Z_I$ and the second uncertainty is the combined systematic
uncertainty associated with variation in fit window described in
Sec.~\ref{sec:analysis},~\ref{sec:npr} and Appendix~\ref{app:stat_modavg}.
Quark mass effects lead to negligible systematic uncertainties because of the nearly physical
pion mass used~\cite{Blum:2014tka}\footnote{
  The renormalization constants require taking the limit $m_{u/d,s}\to0$ for matching to their
  exact perturbative counterparts.
  The associated uncertainty is estimated in Sec.~\ref{sec:npr} to be small and is neglected in
  the present study.}.
Uncertainties in the determination of the lattice spacing in Ref.~\cite{Blum:2014tka} are
negligible compared to the fitting uncertainties in Eq.~(\ref{eq:M2GeV}).
Finite-volume effects have been estimated in chiral perturbation theory to be $\lesssim 1\%$
effects for the volume used for this study~\cite{Bijnens:2017xrz}.
Discretization effects are expected to be the largest unquantified systematic uncertainty that
are neglected in this work.
Chiral symmetry leads to $O(a)$ improvement of the fermion action, and discretization effects on
meson observables for these configurations have been seen to be
percent-level~\cite{Blum:2014tka}.
Discretization effects will be studied and removed from future calculations with multiple
lattice spacing.

Final results for the $\nnbar$ transition matrix elements with statistical and systematic
uncertainties added in quadrature and given in Tab.~\ref{tab:renorm_me}.
These results can be directly compared with MIT bag model results previously used to relate
experimental results to BSM couplings~\cite{Rao:1982gt} as shown in Tab.~\ref{tab:renorm_me}.
Two different sets of MIT bag model parameters are used
to calculate $\nnbar$ transition matrix elements in Ref.~\cite{Rao:1982gt}:
in fit A the up and down quark masses are set to zero,
while in fit B the up and down quark masses are set to 108 MeV
and a different value is used for the ``bag radius'' parameter.
MIT bag model results for both fit A and fit B are compared to LQCD results in
Tab.~\ref{tab:renorm_me}.
In LQCD, the electroweak-nonsinglet matrix element $\mcM_5$ is more than an order of magnitude
smaller than the electroweak-singlet matrix elements.
This feature is captured by the MIT bag model, although the sign of $\mcM_5$ differs
between the two bag model parametrizations.
LQCD results for the electroweak-singlet operator matrix elements $\mcM_1$,
$\mcM_2$, and $\mcM_3$ are larger than MIT bag model results with both
parametrizations by factors of 4-8.
This difference between LQCD and MIT bag model results is significantly larger than
the differences between MIT bag model results with different parameter values.

The effective Lagrangian for $\nnbar$ oscillations given in Eq.~(\ref{eqn:Meft}) can be used to
parameterize the $\nnbar$ vacuum transition rate for a generic BSM theory as
\begin{equation}
\label{eq:taufull}
\tau_{\nnbar}^{-1} = \big|\mcM_{\nnbar}\big|
  = \frac1{\lbsm^5}\Big|
    \sum_{I=1,2,3} \Big( \wtC_I - \eta \wtC_I^{\mcP} \Big)  \mcM_I
    + \Big( \eta^2 \wtC_5 - \eta \wtC_5^\mcP \Big) \mcM_5
  \Big| \,,
\end{equation}
where $\eta = v^2/\lbsm^2$ is the ratio of the Higgs v.e.v. and the BSM scale squared.
Both the matrix elements $\mcM$ and the Wilson coefficients $\wtC^{(\mcP)}$ are
scheme- and scale-dependent, and these dependencies must cancel in $\tau_{\nnbar}$.
Below we present results with coefficients $\wtC$ defined in $\MSbar$ scheme.
The Wilson coefficients in Eq.~(\ref{eq:taufull}) are predicted to be non-zero in various BSM
theories, see Refs.~\cite{Mohapatra:2009wp,Babu:2013yww,Phillips:2014fgb} for reviews and further
references, and are calculable at tree-level in QCD at BSM scales $\mu = \lbsm$.
The $\nnbar$ vacuum transition rate is given in terms of the above results by
\begin{equation}
\label{eq:taufid}
\begin{split}
\tau_{\nnbar}^{-1}
  &= ( 10^{-9} \text{ s}^{-1} ) \left( \frac{700\text{ TeV}}{\lbsm} \right)^5  \Big|\,
      4.2(1.1) \Big( \wtC_1^{\MSbar}(\mu) - \eta \wtC_1^{\MSbar,\mcP}(\mu) \Big)
    - 8.6(1.5) \Big( \wtC_2^{\MSbar}(\mu) - \eta \wtC_2^{\MSbar,\mcP}(\mu) \Big)
  \\ &\quad\quad
    + 4.5(1.1) \Big( \wtC_3^{\MSbar}(\mu) - \eta \wtC_3^{\MSbar,\mcP}(\mu) \Big)
    + 0.096(43)\Big( \eta^2 \wtC_5^{\MSbar}(\mu) - \eta \wtC_5^{\MSbar,\mcP}(\mu) \Big)
  \Big|_{\mu = 2\text{ GeV}} .
\end{split}
\end{equation}
To make the prefactor dimensionless, we use the ``reference'' normalization scale of
$700\,\text{TeV}$.
Estimates based on Eq.~(\ref{eq:taufid}) put BSM theories with scales of $\lbsm \sim 700$ TeV and
$O(1)$ matching coefficients within reach of next-generation experiments that will be able to
detect baryon number violation with $\tau_{\nnbar}^{-1} \geq 10^9$
s~\cite{Milstead:2015toa,Frost:2016qzt,Fomin:2018qrq,Hewes:2017xtr}.
To more precisely assess the expected signatures of theories with $B$-violation  at $\lbsm \sim
700$ TeV, the operators can be evolved to $\mu = \lbsm$ using the results of
Refs.~\cite{Caswell:1982qs,Buchoff:2015qwa},
\begin{equation}
   \begin{split}
      \mcM_1^{\MSbar}(700\text{ TeV}) &= -26(7)(1) \times 10^{-5} \text{ GeV}^{6} \\
      \mcM_2^{\MSbar}(700\text{ TeV}) &= 144(23)(11) \times 10^{-5} \text{ GeV}^{6} \\
      \mcM_3^{\MSbar}(700\text{ TeV}) &= -47(9)(6) \times 10^{-5} \text{ GeV}^{6} \\
      \mcM_5^{\MSbar}(700\text{ TeV}) &= -0.23(10)(3) \times 10^{-5} \text{ GeV}^{6}.
   \end{split}\label{eq:Mlbsm}
\end{equation}
Leading-order one-loop running from 2 GeV to 700 TeV modifies the dominant matrix elements
$M_{1,3}$ by up to 59\%.
At next-to-leading-order (NLO), two-loop running modifies $M_{1,2,3}$ by $4-9\%$ and one-loop
scheme matching modifies them by $2-8\%$.
Neglected next-to-next-to-leading-order perturbative renormalization effects lead to unknown
systematic uncertainties estimated to be at the level of $1-3\%$ (as the square of the relative
NLO effects).
The $\nnbar$ transition rate can be expressed in terms of the matrix elements at this scale as
\begin{equation}
\label{eq:taubsm}
\begin{split}
\tau_{\nnbar}^{-1}
  &= ( 10^{-9} \text{ s}^{-1} ) \left( \frac{700\text{ TeV}}{\lbsm} \right)^5 \Big|\,
      2.4(0.7) \Big( \wtC_1^{\MSbar}(\mu) - \eta \wtC_1^{\MSbar,\mcP}(\mu) \Big)
    - 12.9(2.3)\Big( \wtC_2^{\MSbar}(\mu) - \eta \wtC_2^{\MSbar,\mcP}(\mu) \Big)
  \\& \quad\quad
    + 4.2(1.0) \Big( \wtC_3^{\MSbar}(\mu) - \eta \wtC_3^{\MSbar,\mcP}(\mu) \Big)
    + 0.021(9) \Big( \eta^2 \wtC_5^{\MSbar}(\mu) - \eta \wtC_5^{\MSbar,\mcP}(\mu) \Big)
  \Big|_{\mu = 700\text{ TeV}} .
\end{split}
\end{equation}
This result can be combined with tree-level BSM matching results for $C_I^{\MSbar}(700\text{
TeV})$ to extract constraints on BSM theory parameters from experimental constraints on $\nnbar$
oscillations.

\begin{table}[t!]
\begin{tabular}{||c||c|c|c|c||}
\hline
\text{Operator}
& $\mcM_I^{\MSbar}(2\text{ GeV}) , $
& $\mcM_I^{\MSbar}(700\text{ TeV}) , $
   & $\frac{ \mcM_I^{\MSbar}(2\text{ GeV}) }{\text{MIT bag A}}$ &
   $\frac{ \mcM_I^{\MSbar}(2\text{ GeV}) }{\text{MIT bag B}}$ \\\hline
$Q_1$ &  $-46(13) \times 10^{-5} \text{ GeV}^6$ & $-26(7) \times 10^{-5} \text{ GeV}^6$ & 4.2  & 5.2    \\\hline
$Q_2$ &  $95(17)\times 10^{-5} \text{ GeV}^6$ & $144(26)\times 10^{-5} \text{ GeV}^6$ & 7.5 &  8.7    \\\hline
$Q_3$ &  $-50(12)\times 10^{-5} \text{ GeV}^6$ &  $-47(11)\times 10^{-5} \text{ GeV}^6$ & 5.1 & 6.1    \\\hline
$Q_5$ &  $-1.06(48)\times 10^{-5} \text{ GeV}^6$ & $-0.23(10)\times 10^{-5} \text{ GeV}^6$ & -0.84 & 1.6    \\\hline
\end{tabular}
\caption{
  Matrix element results for the chiral basis operators with independent non-zero matrix
  elements in the isospin limit.
  The second column shows the renormalized matrix elements at a scale of 2 GeV and total
  uncertainty including statistical and systematic uncertainties from the bare matrix elements and
  non-perturbative renormalization factor added in quadrature.
  Renormalized results use the $\MSbar$ scheme with $N_f = 4$ active quark flavors and are
  obtained through nonperturbative RI-MOM renormalization and perturbative matching to $\MSbar$.
  The third column shows the corresponding $\MSbar$ renormalized matrix elements and
  uncertainties after renormalization group evolution from 2 GeV to a higher scale of 700 TeV.
  The fourth and fifth columns show comparisons with the results of the same matrix elements in
  the MIT bag model from Ref.~\cite{Rao:1982gt} as described in the main text.
  \label{tab:renorm_me}}
\end{table}

%%%%%%%%%%%%%%%%%%%%%%%%%%%%%%%%%%%%%%%%%%%%%%%%%%%%%%%%%%%%%%%%%%%%%%%%%%%%%%
%%%%%%%%%%%%%%%%%%%%%%%%%%%%%%%%%%%%%%%%%%%%%%%%%%%%%%%%%%%%%%%%%%%%%%%%%%%%%%
\section{Conclusion}
\label{sec:conclusion}

We have performed the first lattice QCD calculation of the renormalized neutron-antineutron
transition matrix elements needed to extract BSM physics constraints from $\nnbar$ oscillation
experiments.
The precision of our final results including statistical and most systematic uncertainties is
$15 - 30\%$ for the electroweak-singlet matrix elements $\mcM_1$, $\mcM_2$, and
$\mcM_3$, which can be straightforwardly improved in future calculations.
Several important sources of systematic uncertainty are under control for the first time, most 
importantly non-perturbative renormalization, chiral symmetry violations, excited state
contamination, and quark mass dependence.
The two sources of systematic uncertainty that are not completely controlled in this pioneering
calculation are finite volume and discretization effects.
To summarize our control of common systematic uncertainties in lattice calculations:
\begin{itemize}
\item 
The (nearly exact) physical pion mass  $m_\pi=139.2(4)\text{MeV}$ in our calculation
eliminates the need for chiral extrapolation, which would otherwise introduce systematic
uncertainties associated with low-energy effective theory.
In addition, the large difference of our results from the MIT bag model may have a similar
origin as the strong suppression of proton decay matrix elements found in the chiral bag
model~\cite{Martin:2011nd}, therefore using the realistic light quark masses in our calculation
is arguably the most important systematic effect we have under control.
\item 
The chirally symmetric M\"obius domain wall fermion action used to generate these gauge
field ensembles by the RBC/UKQCD collaborations~\cite{Blum:2014tka} and compute
neutron-antineutron matrix elements in this work ensures that the 14 distinct
$|\Delta B|=2$ operators do not mix with each other and renormalization and conversion of
lattice operators to $\MSbar$ scheme is free from associated uncertainties.
In particular, the nonperturbatively computed operator mixing matrix in RI-MOM scheme is diagonal up to
$O(10^{-3})$ corrections, which are two orders of magnitude below other uncertainties and can be
safely neglected.
The identical action is used for valence quarks, so this is a fully unitary calculation.
\item 
Excited-state effects are accounted for using correlated two-state fits with 10 different
values of $\tsep$ and different combinations of nucleon source and sink smearing.
The energy gaps are extracted from correlated fits to nucleon two-point functions.
Since we have limited statistics for such a large number of $\tau,\tsep$ points included in
correlated fits, we use ``shrinkage'' estimators to obtain well-conditioned covariance matrices.
We obtain systematic errors by varying the fit ranges and averaging their results weighted by
the quality-of-fit figure.
\item 
Renormalization effects are included through NPR in an RI-MOM scheme as described in
Sec.~\ref{sec:npr} and one-loop matching to $\MSbar$ using the results of
Ref.~\cite{Buchoff:2015qwa}. 
Some discretization effects in NPR results such as rotational symmetry breaking and $(ap)^2$
dependence are studied and removed by fitting the lattice data with different quark momentum
scales and orientations, varying scale ranges, and comparing to 1- and 2-loop perturbative QCD
running.
One presently uncontrolled systematic uncertainty in our renormalization procedure is the quark
mass dependence.
However, a rough estimate in Sec.~\ref{sec:npr} suggests that this uncertainty should not exceed
$1.7\%$, which is below our current level of precision; this uncertainty will be studied in the
future.
\item 
Although we do not control finite-volume effects directly in this study on a single ensemble, we
expect them to be small. 
First, finite-volume effects are suppressed with $e^{-m_\pi L}$ where $m_\pi L \sim 3.9$ for the
volume used for this study, which is generally considered sufficiently large for nucleon
structure calculations~\cite{Aoki:2016frl}.
Second, chiral perturbation theory calculations in Ref.~\cite{Bijnens:2017xrz} estimate that
finite-volume effects lead to corrections below $1\%$ to $\mcM_I$ for the volume used in
this study. 
Future lattice calculations at additional volumes could be used to test this prediction and perform
an infinite-volume extrapolation.
\item 
Discretization effects are the least-controlled systematic uncertainty in our current work.
Lattice QCD calculations with finer lattice spacing(s) in the immediate future will be
used to fully quantify and remove discretization effects that are not controlled in this
calculation.
However, it is reasonable to assume that discretization effects are small compared to our
current combined uncertainty from other sources.
First, the chirally-symmetric fermion action that we use is automatically $O(a)$-improved.
Second, the meson decay constants computed on this ensemble (before finite volume and
discretization corrections are applied) are within 0.6\% of the physical values 
($f_\pi=131.1(4),\,f_K=156.4(4)\,\text{GeV}$~\cite{Blum:2014tka} compared to 
PDG values $f_\pi=130.4(2),\,f_K=156.2(7)\,\text{GeV}$~\cite{Agashe:2014kda}).
Finally, the nucleon effective mass and energy dependence on the momentum is in close agreement
with the continuum limit~\cite{Syritsyn:2019vvt}.
\end{itemize}

Our renormalized lattice QCD results for $\nnbar$ transition matrix elements provide a
significant step forward in accuracy and reliability compared to previous results from quark
models and preliminary lattice studies.
The matrix elements predicted by QCD are found to be 4-8 times larger than the predictions of
the MIT bag model for the dominant electroweak-singlet operators.
This difference between our lattice results and previously available bag model results is much
larger than the statistical or systematic uncertainties present in this calculation and is also
much larger than the expected size of finite-volume effects that have not yet been studied
directly.
There is less certainty about the size of discretization artifacts; however, the automatic
$O(a)$ improvement due to the chiral symmetry as well as minuscule discretization corrections
in the meson decay constants, nucleon mass and dispersion relation make large discretization
effects in the $\nnbar$ matrix elements very unlikely.

The difference in $\mcM_I$ between the bag model and our lattice results leads to increased
experimental sensitivity to baryon-number violating interactions that may cause $\nnbar$
oscillations.
Numbers of events that can be observed both in quasi-free neutron oscillation experiments and
underground nuclear decay experiments are proportional to $\tau_{\nnbar}^{-2}\propto
|\mcM_I|^2$, therefore the $\times(4\ldots8)$ larger values of the $\nnbar$ matrix elements
found in our work lead to $\times(16\ldots64)$ increase in the event rates.
Since our results are obtained from ab initio QCD calculations in a model-independent way, they
must be used for more precise assessments of the potential of planned $\nnbar$ oscillation
searches as well as stronger constraints on theories of baryon-number violation and baryogenesis
in the future.

%%%%%%%%%%%%%%%%%%%%%%%%%%%%%%%%%%%%%%%%%%%%%%%%%%%%%%%%%%%%%%%%%%%%%%%%%%%%%%
%%%%%%%%%%%%%%%%%%%%%%%%%%%%%%%%%%%%%%%%%%%%%%%%%%%%%%%%%%%%%%%%%%%%%%%%%%%%%%
\begin{acknowledgments}
The authors would like to express gratitude to Yuri Kamyshkov, Rabi Mohapatra, Martin
Savage, Steve Sharpe, Robert Shrock, Mike Snow, Brian Tiburzi for multiple illuminating
discussions.
We are indebted to Norman Christ, Bob Mawhinney, Taku Izubuchi, Oliver Witzel, and the rest of
the RBC/UKQCD collaboration for access to the physical point domain-wall lattice gauge
configurations used in this work.
This work has been supported by the U.~S.~Department of Energy under grant contract no.
DE-FG02-00ER41132 (INT).
This work was performed also under the auspices of the U.S. Department of Energy by Lawrence
Livermore National Laboratory (Lawrence Livermore National Security, LLC) under contract
DE-AC52-07NA27344;
and Brookhaven National Laboratory supported by the U.~S.~Department of Energy under contract
DE-SC0012704.
ER is supported by the RIKEN Special Postdoctoral Researcher fellowship.
SS is supported by the RHIC Physics Fellow Program of the RIKEN BNL Research Center.
MLW was supported by a MIT Pappalardo Fellowship and acknowledges support by the U.S. Department
of Energy, Office of Science, Office of Nuclear Physics under grant Contract Number
DE-SC0011090.
Quark propagators and contractions were computed using the USQCD computing resources at Fermilab
funded by the Office of Science of the US Department of Energy, as well as computing resources
at the Lawrence Livermore National Laboratory made available through the Institutional Computing
Grand Challenge program.
This research used resources of the Argonne Leadership Computing Facility, which is a DOE Office of Science User Facility supported under Contract DE-AC02-06CH11357
The calculations were performed using Chroma~\cite{Edwards:2004sx} and Qlua~\cite{qlua-software}
software packages, in particular with the efficient MDWF inverter developed by
A.~Pochinsky~\cite{mdwf-lib}.

%\begin{itemize}
%\item discussions with...
%\item USQCD configurations
%\item software: Chroma, Qlua, MDWF inverter(s)
%\item LLNL scicomp resources
%\item funding
%\end{itemize}

%We are indebted to Norman Christ, Bob Mawhinney, Taku Izubuchi, Oliver Witzel, and the rest of
%the RBC/UKQCD collaboration for access to the physical point, domain-wall lattices and
%propagators used in this work.  We would like to thank Yuri Kamyshkov, Rabi Mohapatra, Martin
%Savage, Steve Sharpe, Robert Shrock, Mike Snow, Brian Tiburzi, and hosts of others for
%discussions over the duration of this project.  Computing support for this work was provided in
%part from the LLNL Institutional Computing Grand Challenge program and from the USQCD
%Collaboration, which is funded by the Office of Science of the US Department of Energy.

\end{acknowledgments}

%TC:endignore

%%%%%%%%%%%%%%%%%%%%%%%%%%%%%%%%%%%%%%%%%%%%%%%%%%%%%%%%%%%%%%%%%%%%%%%%%%%%%%
%%%%%%%%%%%%%%%%%%%%%%%%%%%%%%%%%%%%%%%%%%%%%%%%%%%%%%%%%%%%%%%%%%%%%%%%%%%%%%
\bibliography{NNbar_long}

%%%%%%%%%%%%%%%%%%%%%%%%%%%%%%%%%%%%%%%%%%%%%%%%%%%%%%%%%%%%%%%%%%%%%%%%%%%%%%
%%%%%%%%%%%%%%%%%%%%%%%%%%%%%%%%%%%%%%%%%%%%%%%%%%%%%%%%%%%%%%%%%%%%%%%%%%%%%%

%%%%%%%%%%%%%%%%%%%%%%%%%%%%%%%%%%%%%%%%%%%%%%%%%%%%%%%%%%%%%%%%%%%%%%%%%%%%%%

%\input{nnbar_appendix.tex}
% vim: tw=100 sw=2 sts=2 et

\appendix

%%%%%%%%%%%%%%%%%%%%%%%%%%%%%%%%%%%%%%%%%%%%%%%%%%%%%%%%%%%%%%%%%%%%%%%%%%%%%%
%%%%%%%%%%%%%%%%%%%%%%%%%%%%%%%%%%%%%%%%%%%%%%%%%%%%%%%%%%%%%%%%%%%%%%%%%%%%%%
\section{$\mcC$-, $\mcP$-, $\mcT$-symmetries and nucleon states}\label{app:CPT}

Fermion field transformations under $\mcC$, $\mcP$, and $\mcT$ are given by
\begin{align}
\mcP\psi_x\mcP^{-1}     &= \np\eta_P \gamma_4 \psi_{\mcP(x)}\,, &
\mcP\bar\psi_x\mcP^{-1} &= \np\eta_P^* \bar\psi_{\mcP(x)}\gamma_4\,, \\
\mcC\psi_x\mcC^{-1}     &= \np\eta_C C \bar\psi_x^T\,, &
\mcC\bar\psi_x\mcC^{-1} &= \np\eta_C^* \psi_x^T C\,, \\
\mcT\psi_x\mcT^{-1}     &= \np\eta_T T \psi_{T(x)} \,, &
\mcT\bar\psi_x\mcT^{-1} &= -\eta_T^* \bar\psi_{\mcT(x)} T\,,
\end{align}
where $C$ is given in Eq.~(\ref{eqn:conjmatr}) and the spin matrix $T$ is
\begin{equation}
T = [\gamma_1\gamma_3]_\EucConv =T^* =-T^T =-T^\dag=-T^{-1}\,,
\end{equation}
and has the property
\begin{equation}
T \gamma_\mu^* T^\dag = \gamma_\mu\,,
\quad T \sigma_{\mu\nu}^* T^{-1}=-\sigma_{\mu\nu}\,.
\end{equation}
Both the color-symmetric and antisymmetric quark bilinears transform as
\begin{align}
\mcP (\psi^T C P_{R,L} \psi) \mcP^{-1} &= -\eta_P^2 (\psi^T C P_{L,R} \psi)\,,\\
\mcC (\psi^T C P_{R,L} \psi) \mcC^{-1}
  &= \eta_C^2 (\bar\psi C P_{L,R} \bar\psi^T)
  = \pm\eta_C^2 (\psi^T C P_{R,L} \psi)^{\dag} \,,\\
\mcT (\psi^T C P_{R,L} \psi) \mcT^{-1} &= \eta_T^2 (\psi^T C P_{L,R} \psi)\,,
\end{align}
from which the transformation properties for the 6-quark
operators~(\ref{eqn:op_Ptransf}-\ref{eqn:op_Ttransf}) follow.

In order to find the effect of these symmetries on the nucleon states and matrix elements of the
operators, we spell out explicitly the neutron interpolating operators,
\begin{equation}
\label{eqn:nnbardef}
\begin{aligned}
n^{(\pm)}_\alpha &=  \varepsilon^{ijk} (u_i^T C \gamma_5 \frac{1\pm\gamma_4}{2} d_j) \, d_k\,,
\\
\bar{n}^{(\pm)}_\alpha &= (n^\dag \gamma_4)_\alpha
  = \varepsilon^{ijk} \bar d_k \, (\bar{d}_j C^\dag \gamma_5 \frac{1\pm\gamma_4}{2} \bar{u}^T_i)\,,
\end{aligned}
\end{equation}
which transform as
\begin{equation}
\begin{aligned}
\mcP n^{(\pm)}_x \mcP^{-1}        &= \np\eta_P^3 \gamma_4  n^{(\pm)}_{\mcP(x)}\,, &
\mcP \bar{n}^{(\pm)}_x \mcP^{-1}  &= \np\eta_P^{*3} \bar{n}^{(\pm)}_{\mcP(x)}\gamma_4\,, \\
\mcC n^{(\pm)}_x \mcC^{-1}        &= -\eta_C^3 C \bar{n}^{(\mp)T}_x\,, &
\mcC \bar{n}^{(\pm)}_x \mcC^{-1}  &= -\eta_C^{*3} n^{(\mp)T}_x C\,, \\
\mcT n^{(\pm)}_x \mcT^{-1}        &= \np\eta_T^3 T  n^{(\pm)}_{T(x)} \,, &
\mcT \bar{n}^{(\pm)}_x \mcT^{-1}  &= -\eta_T^{*3} \bar n^{(\pm)}_{\mcT(x)} T\,.
\end{aligned}
\end{equation}
which are used to construct (anti)neutron states on a lattice.
This construction is more natural in the standard (Dirac-Pauli) basis in which the $\gamma_4$ matrix
is diagonal.
It is related to the de Grand--Rossi basis commonly used in lattice calculations by the
transformation
\begin{equation}
\psi_\text{std} = \frac1{\sqrt2}\left(\begin{array}{rrrr}
   & -1 &    & -1 \\
 1 &    &  1 &    \\
   &  1 &    & -1 \\
-1 &    &  1 &
\end{array}\right)\psi_\text{dGR}
\end{equation}
The operators~(\ref{eqn:nnbardef}) create the neutron and antineutron states with definite $\hat
z$-spin as
\begin{equation}
\label{eqn:nnbar_states}
\begin{aligned}
  \ket{\neut++} &= \np n_1^{(+)\dag}    \ket{\text{vac}}\,,
& \ket{\nbar-+} &= -n^{(-)}_4           \ket{\text{vac}}\,,
\\\ket{\neut+-} &= \np n_2^{(+)\dag}    \ket{\text{vac}}\,,
& \ket{\nbar--} &= \np \bar{n}^{(-)}_3  \ket{\text{vac}}\,,
\end{aligned}
\end{equation}
which can be found to transform as
\begin{equation}
\label{eqn:nnbar_states_CPT}
\begin{aligned}
\mcP \ket{\neut+\pm} &= \np\eta_P^{*3} \ket{\neut+\pm}\,, &
\mcP \ket{\nbar-\pm} &= -\eta_P^3 \ket{\nbar-\pm}\,, \\
\mcC \ket{\neut+\pm} &= -\eta_C^{*3} \ket{\nbar-\pm}\,, &
\mcC \ket{\nbar-\pm} &= -\eta_C^3 \ket{\neut+\pm}\,, \\
\mcT \ket{\neut+\pm} &= \mp\eta_T^{*3} \ket{\neut+\mp}\,, &
\mcT \ket{\nbar-\pm} &= \mp\eta_T^3 \ket{\nbar-\mp}\,.\\
\end{aligned}
\end{equation}
These states are used to determine the properties of the $\nnbar$ matrix elements in
Sec.\ref{sec:CPT} and define them in terms of three-point functions in Sec.~\ref{sec:lattice_setup}.

%%%%%%%%%%%%%%%%%%%%%%%%%%%%%%%%%%%%%%%%%%%%%%%%%%%%%%%%%%%%%%%%%%%%%%%%%%%%%%
%%%%%%%%%%%%%%%%%%%%%%%%%%%%%%%%%%%%%%%%%%%%%%%%%%%%%%%%%%%%%%%%%%%%%%%%%%%%%%
\section{Statistical analysis}

This work uses techniques such as bootstrap resampling that are common to lattice calculations as
well as tools that are less common and detailed below: shrinkage estimation of covariance matrices,
VarPro $\chi^2$-minimization, and weighted averaging of multiple fits with different numbers of
degrees of freedom.

%%%%%%%%%%%%%%%%%%%%%%%%%%%%%%%%%%%%%%%%%%%%%%%%%%%%%%%%%%%%%%%%%%%%%%%%%%%%%%
\subsection{Shrinkage estimation of covariance matrices}
\label{app:shrinkage_cov}

Correlated $\chi^2$-fits require sample covariance matrices that are difficult to estimate when the
number of data samples $N$ is limited compared to the number of data points $K$, as in our case.
In order to estimate covariance matrices that can be safely inverted, we use the
``optimal shrinkage estimator'' described in Refs.~\cite{LEDOIT2004365}.
Shrinkage involves replacing the covariance matrix with a linear combination of a well-conditioned
``shrinkage target'' and the original covariance matrix.
It has been shown that expectation values of ``shrunk'' covariance matrices are closer to the true
covariance matrix than the sample covariance matrix~\cite{stein1956}.
The condition number of the covariance matrix is also improved by shrinkage and estimates of
$\chi^2$ relying on the inverse covariance matrix are more robust.
Shrinkage targets that better approximate the true covariance matrix naturally lead to better
estimates of the true covariance matrix from a finite sample, but any prescription for defining the
``shrinkage parameter'' introduced below that leads to zero shrinkage in the infinite statistics
limit will provide a consistent estimator for the true covariance matrix.

The estimator in Ref.~\cite{LEDOIT2004365} uses a shrinkage target proportional to the $K\times K$
identity matrix $I$ where $K$ is the number of data points.
However, correlation functions in lattice calculations vary over orders of magnitude if a wide range
of $\tsep$ are used for fitting.
To transform the covariance matrix into a form where the shrinkage target of
Ref.~\cite{LEDOIT2004365} more closely resembles the true covariance matrix, we normalize the data
by subtracting the mean and diving by the square root of the variance.
For data points $x_\alpha^i$ where $i=1,\cdots,N$ labels decorrelated statistical samples and
$\alpha = 1,\cdots,K$ labels data points (i.e. $t$ in two-point function fits and $\tau$, $\tsep$ in
three-point function fits),
define normalized data points $y_\alpha^i$ and a normalized sample correlation matrix
$\rho_{\alpha\beta}$ as
\begin{equation}
  \begin{split}
    y_\alpha^i = \frac{x_\alpha^i - \bar{x}_\alpha}{\sqrt{S_{\alpha\alpha}}}, \hspace{20pt} \rho_{\alpha\beta}
    = \frac{S_{\alpha\beta}}{\sqrt{S_{\alpha\alpha}S_{\beta\beta}}}\,,
  \end{split}\label{eq:norm_def}
\end{equation}
where the sample mean and covariance are defined as
\begin{equation}
  \begin{split}
    \bar{x}_\alpha = \frac{1}{N}\sum_{i=1}^N x_\alpha^i, \hspace{20pt} S_{\alpha\beta}
    = \frac{1}{N-1}\sum_{i=1}^N (x_\alpha^i - \bar{x}_\alpha)(x_\beta^i - \bar{x}_\beta)\,.
  \end{split}\label{eq:cov_def}
\end{equation}
The correlation matrix with optimal shrinkage is given by
\begin{equation}
\label{eqn:opt_shrinkage_cov}
\rho^* = \rho(\lambda^*) = \lambda^* \mu I + (1-\lambda^*) \rho\,,
\end{equation}
where $\mu = \frac{1}{K}\Tr[\rho] = 1$ is the mean of the spectrum of $\rho$ and
the optimal shrinkage parameter $\lambda^*$ is defined to minimize the expected Frobenius norm
$||X||=\sqrt{\Tr [X\,X^T]}$ of the difference
$E\big\{\underset{\lambda}{\mathrm{min}}||\rho(\lambda)-\varrho||^2\big\}$
between the estimator $\rho^*$ and the true correlation matrix $\varrho$.
A sample estimator for the optimal shrinkage parameter is given in Ref.~\cite{LEDOIT2004365}
%\begin{equation}
\begin{align}
\lambda^* &= \frac{\mathrm{min}\{\bar{b}^2, d^2\}}{d^2}\,,\\
\bar{b}^2 &= \frac1{N^2}\sum_{n}\sum_{\alpha\beta}
    \big(y^i_\alpha y^i_\beta - \rho_{\alpha\beta}\big)^2\,,\\
d^2 &= \sum_{\alpha\beta}\big(\rho_{\alpha\beta} - \mu \delta_{\alpha\beta}\big)^2\,.
\label{eqn:opt_lambda}
\end{align}
%\end{equation}
The quantity $d^2$ estimates the dispersion of the eigenvalues of the sample correlation matrix
$\rho$, which typically has a wider spectrum and correspondingly larger (worse) condition number
compared to the true correlation matrix $\varrho$.
The optimal estimator~(\ref{eqn:opt_shrinkage_cov}) ``shrinks'' the spectrum by
emphasizing the diagonal elements and makes the matrix $\Sigma^*$ better-conditioned, resulting in
more statistically stable $\chi^2$ values in correlated fits.
Multiplying both sides of Eq.~(\ref{eqn:opt_shrinkage_cov}) by the normalization factor in
Eq.~(\ref{eq:norm_def}) yields the corresponding estimator for the covariance matrix
\begin{equation}
   \begin{split}
      \Sigma^*_{\alpha\beta} &= \sqrt{S_{\alpha\alpha}S_{\beta\beta}} \rho^*_{\alpha\beta}
      = \lambda^*\sqrt{S_{\alpha\alpha}S_{\beta\beta}} \delta_{\alpha\beta}  + (1-\lambda^*)S_{\alpha\beta} \,, \\
      \Sigma^* &= \lambda^* \text{diag}(S) + (1-\lambda^*) S\,.
   \end{split}\label{eqn:opt_sigma_norm}
\end{equation}
This shrinkage prescription is therefore equivalent to an interpolation between a fully correlated
fit with $\lambda^*=0$ (no shrinkage), and an uncorrelated fit with $\lambda^*=1$ (full shrinkage).
Although this prescription does not provide the strictly optimal $\lambda^*$ minimizing the
distance between $\Sigma^*$ and $\Sigma$, it gives a simple practical prescription for a stable and
consistent choice of the shrinkage parameter.
Optimal closeness between $\rho^*$ and $\varrho$ suggests that $\Sigma^*$ should provide an
acceptable approximation of $\Sigma$ that is better-conditioned than $S$.

%%%%%%%%%%%%%%%%%%%%%%%%%%%%%%%%%%%%%%%%%%%%%%%%%%%%%%%%%%%%%%%%%%%%%%%%%%%%%%
\subsection{VarPro $\chi^2$-minimization}\label{app:var_pro}

Fluctuations of the sample mean $G(t) = \frac{1}{N} \sum_i G_i(t)$ of an ensemble of $i=1,\dots,N$
correlation functions become Gaussian distributed as $N\rightarrow \infty$ by central limit
theorems.
The $\chi^2$ value associated with the log-likelihood of the mean correlation function becomes
\begin{equation}
  \begin{split}
    \chi^2(E,Z) &= \sum_{t,t^\prime} \left[ G(t) - \sum_n Z_n f_n(E,t) \right]
    C(t,t^\prime)\left[ G(t^\prime) - \sum_m Z_m f_m(E, t^\prime) \right],
  \end{split}\label{eq:chisq}
\end{equation}
where $f_n(E_n,t) = e^{-E_n t}$ for a two-point correlation function and more complicated
correlation functions differ only in that $f_n$ has more parameters.
The term quadratic in $Z_n Z_m$ can be turned into a sum of squares by transforming to the
eigenbasis of
\begin{equation}
  \begin{split}
    V_{nm}(E) = \sum_{t,t^\prime} f_n(E,t) C^{-1}(t,t^\prime) f_m(E,t^\prime)
  \end{split}\label{eq:Vdef}
\end{equation}
Since $V_{mn}$ is a symmetric positive-definite matrix, it can be diagonalized an orthogonal
transformation, which is equivalent to a change of variables in the likelihood function with a
trivial Jacobian, making the eigenvalues of $V_{mn}$ new independent fit variables.
The minimum of $\chi^2$ is determined by vanishing derivatives with respect to these eigenvalues.
This provides a system of  constraints that can be solved to determine the overlap factors
\begin{equation}
  \begin{split}
    Z_n = \sum_{t,t^\prime} G(t)C^{-1}(t,t^\prime) \sum_m f_m(t^\prime, E) V_{mn}(E).
  \end{split}\label{eq:zprime}
\end{equation}
This solution can now be substituted back into Eq.~(\ref{eq:chisq}) to give the VarPro $\chi^2$
function of the energies $E$ only,
\begin{equation}
  \begin{split}
    \chi_{VP}^2(E)  &= \sum_{t,t^\prime} G(t)
      \left[ C^{-1}(t,t^\prime) - \sum_{n,m} \sum_{t^{\prime\prime}} f_n(t, E) C^{-1}(t,t^{\prime\prime})
      V^{-1}_{nm} C^{-1}(t^{\prime\prime}, t^\prime) f_m(t^\prime, E)  \right] G(t^\prime).
  \end{split}\label{eq:chisqVP}
\end{equation}
The $E_n$ and $Z_n$ minimizing $\chi^2$ in Eq.~(\ref{eq:chisq}) can be obtaining by determining the
$E_n$ that minimize $\chi_{VP}^2$ in Eq.~(\ref{eq:chisqVP}) and then solving for $Z_n$ from
Eq.~(\ref{eq:zprime}).
More details and general discussion can be found in Refs.~\cite{varpro0,varpro1}.

%%%%%%%%%%%%%%%%%%%%%%%%%%%%%%%%%%%%%%%%%%%%%%%%%%%%%%%%%%%%%%%%%%%%%%%%%%%%%%
\subsection{Averaging over fits}
\label{app:stat_modavg}
Our analysis of nonperturbative renormalization and the ground-state matrix elements involves
fits over different ranges of data points.
The $\chi^2$ values of these fits cannot be directly compared due to different numbers of degrees of
freedom $N_\text{dof}$; instead, the quality of each fit $Q$ can be assessed with its $p$-value,
\begin{equation}
p = \mathrm{Prob}(\eta < \chi^2)\,,
\quad \eta \sim \chi^2_{N_\text{dof}}\,.
\end{equation}
In order to compare and average values from a family of fits as well as estimate their stochastic
and systematic uncertainties in a ``blind'' fashion, we use the $p$-value above as a proxy for the
likelihood that these fits describe data.
Thus, for any parameter $x$ extracted from a particular fit $m$ as $x_m \pm \delta x_m^\text{stat}$,
we use the combination of its statistical uncertainty (estimated with bootstrap or jackknife) and
the $p$-value of the fit $p_m$ as the weight
\begin{equation}
\label{eqn:wQweight}
w_m \propto p_m\,\big(\delta x_m^\text{stat}\big)^{-2}\,,
\end{equation}
to compute the ``global'' average value $\hat x$ and its statistical fluctuation
$\delta\hat x^\text{stat}$,
\begin{align}
\label{eqn:modavgval}
\hat{x}
  &= \langle x\rangle_{w} = \frac{\sum_m w_m x_m}{\sum_m w_m}
  = \frac{\sum_m p_m\,\big(\delta x^\text{stat}_m\big)^{-2}\,x_m}
         {\sum_m p_m\,\big(\delta x^\text{stat}_m\big)^{-2}} \,,\\
\label{eqn:modavgerrstat}
(\delta \hat x^\text{stat})^2
  &= \langle\big(\delta x^\text{stat}\big)^2\rangle_{w}
  = \frac{\sum_m w_m \big(\delta x^\text{stat}_m\big)^2}{\sum_m w_m}
  = \left(\frac{\sum_m p_m\,\big(\delta x^\text{stat}_m\big)^{-2}}{\sum_m p_m}\right)^{-1}\,.
\end{align}
while the weighted deviation from the total average serves as the estimate of the systematic
uncertainty $\delta x^\text{sys}$,
\begin{equation}
\label{eqn:modavgerrsys}
(\delta \hat x^\text{sys})^2
  = \langle(x - \hat x)^2\rangle_w
  = \frac{\sum_m w_m (x_m - \hat x)^2}{\sum_m x_m}
\,.
\end{equation}

The rationale for using the weight~(\ref{eqn:wQweight}) is that it penalizes both bad fits (small
$p_m$) and unconstraining fits (large statistical uncertainty $\delta x^\text{stat}$ typical of
overfitting).
The $(\delta x^\text{stat})^{-2}$ factor is motivated by similarity to a weighted average
of independent data.
However, since all these fits are performed on the same data set, the ``global'' stochastic
uncertainty is computed as the inverse-squared \emph{average} (instead of the sum) of individual fit
uncertainties.
The resemblance is especially evident if the ``likelihoods'' of all the fits are the same,
$p_m=\text{const}$.
Thus, the estimator~(\ref{eqn:modavgerrstat}) is also somewhat robust with respect to having similar
or (nearly-)duplicated fits in the set.

\end{document}